%% file: main.tex
\documentclass[11pt]{article}
\input{preamble}
\usepackage{booktabs}   
\usepackage{graphicx}   
\usepackage{amsmath}    
\usepackage{authblk}
\usepackage{CJKutf8}

\title{\LARGE\bfseries The Male Abstraction Bias in Internet Culture and Large Language Models}
\author[1,*]{Douglas Guilbeault}
\author[2]{Bhargav Srinivasa Desikan}

\affil[1]{Stanford University, Stanford, CA, USA}
\affil[2]{Oxford Internet Institute, Oxford, UK}
\date{}

\begin{document}
\maketitle

\begin{NoHyper}
\renewcommand{\thefootnote}{*}
\footnotetext{Corresponding author: dguilb@stanford.edu}
\end{NoHyper}

\begin{abstract}
\noindent

Conceptual abstraction and gender are two equally central aspects of the human mind, yet they have traditionally been studied in isolation from one another, as though they were fundamentally separate. Here, we find gender and abstraction are systematically linked throughout the space of concepts spanning single (\emph{N} = 40,000) and multi-word expressions (\emph{N} = 67,000): across billions of words and millions of images on the internet ---  and across numerous measurement techniques, including human and AI classifications --- more abstract concepts are more associated with men, a pattern we show replicates in languages around the world. This male abstraction bias not only holds across semantic domains, but it also maps onto grammatical structure: verbs and adverbs are more abstract and more male-coded, revealing a system-wide trace of stereotypes accentuating agentic qualities in men. Furthermore, we show this bias readily emerges in the free associations humans and AIs produce in controlled online tasks, indicating its psychological internalization and its algorithmic reproduction. These findings identify a latent structure in conceptual space that unifies a range of stereotypes: for example, framings of women as more embodied, emotional, and warm, versus men as more cognitive, rational, and competent; as well as erroneous views of men as better at mathematics, science, and engineering. More broadly, by analyzing an unprecedently large, multimodal sample of online data, we provide culture-wide evidence of gender as a core organizing principle shaping conceptual space far beyond the characterization of people and social roles.

\vspace{0.6em}
\noindent\textbf{Keywords:} gender; abstraction; bias; stereotypes; large language models; multimodal; images
\end{abstract}

\section*{Significance}

Gender stereotypes are primarily studied as biased beliefs about people and social roles: who is warm, who is competent, who belongs in science. Here, we show that gender runs far deeper, organizing the structure of concepts themselves. Across billions of words and millions of images online, spanning 100,000 common words and phrases, we find that more abstract (concrete) concepts are systematically more associated with men (women). This male abstraction bias replicates across semantic domains and languages, maps onto fundamental grammar (verbs and adverbs are more abstract and male-coded), and surfaces in the free associations of humans and large language models. These findings reveal a hidden structure in conceptual space that unifies many stereotypes, a structure AI is learning and reproducing.

\section*{Main}

Does social identity shape the foundational structure of concepts? Traditionally, cognitive science has examined the structure of human concepts by identifying computational principles and constraints operating in the individual's mind \cite{rosch1976basic,rosch1978principles,murphy2002, tenenbaum2011, griffiths2026}, treating social factors as inputs that shape concept learning --- e.g., by influencing exposure to and adoption of concepts --- rather than as constituents of conceptual structure itself. This is especially the case when considering the most foundational dimensions of concepts like abstractness, taxonomic hierarchy, and statistical inference. For example, the abstractness of concepts is often defined by how easy it is to perceive and physically interact with the referents for a given category (e.g., ``chairs'' are more concrete and perceivable, whereas ``beliefs'' and ``nations'' are more abstract), such that decades of work have argued that various individual-level factors relating to cognitive efficiency skew concepts toward an optimal middle-range (``basic level'') of abstraction \cite{rosch1976basic, rosch1978principles, murphy1982basic, corter1992explaining, iordan2015basic}. While recent accounts explore the social grounding of abstract concepts, they focus on social interaction as the route by which abstract concepts are acquired and stabilized\cite{borghi2025}, rather than on whether social information (e.g., relating to identity) shapes conceptual abstraction itself. By contrast, aspects of human cognition relating to personal identity, such as gender and race, are often viewed as socially constructed and historically contingent, and as thus belonging to emergent communication and cultural dynamics, rather than to the foundations of conceptual structure\cite{wood2002,ridgeway2011}. However, recent work has discovered numerous ways in which gender influences various cognitive processes, from logical reasoning and emotion to the perception of everyday objects\cite{bian2017,plant2000,wardle2022}, suggesting that gender may play a more foundational role in conceptual structure than has been previously recognized \cite{martin2021}. Here, we investigate whether abstraction and gender are systematically and structurally linked throughout the space of human concepts. 

A key challenge in the study of concept abstractness is the disconnect between formal definitions and human judgements. Initially, cognitive science defined abstractness via hierarchical generality\cite{collins1969,rosch1976}; a superset like ``animal'' is more abstract than its subset ``dog'', which is more abstract than its subset ``border collies'', with each step down the tree restricting the range of referents. However, a large body of work has since found that human judgments depart from this hierarchical definition of concept generality\cite{bolognesi2020,muraki2026,iliev2017}. In a now classic study, Brysbaert et~al.\cite{brysbaert2014} recruited many human participants to evaluate the 40,000 most common concepts in the English language in terms of how concrete or abstract they are. Not infrequently, general concepts like ``animal'' were rated as concrete, whereas specific concepts like ``string theory'' (a narrow subset of all scientific theories) were evaluated as abstract. This led cognitive scientists to distinguish hierarchical generality (`concept specificity') from the concreteness--abstractness continuum\cite{bolognesi2020,muraki2026}. In the current view, concrete concepts are those whose referents are easily perceivable and imageable (e.g., a ``rock'' or a ``house''), while abstract concepts lack clear and perceivable sensory structure (e.g., ``justice'' and ``knowledge'')\cite{brysbaert2014,paivio1986,barsalou2005,borghi2017}. The limitations of relying on human subjective judgments notwithstanding\cite{pollock2018}, research spanning over a decade has found that judgments of concept concreteness--abstractness are not only reliable across raters and replicable across languages\cite{brysbaert2014dutch,thompson2018}, and closely approximated by large language models\cite{martinez2025}, but are also predictive of several conceptual features and processes, including language acquisition\cite{gleitman2005}, sentence processing\cite{schwanenflugel1983}, emotion\cite{kousta2011}, metaphor\cite{lakoff1980,jamrozik2016}, and persuasion\cite{hansen2010,packard2021}.

Despite the minimal cross-pollination between the study of conceptual abstraction and gender, recurring themes across widespread stereotypes suggest that concept concreteness and abstractness may be deeply tied to gender associations. A number of popular gender stereotypes skew toward associating women with more concrete, embodied qualities, such as emotions and interpersonal warmth\cite{plant2000,eagly1984,fiske2002,ellemers2018,leshin2025}, while associating men with more abstract qualities like reasoning, problem solving, and competence\cite{bian2017,fiske2002,ellemers2018,leslie2015}. This generalizes to stereotypes related to occupational sorting in which men are (falsely) assumed to be more competent in math and STEM disciplines than women\cite{nosek2002,hyde2008,mossracusin2012}, skewing people toward assuming men are more qualified for highly cognitive, knowledge work. Importantly, prior work has focused on a definition of stereotypes as directly connected to social roles, with a focus on domain-specific biases and patterns of discrimination (e.g., hiring biases in STEM fields). However, this leaves open the question of whether stereotypes are primarily byproducts of role dynamics in these social contexts (e.g., normative biases in the kinds of work women and men are expected and hired to do), or whether these domain-specific role judgments are extensions of deeper, more general gender biases in the structure of concepts themselves, and specifically in systematic differences between concrete and abstract concepts. This question is important because while stereotyped role dynamics may be possible to shift in particular contexts (e.g., by increasing women's leadership in organizations \cite{lawson2022hiring}), if gender is structurally entangled with abstraction as a fundamental structure throughout human concepts, including our most basic and common notions, this suggests the challenge of achieving enduring changes in gender inequality may require far more fundamental shifts in human thought and culture. 

In this direction, a developing stream of research in social psychology has begun to identify ways in which abstractness may be related to gender. However, much of this literature is based on construal level theory (CLT)\cite{trope2010,semin1988} which defines abstractness as psychological distance -- e.g., via the level of spatial, temporal, or social distance between an observer and what is being described -- and thus is not directly related to or engaged with questions about concept abstractness and conceptual structure specifically, since it relates more to the stance of the observer rather than to the concepts themselves. Nevertheless, using CLT, Joshi et~al.\cite{joshi2020} show---across a meta-analysis of experiments, a corpus of blog posts, and the floor speech of members of the US Congress---that men tend to describe their experiences and work in more abstract language than women. Dodson et~al.\cite{dodson2025} similarly draw from CLT across six preregistered studies to show that people implicitly associate women's names with low-construal terms and men's names with high-construal terms, and that they describe women as more concrete than abstract and men as more abstract than concrete, both in general and across 48 occupations---although in one study women were also described as more abstract than men. Otherwise, in a recent opinion piece, Bailey and Leshin\cite{bailey2026} argue that, absent an explicit gender frame, people represent women specifically as women but represent men as generic people, such that men come to be associated with a wider variety of referents and attributes\cite{bailey2022}; although the authors do not frame this asymmetry in terms of abstractness, it maps onto concept specificity---the breadth of referents a category subsumes---which is related to, but empirically separable from, rated concreteness\cite{bolognesi2020,muraki2026}. Notably, Bailey and Leshin\cite{bailey2026} evaluate the construal-level stereotypes\cite{dodson2025} from prior work as an instance of `balanced' gender representation---women as one thing, men as its complement---that is distinct from the categorical asymmetry they propose, with women represented as one thing and men as many. 

Overall, despite these suggestive developments, the link between gender and abstraction at the level of conceptual structure remains unclear. In addition to the fact that CLT focuses on observers and not concepts, the methodological variation and modest scale of prior studies limit their ability to examine whether gender is systematically entangled with conceptual structure. Joshi et~al.\cite{joshi2020} focus on the supply side---the abstraction of the language that men and women produce---rather than on the gendered associations that concepts themselves carry, and their finding that the gender gap closes when audience construal distance is made salient leaves open whether abstraction is male-coded in the semantics of concepts or only for the people using these concepts to particular audiences. Bailey and Leshin's study\cite{bailey2026} is an opinion piece without new empirical analyses or data; and Dodson et al.'s studies\cite{dodson2025} rely on survey samples of a few hundred participants who describe or rate individual men and women, with a primary focus on descriptions of men and women in different occupations. Altogether, it remains open whether gendered abstraction merely reflects (i) supply-side differences in how women and men speak, (ii) differences in psychological distance unrelated to concepts themselves, (iii) role dynamics shaping the representation of men and women in limited, specific contexts (e.g., the representation of occupations and biases relating to labor market sorting), (iv) a difference in the breadth and specificity of the categories `men' and `women', unrelated to concreteness-abstractness specifically, or (v) a far more general and fundamental gendering of how concrete and abstract concepts are represented across domains.

Another feature constraining prior work is its focus on textual data as the modality through which gender associations and concept abstractness are measured. Yet, examining the concrete vs. abstract dimension of concepts benefits from examining more immediately embodied communication modalities, such as images, in which concepts are frequently represented through concrete, visual anchors. Indeed, concreteness is often defined with respect to the imageability of concepts. Decades of work in cognitive science have not only shown that human conceptual knowledge is richly multimodal, involving both symbolic and image-based representations\cite{paivio1986,barsalou1999,barsalou2008}, but also more recent work demonstrates that images are an especially psychologically potent modality that amplifies gender stereotypes \cite{guilbeault2024,vlasceanu2022,guilbeault2025}. A full treatment of how gender is linked to the concreteness-abstractness dimension of concepts should therefore incorporate analyses of more directly embodied and concrete representations of concepts like images. 

In this study, we address this gap by conducting a large-scale examination of the connection between gender and conceptual abstractness across the entire space of human concepts using billions of words and millions of images from the internet, as well as using behavioral response data of humans and large language models in controlled online tasks \cite{caliskan2017,garg2018,lewis2020,charlesworth2021, dedeyne2019,abramski2025}. The scale of our approach -- spanning conceptual domains and communication modalities -- allows us to examine not only whether gender and concept abstractness are associated across semantic domains, but also to examine which aspects of concepts, e.g., their grammatical form, account for underlying patterns organizing these associations. By examining the gender associations of the same concepts across both text and image, we are able to further test whether the male abstraction bias manifests both in how concepts are described, and also in how they are captured by the concrete, visual representation of concepts, which includes the use of concrete referents in images to metaphorically depict more abstract and difficult to perceive concepts \cite{lakoff1980, jamrozik2016, guilbeault2020}. 

We begin by comparing crowdsourced concreteness judgments of the 103,428 most common concepts in English (comprising single \cite{brysbaert2014} and multi-word expressions \cite{muraki2023}) with their corresponding gender associations (female or male), which we identified using the `geometry of culture' method \cite{kozlowski2019, guilbeault2024} applied to the co-occurrence patterns among billions of words on the internet according to 12 distinct language models, spanning static embedding models and frontier generative models, both closed and open source (Supplementary Table~\ref{tab:embedding-models}; see `Language Models' and `Measuring Gender Associations' in Materials and Methods). We find that increasingly abstract concepts carry increasingly male gender associations across all models, and that this pattern is stable across semantic domains according to WordNet's model of English concepts (`Identifying Semantic Domains' in Materials and Methods). Given our ability to decompose the male abstraction bias across a large corpus of word forms, we further identify a clear underlying pattern across conceptual domains, namely that the male abstraction bias is tightly mapped onto grammatical word form: regardless of topic domain, \emph{verbs} and \emph{adverbs} are more abstract and more male-coded compared to \emph{nouns}, suggesting that established stereotypes ascribing more agentic, action-oriented qualities to men \cite{eagly1984gender, diekman2000stereotypes,rudman2001prescriptive, abele2007agency, koenig2011leader, hentschel2019multiple, stuhler2024gender} may go beyond role dynamics in specific contexts and may reach into the grammatical foundations of language. Moreover, these findings are not unique to English. Supplementary analyses show that the male abstraction bias replicates across distinct language families that vary in whether they explicitly grammaticalize gender. We replicate our findings by analyzing extensive concreteness norms provided by native speakers of Mandarin Chinese, Persian, Dutch, and Estonian, which we map to gender associations identified using embedding models trained specifically on each language (Supplementary Sections~\ref{si:dutch} to~\ref{si:persian}, Supplementary Figs.~\ref{fig:S-dutch} to~\ref{fig:S-persian} and Supplementary Tables~\ref{tab:S-dutch-spec} to~\ref{tab:S-persian-poles}). Smaller sets of concreteness norms for five further languages point in the same direction (Supplementary Section~\ref{si:forest}, Supplementary Fig.~\ref{fig:S-forest} and Supplementary Table~\ref{tab:S-forest}). 

Next, we test whether this clear correlation between concept abstractness and masculine association holds beyond the textual modality by examining its presence in a large sample of online images collected for this study. This multimodal approach is a valuable test of our theory given widespread concerns that textual representations are a limited measurement of both cultural associations and human thought \cite{fedorenko2024language, guilbeault2024}. This is, based on the present review of the literature, the first large-scale examination of the relationship between concreteness judgments and image-based associations. For this analysis, we collected roughly 300 Google images for each of the 40,000 concepts associated with crowdsourced human concreteness judgments in Brysbaert et~al.'s study\cite{brysbaert2014}, yielding 11,346,925 images. We then applied a recent 11 billion parameter, open-source large language model (Llama-3.2-11B-Vision-Instruct) to classify all of the objects in each of the images, including the number of human faces in each image, as well as their associated gender (male or female) \cite{meta2024llama32, grattafiori2024llama3} (`Image Data' and `Image Classification' in Materials and Methods). 

Using this approach, we test whether people are more likely to appear as concrete anchors in online images for increasingly abstract concepts, and moreover, whether people are more likely to be men when functioning as concrete anchors. The evidence shows clearly that the probability of observing people in online images increases for more abstract concepts, and -- following from the established over-representation of men online \cite{guilbeault2024} -- men are increasingly more likely to appear in association with abstract concepts. This sheds mechanistic light on a statistical driver behind the male abstraction bias: as visual representations of people skew male -- and as visual representations of abstract concepts skew toward displaying people -- by statistical sampling alone, abstract concepts become more associated with men. We go further yet to reveal the ability for image analyses to provide a novel empirical test of an outstanding hypothesis in current computational work on gender stereotypes; building on Bailey and Leshin's opinion piece\cite{bailey2026}, we present an empirical test of the hypothesis that women are represented in a more unitary manner, while men are represented with greater breadth and diversity of association; specifically, by examining whether the images associated with more male-coded concepts (according to image and textual measures of gender) are associated with a wider variety of objects (referents) compared to female-coded concepts. Indeed, across all techniques for measuring the gender association of a concept, more abstract concepts are not only more male, but are also associated with more diverse sets of visual referents. 

Given that the analyses presented above rely on large-scale observational data, an open question is whether this bias is instantiated and measurable in the response patterns of individual humans and AI agents. We conclude by testing whether the male abstraction bias is detectable in existing experimental data collected by separate teams to examine the free associations that humans and large language models provided when given a seed concept and requested to state, with no constraint, three concepts that come to mind \cite{dedeyne2019, abramski2025}. Across both datasets, we find that humans and all the large language models involved are significantly more likely to generate male-coded (female-coded) responses when asked to free associate for more abstract (concrete) seed words, despite no explicit prompting or priming relating either to gender or to abstractness. This suggests that the male abstraction bias is not only a statistical pattern observed in large social datasets, but is also implicitly encoded in the spreading activation of free association patterns in human and AI psychology. The implications for the algorithmic amplification of the male abstraction bias online -- in light of the increasingly male-coded discourse surrounding the framing and theorizing of AI itself -- are discussed. 

\input{results}

\input{discussion}

\input{methods}

\bibliography{references}

\clearpage
\beginsupplement
\input{supplementary}

\end{document}

%% file: preamble.tex
\usepackage[T1]{fontenc}
\usepackage[utf8]{inputenc}
\usepackage{lmodern}
\usepackage{microtype}

\usepackage[margin=1in]{geometry}
\usepackage{setspace}
\usepackage{lineno}
\usepackage{amsmath,amssymb}
\usepackage{booktabs}
\usepackage{tabularx}
\usepackage{graphicx}
\graphicspath{{figures/}}
\usepackage[font=small,labelfont=bf]{caption}

\usepackage[super,sort&compress,comma]{natbib}
\usepackage{xcolor}
\PassOptionsToPackage{hyphens}{url}  
\usepackage[colorlinks=true,allcolors=blue!50!black]{hyperref}

\DeclareUrlCommand\code{\urlstyle{tt}}

\newcommand{\beginsupplement}{%
  \setcounter{section}{0}%
  \setcounter{figure}{0}%
  \setcounter{table}{0}%
  \setcounter{equation}{0}%
  \renewcommand{\figurename}{Supplementary Fig.}%
  \renewcommand{\tablename}{Supplementary Table}%
  \renewcommand{\theequation}{S\arabic{equation}}%
  \renewcommand{\theHsection}{S\arabic{section}}%
  \renewcommand{\theHfigure}{S\arabic{figure}}%
  \renewcommand{\theHtable}{S\arabic{table}}%
  \renewcommand{\theHequation}{S\arabic{equation}}%
}

%% file: results.tex
\input{stats_fig1}   

\providecommand{\VLM}{Llama-3.2-11B-Vision-Instruct}

\newenvironment{herefigure}
  {\par\addvspace{\bigskipamount}\captionsetup{type=figure,parbox=none,singlelinecheck=off}\centering}
  {\par\addvspace{\bigskipamount}}

\section*{Results}

\subsection*{The Male Abstraction Bias in Online Text}

We began by examining the relationship between the gender association of a concept and its
concreteness in text. Concreteness was measured using human judgements, on a scale from 1 (abstract)
to 5 (concrete), for the 39,953 most common words in English\cite{brysbaert2014} and for a recent
extension of these norms to 62,889 multiword phrases\cite{muraki2023} (`Concreteness Norms' in
Materials and Methods). The gender association of
each term was measured separately in each of 12 language models that vary in their architecture,
training data and year of release (see `Language Models' and `Measuring Gender Associations' in
Materials and Methods).
Figure~\ref{fig:text}a shows that terms with more masculine associations are systematically more
abstract. Across models, the correlation between the masculine association of a term and its
concreteness was negative (mean $r =$~\FigOneAAllMeanR; 95\% confidence interval (CI) across
models $=$ \FigOneAAllCIlo\ to \FigOneAAllCIhi; range across models $=$ \FigOneAAllMinR\ to
\FigOneAAllMaxR; $t(\FigOneAAllDf) =$~\FigOneAAllT; \textit{P}~\FigOneAAllP; two-tailed one-sample
$t$-test across models), and it was negative and significant after Bonferroni correction in
\FigOneAAllKSigNeg\ of \FigOneAAllKTot\ models ($n =$ \FigOneAAllNmin\ to \FigOneAAllNmax\ terms per
model; two-tailed Pearson's correlations; Supplementary Table~\ref{tab:S-fig1-human}; `Statistical
Analysis of Textual Data and Robustness Tests' in Materials and Methods). Averaged
across models, terms in the most feminine decile had a mean concreteness of \FigOneAAllDecOne,
compared with \FigOneAAllDecTen\ for terms in the most masculine decile. The male abstraction
bias replicates when separately analyzing single words (mean $r =$~\FigOneAWrdMeanR; 95\% CI $=$ \FigOneAWrdCIlo\ to \FigOneAWrdCIhi; negative in
\FigOneAWrdKNeg\ of \FigOneAWrdKTot\ models, with one small reversal) and phrases (mean $r =$~\FigOneAPhrMeanR; 95\% CI $=$ \FigOneAPhrCIlo\ to
\FigOneAPhrCIhi; negative in \FigOneAPhrKNeg\ of \FigOneAPhrKTot\ models) (Supplementary
Section~\ref{si:words-phrases}). 

These associations are robust to lexical properties that are correlated with concreteness.
Controlling for word frequency left the correlation for single words essentially unchanged (mean $\beta = -0.091$; 95\% CI across models $= -0.143$ to $-0.039$), and it remained negative (i) after controlling for word frequency, length, polysemy, age of acquisition, valence, arousal, dominance
and part of speech (mean $\beta = -0.099$; 95\% CI $= -0.143$ to $-0.054$; $t(11) = -4.90$;
\textit{P}~$= 4.7\times10^{-4}$; negative in 11 of 12 models; $n = 13{,}069$ to $13{,}082$ words per model), (ii) when the analysis was restricted to reliably rated words, within each quintile of word frequency, and (iii) after excluding person-denoting nouns, lexically gendered terms and function words (Supplementary Section~\ref{si:text-covariates} and Supplementary Tables~\ref{tab:S-text-cov-words} to~\ref{tab:S-text-composition}). For phrases, the frequency, length and part of speech of the constituent words accounted for roughly half of the association, which nonetheless remained negative in 11 of 12 models with all controls included (mean $\beta = -0.106$; 95\% CI $= -0.150$ to $-0.062$).

\begin{herefigure}
  \centering
  \includegraphics[width=\textwidth]{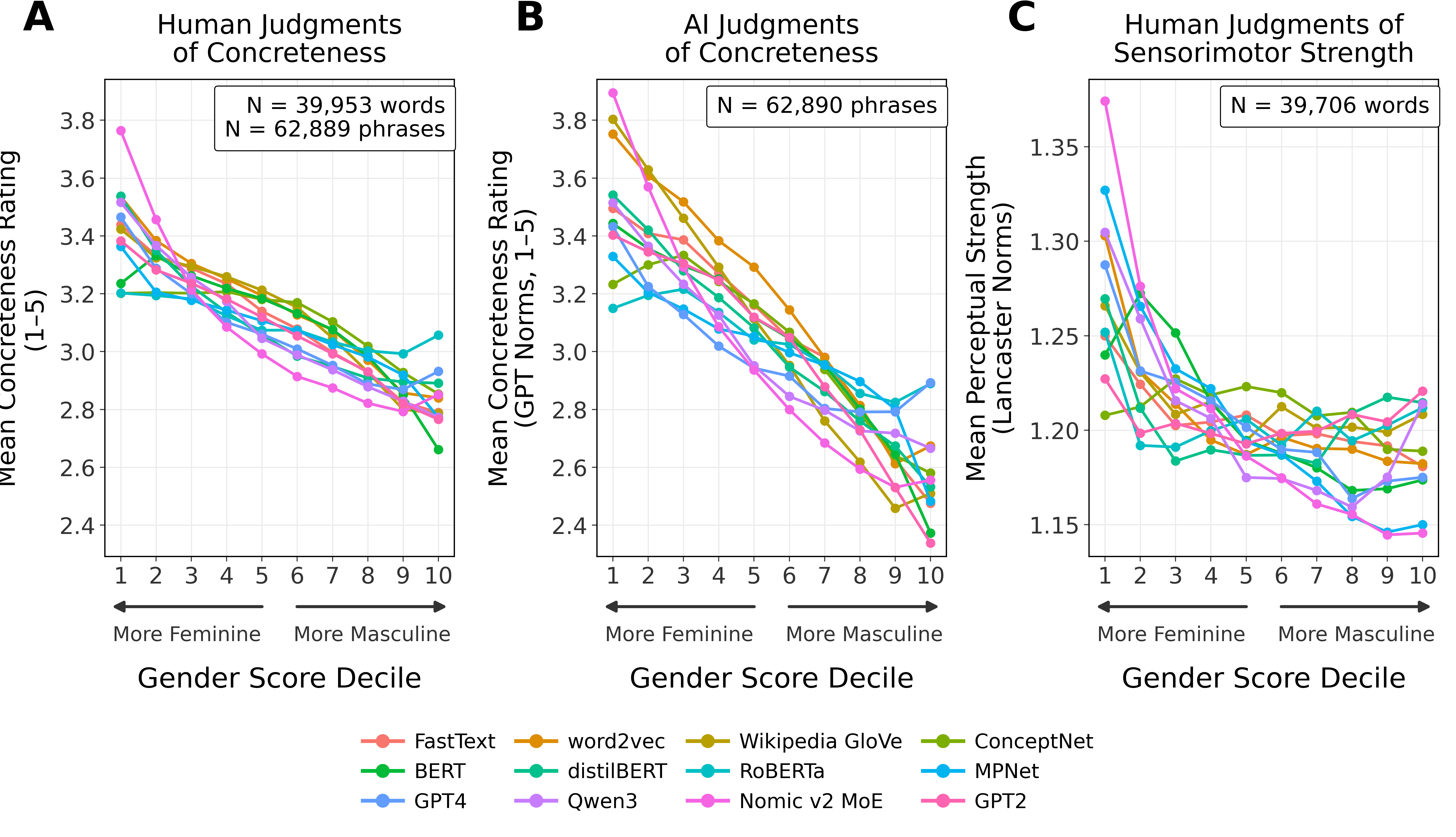}
  \caption{\textbf{More masculine words and phrases are more abstract across 12 language models.}
  \textbf{a}--\textbf{c}, Mean rating of terms within each decile of gender association
  (1, most feminine; 10, most masculine). Deciles were computed separately within each language
  model (colours), and data points show the mean rating of all terms in a decile.
  \textbf{a}, Human judgements of concreteness (1, abstract; 5, concrete) for $n = 39{,}953$
  words\cite{brysbaert2014} and $n = 62{,}889$ phrases\cite{muraki2023}, pooled.
  \textbf{b}, Concreteness ratings generated by GPT\cite{martinez2025} for $n = 62{,}890$ phrases,
  for all 12 models.
  \textbf{c}, Human judgements of perceptual strength from the Lancaster Sensorimotor
  Norms\cite{lynott2020} (mean rating across six perceptual modalities; 0--5) for the $n = 39{,}706$
  words covered by the norms. Sample sizes
  give the number of rated terms; the number of terms scored by each model, and per-model
  correlations with 95\% confidence intervals, are reported in Supplementary
  Tables~\ref{tab:S-fig1-human} and \ref{tab:S-fig1-lanc}.}
  \label{fig:text}
\end{herefigure}

Notably, these results do not depend on the source of the concreteness judgements.
Figure~\ref{fig:text}b replicates our analysis using concreteness ratings generated by
GPT\cite{martinez2025}, which are available for the phrases only ($n = 62{,}890$). GPT's ratings
closely track human ratings of the same phrases ($r = 0.81$; 95\% CI $= 0.81$ to $0.82$;
\textit{P}~$< 2.2\times10^{-16}$; $n = 62{,}889$ phrases; two-tailed Pearson's correlation), and they
reproduce the male abstraction bias across models (mean $r =$~\FigOneBPhrMeanR; 95\% CI
$=$ \FigOneBPhrCIlo\ to \FigOneBPhrCIhi; $t(\FigOneBPhrDf) =$~\FigOneBPhrT;
\textit{P}~\FigOneBPhrP; negative and significant after Bonferroni correction in
\FigOneBPhrKSigNeg\ of \FigOneBPhrKTot\ models; $n =$ \FigOneBPhrNmin\ to \FigOneBPhrNmax\ phrases per
model). 

As an additional robustness test, we examined whether these results hold for human judgments along
a related but distinct dimension of conceptual concreteness: the Lancaster Sensorimotor
Norms\cite{lynott2020}. Whereas concreteness ratings capture whether the referent of a concept can
be experienced through the senses at all, the Lancaster norms capture how strongly a concept is
experienced through each of six perceptual modalities (vision, hearing, touch, taste, smell and
interoception), which we averaged into a single measure of perceptual strength. Perceptual strength is only moderately correlated with people's concreteness judgments ($r = 0.32$; 95\% CI $= 0.31$ to $0.33$;
\textit{P}~$< 2.2\times10^{-16}$; $n = 39{,}706$ words), providing a meaningfully distinct and complementary analysis. Because the norms were only collected for single words, this analysis is restricted to the 39,706 words in the concreteness norms that were rated in the Lancaster norms. Figure~\ref{fig:text}c shows that words with more masculine associations are also lower in perceptual strength. The association is smaller than for concreteness, as expected given the modest correlation between the two measures, but it is consistent across models (mean $r =$~\FigOneCWrdMeanR; 95\% CI $=$ \FigOneCWrdCIlo\ to \FigOneCWrdCIhi; $t(\FigOneCWrdDf) =$~\FigOneCWrdT; \textit{P}~\FigOneCWrdP; negative in \FigOneCWrdKNeg\ of \FigOneCWrdKTot\ models and significantly so after Bonferroni correction in \FigOneCWrdKSigNeg; $n =$ \FigOneCWrdNmin\ to \FigOneCWrdNmax\ words per model). Words in the most feminine decile had a mean perceptual strength of \FigOneCWrdDecOne, compared with \FigOneCWrdDecTen\ for words in the most masculine decile. These results replicate when we extend the norms to the 62,321 phrases by assigning each phrase the mean perceptual strength of its constituent words, both for the phrases alone (mean $r =$~\FigOneCPhrMeanR; 95\% CI $=$ \FigOneCPhrCIlo\ to \FigOneCPhrCIhi; negative in \FigOneCPhrKNeg\ of \FigOneCPhrKTot\ models) and for words and phrases pooled (mean $r =$~\FigOneCAllMeanR; 95\% CI $=$ \FigOneCAllCIlo\ to \FigOneCAllCIhi; Supplementary Section~\ref{si:lancaster} and Supplementary Table~\ref{tab:S-fig1-lanc}).

Finally, we conducted supplementary analyses showing that the male abstraction bias replicates in other languages for which extensive human concreteness judgments from native speakers are publicly available at sufficient scale: Dutch\cite{brysbaert2014dutch}, Estonian\cite{proos2023}, Mandarin Chinese\cite{chan2024,song2021} and Persian\cite{nemati2026}, with gender associations measured in word vectors trained on each of these languages\cite{honnibal2020spacy,grave2018} (Supplementary Sections~\ref{si:dutch} to~\ref{si:persian}, Supplementary Figs.~\ref{fig:S-dutch} to~\ref{fig:S-persian} and Supplementary Tables~\ref{tab:S-dutch-spec} to~\ref{tab:S-persian-poles}). In every language, terms with more masculine associations are rated as more abstract ($r = -0.11$ to $-0.27$; all $P < 2.2 \times 10^{-16}$), and the association survives the same controls and exclusions as in English. While these robustness tests are not sufficient for arguing that the male abstraction bias is a universal pattern across languages, they are sufficient for demonstrating based on available data that the male abstraction bias is not idiosyncratic to the English language and replicates in qualitatively distinct language families that vary along key dimensions, such as whether gender is explicitly grammaticalized in the language's morphology. Smaller sets of native-speaker concreteness norms exist for five further languages -- Turkish, Italian, French, Portuguese, and Indonesian -- and supplementary analyses show that all data present patterns consistent with the male abstraction bias, again while comparing against gendered associations derived from language-specific embedding models (Supplementary Section~\ref{si:forest} and Supplementary Table~\ref{tab:S-forest}).

\subsection*{Cross-domain Stability and Grammatical Grounding}

\begin{herefigure}
  \centering
  \includegraphics[width=\textwidth]{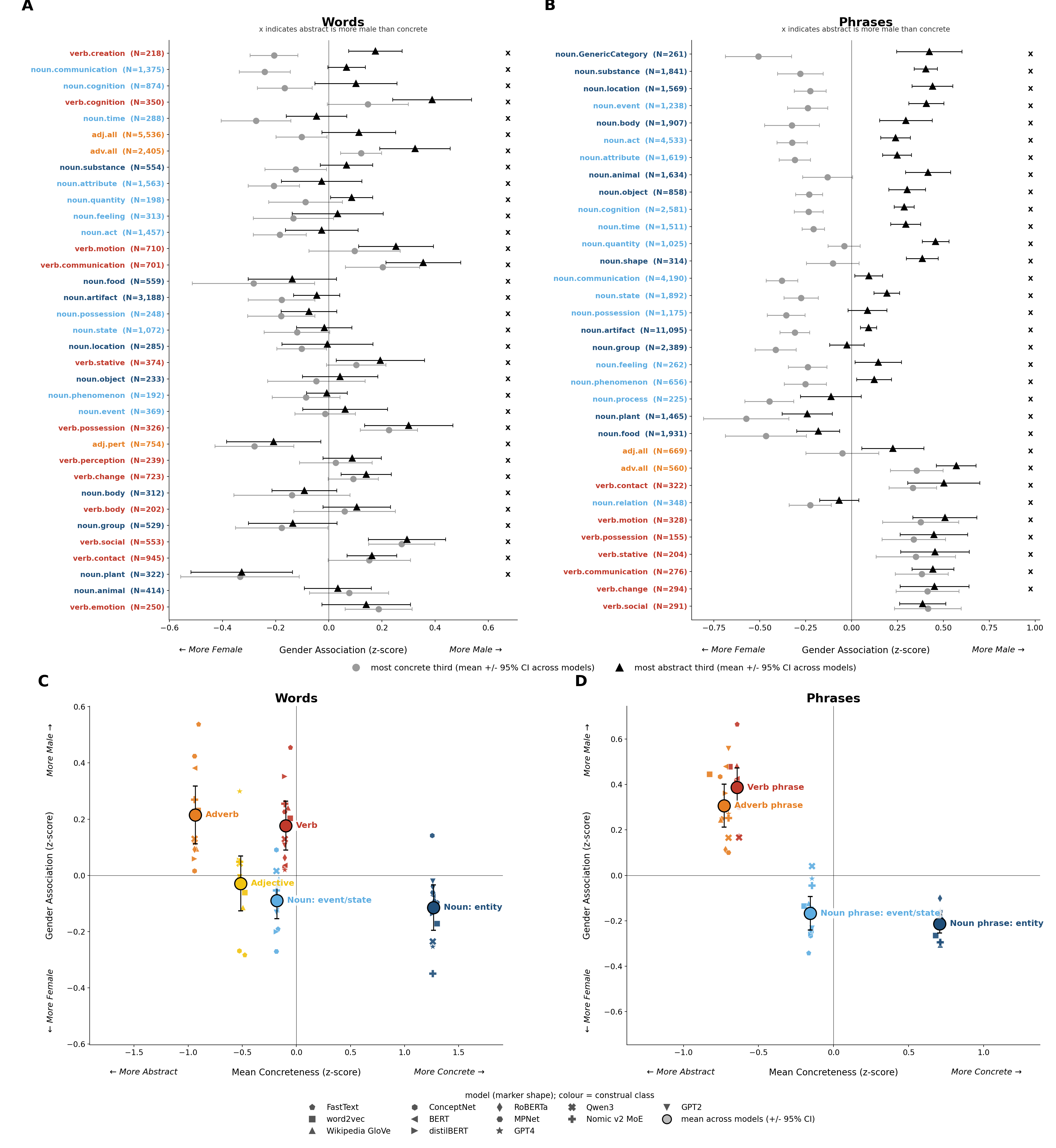}
  \caption{\textbf{The male abstraction bias holds within semantic domains and varies with word form.}
  \textbf{a},\textbf{b}, Within-domain contrasts for words (\textbf{a}; 35 WordNet domains;
  $n = 31{,}591$ words) and phrases (\textbf{b}; 33 domains; $n = 56{,}879$ phrases). Within each
  domain, terms were split into thirds by their human concreteness rating. Grey circles show the
  mean gender association of the most concrete third and black triangles that of the most abstract
  third, averaged with equal weight across the 12 language models of Fig.~\ref{fig:text}; error bars
  show 95\% confidence intervals across models. Domains are ordered by the size of the difference,
  and an `x' marks domains in which the abstract third is more male-associated than the concrete
  third (33 of 35 domains for words; 32 of 33 for phrases; no domain shows a significant difference
  in the opposite direction). The number of terms in each domain is given in parentheses, and label
  colours give the predominant word form in each domain (red, verbs; orange, adverbs and adjectives;
  light blue, nouns for events and states; dark blue, nouns for entities).
  \textbf{c},\textbf{d}, Mean concreteness (horizontal axis) and mean gender association (vertical
  axis) of each word form across all domains, for words (\textbf{c}) and phrases (\textbf{d}). Small
  markers show each language model (marker shape); large points show the mean across models, with
  error bars showing 95\% confidence intervals across models. Gender association and concreteness
  are $z$-scored (gender association within each model). These analyses exclude function words,
  lexically gendered terms and nouns that denote persons (`Identifying Semantic Domains' in Materials
  and Methods).}
  \label{fig:domains}
\end{herefigure}

The above findings provide evidence of a stable male abstraction bias across the most common
words and phrases in the English language. However, these analyses aggregate over a wide range
of concepts, topics and word forms, which raises the question of whether this pattern holds across
conceptual domains or is instead driven by a subset of them. For this reason, we decomposed our results
by semantic domain. We matched each word and phrase in the concreteness norms to
WordNet\cite{miller1995}, a canonical lexical database of English, and assigned it the semantic
domain that WordNet records for its primary sense (for example, \textit{noun.food} or
\textit{verb.communication}); words were matched on their dominant part of speech, and phrases
through their own WordNet entry or, failing that, through their head noun (see `Identifying
Semantic Domains' in Materials and Methods). To ensure that these analyses capture the gendering
of concepts rather than of people, we excluded function words, lexically gendered terms (such as
`waitress') and nouns that denote persons, leaving 31,591 words in 35 domains and 56,879 phrases
in 33 domains. Within each domain, we compared the gender association of the most abstract third
of terms with that of the most concrete third, averaging with equal weight across language models.

Figure~\ref{fig:domains}a,b reveals a stable pattern. For words, the most abstract
third of a domain was more male-associated than its most concrete third in 33 of 35 domains
(\textit{P}~$= 3.7\times10^{-8}$; two-tailed binomial sign test; median difference across domains
$= 0.10$~s.d.; \textit{P}~$< 0.05$ in 15 domains; two-tailed paired $t$-tests across 12 models).
For phrases, this held in 32 of 33 domains (\textit{P}~$= 7.9\times10^{-9}$; median difference
$= 0.40$~s.d.; \textit{P}~$< 0.05$ in 29 domains; 12 models). In no domain, for words or phrases,
was the most concrete third significantly more male-associated. The pattern spans domains
describing physical entities (for example, \textit{noun.substance}: $0.07$~s.d.\ for the abstract
third versus $-0.12$~s.d.\ for the concrete third), events and states (\textit{noun.communication}:
$0.07$ versus $-0.24$) and actions (\textit{verb.creation}: $0.18$ versus $-0.21$). This suggests
that our findings are not driven by particular domains, but instead reflect a system-wide bias
towards associating more abstract concepts with masculine connotations, regardless of topic. Supplementary Section~\ref{si:domain-examples} illustrates this pattern at the level
of individual words: for three example domains (acts of communication, general adjectives and
artifact nouns) and three embedding models, it plots the continuous within-domain association and
highlights the individual words that anchor its two poles---for example, in the domain of acts of communication, we find \emph{crying}, \emph{singing}, and \emph{whisper} at the female--concrete pole against \emph{justify},
\emph{condone} and \emph{vindicate} (all explicitly related to epistemic work) at the male--abstract pole (Supplementary
Fig.~\ref{fig:S-domain-examples} and Supplementary Table~\ref{tab:S-domain-examples}). Supplementary Section~\ref{si:domain-sensitivity} and Supplementary Table~\ref{tab:S-domain-sensitivity} show that these within-domain results are not sensitive to how words are assigned to
WordNet senses and domains.

Our analyses do, however, reveal systematic variation by word form. Figure~\ref{fig:domains}c,d
locates each grammatical word form, across all domains, in terms of its mean concreteness and its
mean gender association. For words, verbs ($n = 6{,}198$) and adverbs ($n = 2{,}622$) are not only
more abstract than nouns that denote entities ($n = 6{,}524$; mean concreteness of 2.91 and 2.04
versus 4.34), but also more male-associated: the mean gender association was $0.18$~s.d.\ for
verbs (95\% CI across models $= 0.09$ to $0.26$; positive in 12 of 12 models) and $0.22$~s.d.\
for adverbs (95\% CI $= 0.11$ to $0.32$; 12 of 12 models), compared with $-0.12$~s.d.\ for
entity nouns (95\% CI $= -0.20$ to $-0.03$). Relative to entity nouns, verbs were $0.29$~s.d.\
more male-associated (95\% CI $= 0.16$ to $0.42$; \textit{P}~$= 0.002$) and adverbs $0.33$~s.d.\
(95\% CI $= 0.19$ to $0.47$; \textit{P}~$= 0.001$; two-tailed Wilcoxon signed-rank tests across
models). The same pattern was roughly twice as strong for phrases: verb phrases ($n = 12{,}601$;
$0.39$~s.d.; 95\% CI $= 0.30$ to $0.47$) and adverb phrases ($n = 2{,}520$; $0.31$~s.d.; 95\% CI
$= 0.21$ to $0.40$) were more male-associated than entity noun phrases ($n = 21{,}234$;
$-0.21$~s.d.; 95\% CI $= -0.25$ to $-0.17$), a difference of $0.60$~s.d.\ between verb phrases and
entity noun phrases (95\% CI $= 0.48$ to $0.71$; positive in 12 of 12 models;
\textit{P}~$< 0.001$). Adjectives and nouns for events and states were also abstract, yet they were
not male-associated (Fig.~\ref{fig:domains}c,d), which indicates that word form shapes gender
associations beyond abstractness alone. That the most male-coded word forms are those that
describe actions, and the manner in which actions are carried out, resonates with established
stereotypes that ascribe agentic and more action-oriented qualities to men more than to
women\cite{eagly1984,fiske2002,ellemers2018, stuhler2024gender}. These findings suggest that gender stereotypes relating to agency may go much deeper than role dynamics (e.g., perceptions of leadership in organizations), and may connect to fundamental grammatical structure differentiating dynamic (verbs and adverbs) and more static (noun-based) concepts. 

\subsection*{The Male Abstraction Bias in Online Images}

\begin{herefigure}
  \centering
  \includegraphics[width=\textwidth]{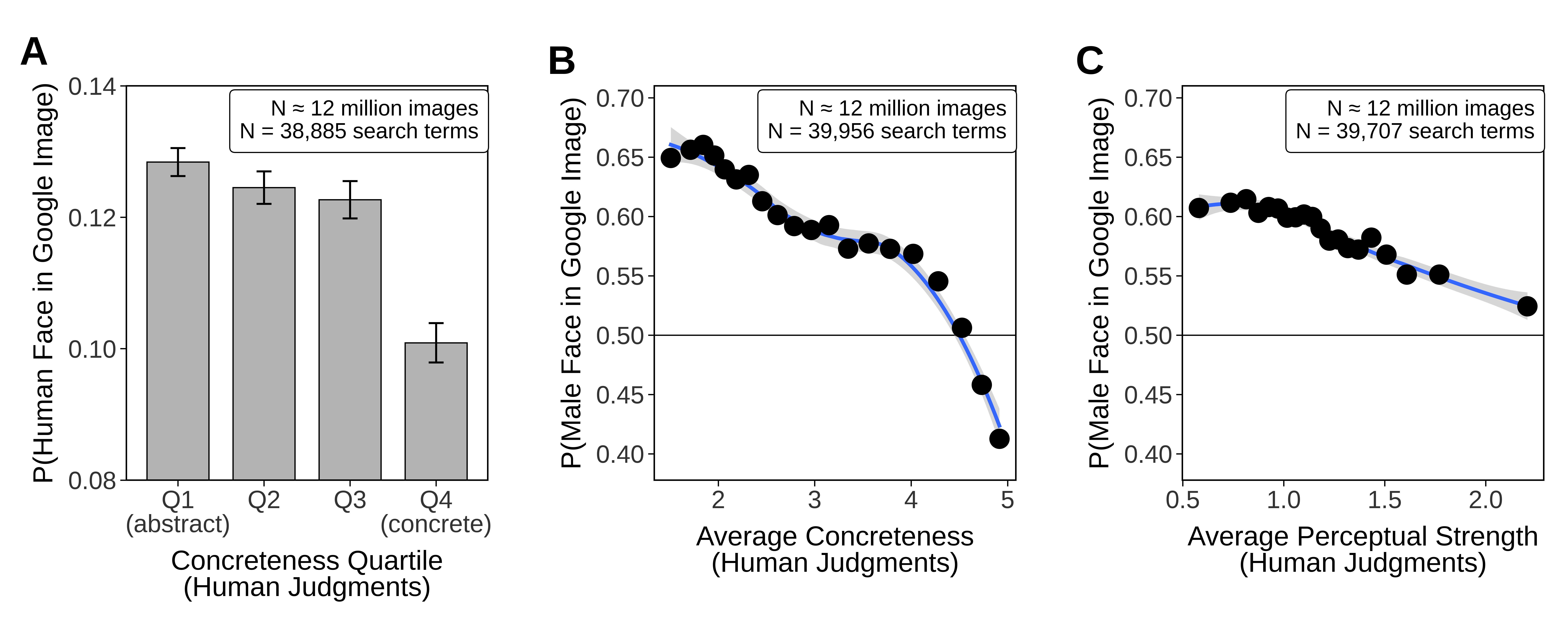}
  \caption{\textbf{More abstract concepts are depicted with people, and with men, in Google
  Images.} \textbf{a}, The probability that an image retrieved for a search term contains a human
  face, of either gender, by quartile of human judgements of the term's concreteness (Q1, most
  abstract, to Q4, most concrete; $n = 38{,}885$ search terms for which the number of retrieved
  images is recorded). Bars show the fraction of a term's images in which a face was detected,
  averaged over the terms in each quartile; error bars show 95\% confidence intervals of the mean.
  The vertical axis does not start at zero. \textbf{b},\textbf{c}, The fraction of male faces among
  the faces detected in the Google Images retrieved for each search term, as a function of human
  judgements of the term's concreteness (\textbf{b}; $n = 39{,}956$ search terms) and of its
  perceptual strength according to the Lancaster Sensorimotor Norms (\textbf{c}; $n = 39{,}707$
  search terms), across approximately 12 million images. Data points show mean values for 20
  equally sized bins of search terms; the blue line shows a locally estimated scatterplot smoothing
  (LOESS) fit, with 95\% confidence bands in grey. The horizontal line marks 0.5. Search terms whose
  images contained no faces ($n = 4{,}043$) are included in \textbf{b} and \textbf{c} with a value
  of zero; Supplementary Tables~\ref{tab:S-image-agg} and \ref{tab:S-image} report these
  associations for the search terms whose images contained at least one face, with statistical
  controls.}
  \label{fig:images}
\end{herefigure}

Next, we examined whether the male abstraction bias generalizes beyond text to the visual
modality, in which we examine the role of people as concrete anchors in the representation of both concrete and abstract concepts. For each of the 39,956 terms in the single-word concreteness norms, we collected the top images returned by Google Image search (approximately 300 images per term and 12 million images in total; see `Image Data' in Materials and Methods), which allowed us to map the features of images onto the concreteness of the concepts they depict. We used a recent multimodal language model (\VLM\cite{meta2024llama32}) to automatically detect the faces in each image and to classify their gender. From these classifications we derived two measures for each term: the probability that one of its images depicts a person, that is, the fraction of its images in which a face was detected, and its male association, that is, the fraction of male faces among all faces detected in its images (`Image Classification' and `Image Analyses' in Materials and Methods).

We first examined whether abstract and concrete concepts differ in their rate of depicting people in general. Figure~\ref{fig:images}a shows that more abstract search terms are more
likely to return images of people than more concrete terms. Among the most abstract quartile of
terms, 12.8\% of a term's images contained a human face on average (95\% CI $= 12.6\%$ to
$13.0\%$; $n = 9{,}722$ terms), compared with 10.1\% among the most concrete quartile (95\% CI
$= 9.8\%$ to $10.4\%$; $n = 9{,}721$ terms; difference $= 2.7$ percentage points; 95\% CI $= 2.4$
to $3.1$; Welch's $t = 14.65$; d.f.\ $= 17{,}513$; \textit{P}~$= 2.4\times10^{-48}$). Across all
terms, the probability that an image depicts a person declined with concreteness ($r = -0.09$;
95\% CI $= -0.10$ to $-0.08$; \textit{P}~$< 2.2\times10^{-16}$; $n = 38{,}885$ search terms;
two-tailed Pearson's correlation). This pattern is consistent with the long-standing argument that abstract concepts are understood through reference to concrete anchors from embodied experience \cite{lakoff1980, barsalou2008,borghi2017,borghi2025}; we add to this the use of large-scale visual data to examine the role of people as concrete, embodied anchors leveraged in the representation of abstract concepts. If human bodies are the embodied anchors of abstract concepts, this raises the question of who is more likely to appear when people are used to represent them.

Consistent with prior work revealing the over-representation of men in online images \cite{vlasceanu2022, guilbeault2024}, our analyses identify a clear statistical bias toward displaying men when images of people appear in general. Across all images from all concepts, 67.26\% are male, indicating a strong and significant bias toward male representation (815{,}096 male versus 396{,}719 female; Wilson 95\% CI $= 67.2\%$ to $67.4\%$; one-sample $z = 405.0$; \textit{P}~$< 10^{-300}$ against parity; two-tailed), and the male dominance remains far above parity when search terms rather than faces are treated as the unit of sampling (cluster-robust 95\% CI $= 67.1\%$ to $67.5\%$; $z = 156.0$;
\textit{P}~$< 10^{-300}$; $n = 35{,}913$ search terms with at least one face). This points toward a simple and consequential statistical mechanism leading toward the over-representation of men as concrete anchors in visual depictions of abstract concepts. As visual representations of people skew male -- and as visual representations of abstract concepts skew toward displaying people -- it follows that by statistical bias alone, it is expected that abstract concepts will be more associated with men. 

Indeed, Fig.~\ref{fig:images}b,c shows that, when people are depicted, they are disproportionately men,
and increasingly so for more abstract concepts. More abstract concepts are systematically more
likely to be depicted with male faces, both when abstraction is measured using human judgements of
concreteness (Fig.~\ref{fig:images}b; $r = -0.22$; 95\% CI $= -0.23$ to $-0.22$;
\textit{P}~$< 2.2\times10^{-16}$; $n = 39{,}956$ search terms) and when it is measured using
perceptual strength (Fig.~\ref{fig:images}c; $r = -0.08$; 95\% CI $= -0.09$ to $-0.07$;
\textit{P}~$< 2.2\times10^{-16}$; $n = 39{,}707$ search terms; two-tailed Pearson's correlations).
This association remains when examining only the 35,913 terms whose images contained at least one face ($r = -0.07$; 95\% CI $= -0.08$ to $-0.06$; \textit{P}~$< 2.2\times10^{-16}$); it is also robust to (i) controlling for the number of faces associated with each term (that is, the number of its images in which a face was detected), word frequency, the confidence of the gender classifier and part of speech ($\beta = -0.010$; s.e.\ $= 0.001$; $t = -8.99$; \textit{P}~$< 2.2\times10^{-16}$; ordinary least squares regression with robust standard errors), (ii) examining only terms associated with at least ten faces, and (iii) modeling the counts of male and female faces directly; all of these results replicate using perceptual strength as an alternative measure of concept abstractness (Supplementary Section~\ref{si:image-robustness} and Supplementary Tables~\ref{tab:S-image-agg} and \ref{tab:S-image}). The direction of the association is likewise robust to how terms without faces are treated---whether they are retained with an indicator for returning any face, excluded, or scored as neutral rather than as 0\% male---and to requiring at least 5, 10 or 20 faces per term, although its magnitude is an upper bound: among terms whose images contain faces, a one-standard-deviation increase in concreteness or in perceptual strength is associated with a decrease of approximately 0.05 standard deviations in the fraction of faces that are male (Supplementary Table~\ref{tab:S-image-face-controls}). 

\begin{herefigure}
  \centering
  \includegraphics[width=\textwidth]{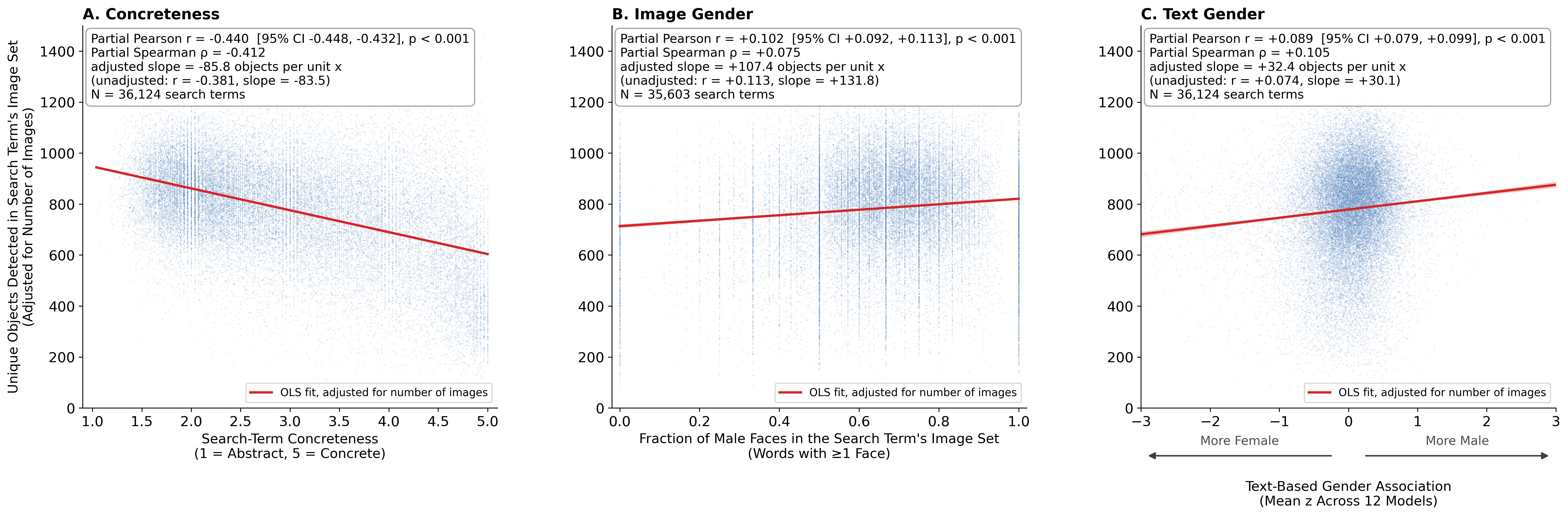}
  \caption{\textbf{More abstract and more male-typed concepts are depicted through a wider range
  of objects.} \textbf{a}--\textbf{c}, The number of unique objects detected across the Google
  Images retrieved for each search term, as a function of human judgements of the term's
  concreteness (\textbf{a}; $n = 36{,}124$ search terms), the fraction of male faces in the term's
  images (\textbf{b}; $n = 35{,}603$ search terms whose images contained at least one face) and the
  term's gender association in text, measured as the mean $z$-scored gender association across the 12
  language models of Fig.~\ref{fig:text} (\textbf{c}; $n = 36{,}124$ search terms; horizontal axis
  truncated at $\pm3$). All panels control for the number of images retrieved for each term: each
  data point is a search term, with its number of unique objects adjusted to the mean number of
  images per term, and red lines show ordinary least squares fits with 95\% confidence bands from
  regressions that include the log-transformed number of images as a covariate. Insets report
  two-tailed partial Pearson's correlations controlling for the number of images, with 95\%
  confidence intervals, partial Spearman's rank correlations and the slopes of the fitted lines
  (Supplementary Section~\ref{si:many-robustness}).}
  \label{fig:many}
\end{herefigure}

We next combined our text and image data to provide an empirical test of a recent hypothesis
by Bailey and Leshin\cite{bailey2026}, namely that people think of men as many things and of women
as one: men are conceptualized in varied and wide-ranging ways, whereas women are conceptualized
in a more unitary way. If the male abstraction bias is connected to this asymmetry, then concepts
that are more abstract, and concepts that are more male-typed, should be depicted through a
wider range of visual content. To test this, we used \VLM\ to classify the full set of objects that appear in each image, and we counted the number of unique objects detected across the images associated with each term. Because the number of unique objects grows with the number of images examined, and the number of images retrieved varied across terms (median $= 265$; mean $= 256$; s.d.\ $= 39$), all estimates in Fig.~\ref{fig:many} control for the number of images in each term's image set: we regressed the number of unique objects on each predictor together with the log-transformed number of images, and we report the partial correlation and slope for the predictor, with each term's count adjusted to the mean number of images in the figure (Supplementary Section~\ref{si:many-robustness}).

Figure~\ref{fig:many}a shows that more abstract concepts are depicted through a substantially
wider range of objects (partial $r = -0.44$; 95\% CI $= -0.45$ to $-0.43$;
\textit{P}~$< 2.2\times10^{-16}$; partial Spearman's $\rho = -0.41$; $n = 36{,}124$ search terms),
such that each one-point decrease in concreteness corresponds to 85.8 additional unique objects at a given number of images. More male-typed concepts, which are more abstract, are likewise depicted through a wider range of objects. This holds when the male association of a term is measured using its images (Fig.~\ref{fig:many}b; partial $r = 0.10$; 95\% CI $= 0.09$ to $0.11$; \textit{P}~$< 2.2\times10^{-16}$; $\rho = 0.07$; $n = 35{,}603$ search terms with at least one face) and when it is measured using its gender association in text (Fig.~\ref{fig:many}c; partial
$r = 0.09$; 95\% CI $= 0.08$ to $0.10$; \textit{P}~$< 2.2\times10^{-16}$; $\rho = 0.10$;
$n = 36{,}124$ search terms; two-tailed partial Pearson's correlations controlling for the
log-transformed number of images). All three associations replicate when the number of images is not controlled for: $r = -0.38$, $0.11$ and $0.07$ for Fig.~\ref{fig:many}a--c, respectively, each with
\textit{P}~$< 2.2\times10^{-16}$; they also replicate under alternative specifications of the
control, when additionally controlling for the total number of objects detected, and when
examining only terms with at least 200 images; Supplementary Section~\ref{si:many-robustness} and
Supplementary Table~\ref{tab:S-many}.  Importantly, recent work shows that the representation of people in online images has measurable effects on strengthening gender biases in people's representation of concepts \cite{guilbeault2024, guilbeault2025}. We thus proceed to test whether the male abstraction bias is indeed measurable in people's implicit associations with a wide variety of concepts. 

\subsection*{Evidence from the Implicit Free Associations of Humans and AI}

\begin{herefigure}
  \centering
  \includegraphics[width=0.88\textwidth]{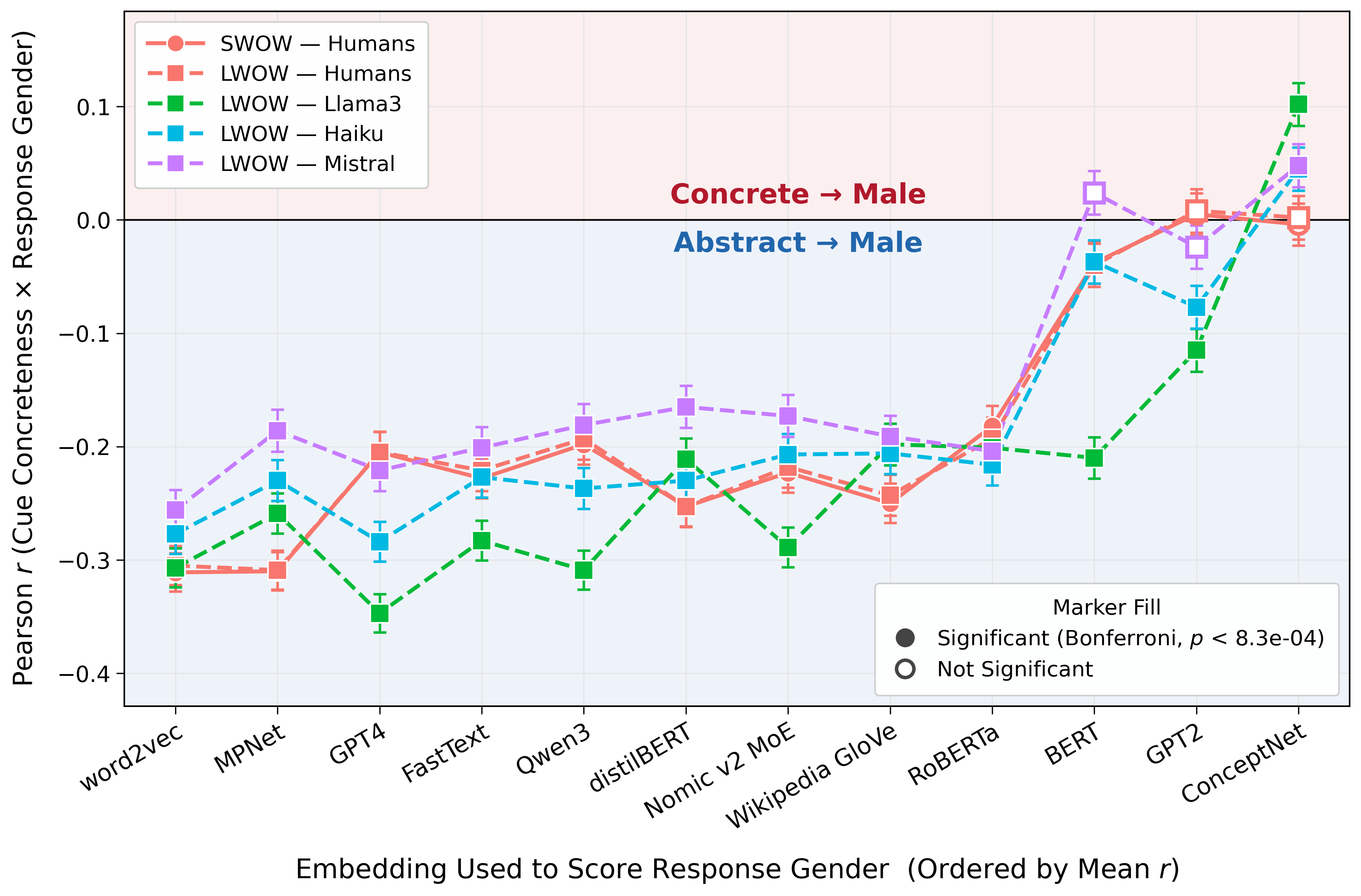}
  \caption{\textbf{The concreteness of a cue word predicts the gender association of the free
  associations it elicits from humans and from large language models.} Pearson's correlation
  between the human-rated concreteness of each cue word and the mean gender association of the
  words that respondents freely associated with it. Negative values indicate that more abstract
  cues elicit more masculine associations (equivalently, that more concrete cues elicit more
  feminine associations). Correlations are shown separately for each
  source of responses (colours): human participants in the Small World of Words dataset
  (SWOW)\cite{dedeyne2019}, and human participants and three large language models (Llama~3,
  Claude~3.5 Haiku and Mistral) in the LLM World of Words dataset (LWOW)\cite{abramski2025}. The
  horizontal axis gives the language model used to score the gender association of the responses,
  ordered by the mean correlation. Error bars show 95\% confidence intervals. Filled markers are
  significant after Bonferroni correction for all 60 tests (\textit{P}~$< 8.3\times10^{-4}$;
  two-tailed); hollow markers are not. $n = 11{,}179$ cues for SWOW and $n = 10{,}495$ to
  $10{,}527$ cues for LWOW.}
  \label{fig:freeassoc}
\end{herefigure}

So far, all of our findings rely on aggregate patterns in large observational datasets. This raises the
question of whether the male abstraction bias is merely an emergent statistical regularity of
cultural data, or whether it is also an implicit bias in the minds of individuals. Asking this question amounts to an out-of-sample test of whether patterns discovered in large-scale cultural data yield insights into individual cognition. This applies as well to AI agents, which are trained on this large-scale cultural data, suggesting the ability for biases therein to propagate through the AI and its responses. To this end, we analyzed existing data on the free associations that both humans and AI agents generate in response to a wide range of cue words: the Small World of Words dataset (SWOW)\cite{dedeyne2019}, in which human participants listed the first three words that came to mind for each of more than 12,000 cues, and the LLM World of Words dataset (LWOW)\cite{abramski2025}, which administered the same task to three large language models (Llama~3, Claude~3.5 Haiku and Mistral) alongside the human responses. These datasets were collected by different teams for different purposes, with neither concreteness nor gender in mind, and thus provide a conservative test. We matched each cue word to its human-rated concreteness, and we measured the gender association of every freely associated word using each of our 12 language models, which yielded the mean gender association of the responses to each cue (`Free Association Data' in Materials and Methods).

Figure~\ref{fig:freeassoc} shows that the more concrete (abstract) a cue word is, the more
feminine (masculine) the associations that humans and AI agents generate. For example, when
the gender association of responses is measured using OpenAI's embedding model, the correlation
between cue concreteness and response gender was $r = -0.21$ for humans in SWOW (95\% CI
$= -0.22$ to $-0.19$; $n = 11{,}179$ cues) and $r = -0.21$ for humans in LWOW (95\% CI $= -0.22$ to
$-0.19$; $n = 10{,}495$), and $r = -0.35$ for Llama~3 (95\% CI $= -0.36$ to $-0.33$;
$n = 10{,}496$), $r = -0.28$ for Claude~3.5 Haiku (95\% CI $= -0.30$ to $-0.27$; $n = 10{,}527$)
and $r = -0.22$ for Mistral (95\% CI $= -0.24$ to $-0.20$; $n = 10{,}527$; all
\textit{P}~$< 2.2\times10^{-16}$; two-tailed Pearson's correlations). These results are robust to
the method used to measure the gender association of responses. Averaging across all 12 scoring
models, the mean correlation was $-0.18$ for humans in SWOW (95\% CI across models $= -0.25$ to
$-0.11$; $t(11) = -5.73$; \textit{P}~$= 1.3\times10^{-4}$), $-0.18$ for humans in LWOW ($-0.25$ to
$-0.11$; $t(11) = -5.68$; \textit{P}~$= 1.4\times10^{-4}$), $-0.22$ for Llama~3 ($-0.30$ to $-0.14$;
$t(11) = -6.34$; \textit{P}~$= 5.5\times10^{-5}$), $-0.18$ for Claude~3.5 Haiku ($-0.25$ to $-0.12$;
$t(11) = -6.17$; \textit{P}~$= 7.0\times10^{-5}$) and $-0.14$ for Mistral ($-0.21$ to $-0.08$;
$t(11) = -4.97$; \textit{P}~$= 4.2\times10^{-4}$; two-tailed one-sample $t$-tests across models).
Of the 60 correlations examined, 51 were negative and significant after Bonferroni correction,
six were not significant, and three were significantly positive; the latter all involved a single
scoring model (ConceptNet) applied to the responses of AI agents ($r \leq 0.10$). One possible reason that ConceptNet deviates from the other models is that it is distinctly designed to factor in expert-coded taxonomic knowledge about the relationships among concepts that are not learned via population-level word association patterns. Overall, 9 of the 12 scoring models yielded a significant negative correlation for every source of responses (Fig.~\ref{fig:freeassoc}). These findings provide evidence that both humans and AI agents have internalized the male abstraction bias at an implicit level, as indicated by their patterns of free association.

%% file: stats_fig1.tex
\newcommand{\FigOneAAllMeanR}{$-0.174$}
\newcommand{\FigOneAAllCIlo}{$-0.205$}
\newcommand{\FigOneAAllCIhi}{$-0.142$}
\newcommand{\FigOneAAllMinR}{$-0.233$}
\newcommand{\FigOneAAllMaxR}{$-0.066$}
\newcommand{\FigOneAAllKNeg}{12}
\newcommand{\FigOneAAllKSigNeg}{12}
\newcommand{\FigOneAAllKSigPos}{0}
\newcommand{\FigOneAAllKTot}{12}
\newcommand{\FigOneAAllT}{$-12.03$}
\newcommand{\FigOneAAllDf}{11}
\newcommand{\FigOneAAllP}{$= 1.1\times10^{-7}$}
\newcommand{\FigOneAAllWilcP}{$= 4.9\times10^{-4}$}
\newcommand{\FigOneAAllDecOne}{$3.42$}
\newcommand{\FigOneAAllDecTen}{$2.83$}
\newcommand{\FigOneAAllNmin}{85{,}229}
\newcommand{\FigOneAAllNmax}{102{,}842}

\newcommand{\FigOneAWrdMeanR}{$-0.085$}
\newcommand{\FigOneAWrdCIlo}{$-0.131$}
\newcommand{\FigOneAWrdCIhi}{$-0.039$}
\newcommand{\FigOneAWrdMinR}{$-0.209$}
\newcommand{\FigOneAWrdMaxR}{$0.037$}
\newcommand{\FigOneAWrdKNeg}{11}
\newcommand{\FigOneAWrdKSigNeg}{11}
\newcommand{\FigOneAWrdKSigPos}{1}
\newcommand{\FigOneAWrdKTot}{12}
\newcommand{\FigOneAWrdT}{$-4.10$}
\newcommand{\FigOneAWrdDf}{11}
\newcommand{\FigOneAWrdP}{$= 0.002$}
\newcommand{\FigOneAWrdWilcP}{$= 0.002$}
\newcommand{\FigOneAWrdDecOne}{$3.31$}
\newcommand{\FigOneAWrdDecTen}{$3.02$}
\newcommand{\FigOneAWrdNmin}{31{,}609}
\newcommand{\FigOneAWrdNmax}{39{,}953}

\newcommand{\FigOneAPhrMeanR}{$-0.252$}
\newcommand{\FigOneAPhrCIlo}{$-0.295$}
\newcommand{\FigOneAPhrCIhi}{$-0.210$}
\newcommand{\FigOneAPhrMinR}{$-0.324$}
\newcommand{\FigOneAPhrMaxR}{$-0.108$}
\newcommand{\FigOneAPhrKNeg}{12}
\newcommand{\FigOneAPhrKSigNeg}{12}
\newcommand{\FigOneAPhrKSigPos}{0}
\newcommand{\FigOneAPhrKTot}{12}
\newcommand{\FigOneAPhrT}{$-13.10$}
\newcommand{\FigOneAPhrDf}{11}
\newcommand{\FigOneAPhrP}{$= 4.7\times10^{-8}$}
\newcommand{\FigOneAPhrWilcP}{$= 4.9\times10^{-4}$}
\newcommand{\FigOneAPhrDecOne}{$3.55$}
\newcommand{\FigOneAPhrDecTen}{$2.74$}
\newcommand{\FigOneAPhrNmin}{51{,}266}
\newcommand{\FigOneAPhrNmax}{62{,}889}

\newcommand{\FigOneBPhrMeanR}{$-0.204$}
\newcommand{\FigOneBPhrCIlo}{$-0.244$}
\newcommand{\FigOneBPhrCIhi}{$-0.163$}
\newcommand{\FigOneBPhrMinR}{$-0.306$}
\newcommand{\FigOneBPhrMaxR}{$-0.086$}
\newcommand{\FigOneBPhrKNeg}{12}
\newcommand{\FigOneBPhrKSigNeg}{12}
\newcommand{\FigOneBPhrKSigPos}{0}
\newcommand{\FigOneBPhrKTot}{12}
\newcommand{\FigOneBPhrT}{$-11.07$}
\newcommand{\FigOneBPhrDf}{11}
\newcommand{\FigOneBPhrP}{$= 2.7\times10^{-7}$}
\newcommand{\FigOneBPhrWilcP}{$= 4.9\times10^{-4}$}
\newcommand{\FigOneBPhrDecOne}{$3.50$}
\newcommand{\FigOneBPhrDecTen}{$2.58$}
\newcommand{\FigOneBPhrNmin}{51{,}266}
\newcommand{\FigOneBPhrNmax}{62{,}890}

\newcommand{\FigOneCWrdMeanR}{$-0.067$}
\newcommand{\FigOneCWrdCIlo}{$-0.095$}
\newcommand{\FigOneCWrdCIhi}{$-0.039$}
\newcommand{\FigOneCWrdMinR}{$-0.151$}
\newcommand{\FigOneCWrdMaxR}{$-0.005$}
\newcommand{\FigOneCWrdKNeg}{12}
\newcommand{\FigOneCWrdKSigNeg}{11}
\newcommand{\FigOneCWrdKSigPos}{0}
\newcommand{\FigOneCWrdKTot}{12}
\newcommand{\FigOneCWrdT}{$-5.26$}
\newcommand{\FigOneCWrdDf}{11}
\newcommand{\FigOneCWrdP}{$= 2.7\times10^{-4}$}
\newcommand{\FigOneCWrdWilcP}{$= 4.9\times10^{-4}$}
\newcommand{\FigOneCWrdDecOne}{$1.28$}
\newcommand{\FigOneCWrdDecTen}{$1.19$}
\newcommand{\FigOneCWrdNmin}{31{,}446}
\newcommand{\FigOneCWrdNmax}{39{,}706}

\newcommand{\FigOneCPhrMeanR}{$-0.119$}
\newcommand{\FigOneCPhrCIlo}{$-0.149$}
\newcommand{\FigOneCPhrCIhi}{$-0.089$}
\newcommand{\FigOneCPhrMinR}{$-0.190$}
\newcommand{\FigOneCPhrMaxR}{$-0.018$}
\newcommand{\FigOneCPhrKNeg}{12}
\newcommand{\FigOneCPhrKSigNeg}{12}
\newcommand{\FigOneCPhrKSigPos}{0}
\newcommand{\FigOneCPhrKTot}{12}
\newcommand{\FigOneCPhrT}{$-8.80$}
\newcommand{\FigOneCPhrDf}{11}
\newcommand{\FigOneCPhrP}{$= 2.6\times10^{-6}$}
\newcommand{\FigOneCPhrWilcP}{$= 4.9\times10^{-4}$}
\newcommand{\FigOneCPhrDecOne}{$1.47$}
\newcommand{\FigOneCPhrDecTen}{$1.33$}
\newcommand{\FigOneCPhrNmin}{50{,}214}
\newcommand{\FigOneCPhrNmax}{62{,}321}

\newcommand{\FigOneCAllMeanR}{$-0.073$}
\newcommand{\FigOneCAllCIlo}{$-0.101$}
\newcommand{\FigOneCAllCIhi}{$-0.044$}
\newcommand{\FigOneCAllMinR}{$-0.138$}
\newcommand{\FigOneCAllMaxR}{$-0.002$}
\newcommand{\FigOneCAllKNeg}{12}
\newcommand{\FigOneCAllKSigNeg}{11}
\newcommand{\FigOneCAllKSigPos}{0}
\newcommand{\FigOneCAllKTot}{12}
\newcommand{\FigOneCAllT}{$-5.63$}
\newcommand{\FigOneCAllDf}{11}
\newcommand{\FigOneCAllP}{$= 1.5\times10^{-4}$}
\newcommand{\FigOneCAllWilcP}{$= 4.9\times10^{-4}$}
\newcommand{\FigOneCAllDecOne}{$1.38$}
\newcommand{\FigOneCAllDecTen}{$1.28$}
\newcommand{\FigOneCAllNmin}{84{,}003}
\newcommand{\FigOneCAllNmax}{102{,}027}

\newcommand{\TabFigOneHuman}{%
\begingroup\scriptsize\setlength{\tabcolsep}{1.9pt}%
\begin{tabular}{@{}lrcrrcrrcr@{}}
\toprule
 & \multicolumn{3}{c}{Pooled (Fig.~1a)} & \multicolumn{3}{c}{Words only} & \multicolumn{3}{c}{Phrases only} \\
\cmidrule(lr){2-4}\cmidrule(lr){5-7}\cmidrule(lr){8-10}
Model & $r$ & 95\% CI & $n$ & $r$ & 95\% CI & $n$ & $r$ & 95\% CI & $n$ \\
\midrule
FastText & $-0.200$ & [$-0.206$, $-0.194$] & 93{,}742 & $-0.070$ & [$-0.080$, $-0.059$] & 34{,}132 & $-0.317$ & [$-0.324$, $-0.310$] & 59{,}610 \\
word2vec & $-0.208$ & [$-0.215$, $-0.202$] & 85{,}229 & $-0.115$ & [$-0.126$, $-0.105$] & 33{,}963 & $-0.322$ & [$-0.330$, $-0.315$] & 51{,}266 \\
Wikipedia GloVe & $-0.200$ & [$-0.207$, $-0.194$] & 89{,}877 & $-0.076$ & [$-0.087$, $-0.065$] & 31{,}609 & $-0.324$ & [$-0.331$, $-0.316$] & 58{,}268 \\
ConceptNet & $-0.093$ & [$-0.099$, $-0.086$] & 97{,}080 & $0.037$ & [$0.027$, $0.047$] & 38{,}341 & $-0.189$ & [$-0.197$, $-0.181$] & 58{,}739 \\
BERT & $-0.165$ & [$-0.171$, $-0.159$] & 102{,}842 & $-0.113$ & [$-0.123$, $-0.103$] & 39{,}953 & $-0.267$ & [$-0.274$, $-0.260$] & 62{,}889 \\
distilBERT & $-0.187$ & [$-0.192$, $-0.181$] & 102{,}842 & $-0.042$ & [$-0.052$, $-0.032$] & 39{,}953 & $-0.278$ & [$-0.285$, $-0.271$] & 62{,}889 \\
RoBERTa & $-0.066$ & [$-0.072$, $-0.060$] & 102{,}842 & $-0.026$ & [$-0.035$, $-0.016$] & 39{,}953 & $-0.108$ & [$-0.116$, $-0.100$] & 62{,}889 \\
MPNet & $-0.158$ & [$-0.164$, $-0.152$] & 99{,}876 & $-0.057$ & [$-0.067$, $-0.048$] & 39{,}953 & $-0.235$ & [$-0.243$, $-0.228$] & 59{,}923 \\
GPT4 & $-0.165$ & [$-0.171$, $-0.159$] & 99{,}876 & $-0.150$ & [$-0.160$, $-0.141$] & 39{,}953 & $-0.180$ & [$-0.188$, $-0.172$] & 59{,}923 \\
Qwen3 & $-0.223$ & [$-0.229$, $-0.217$] & 99{,}876 & $-0.209$ & [$-0.219$, $-0.200$] & 39{,}953 & $-0.232$ & [$-0.239$, $-0.224$] & 59{,}923 \\
Nomic v2 MoE & $-0.233$ & [$-0.239$, $-0.227$] & 99{,}876 & $-0.182$ & [$-0.192$, $-0.173$] & 39{,}953 & $-0.267$ & [$-0.274$, $-0.259$] & 59{,}923 \\
GPT2 & $-0.186$ & [$-0.192$, $-0.180$] & 99{,}876 & $-0.017$ & [$-0.027$, $-0.007$] & 39{,}953 & $-0.307$ & [$-0.314$, $-0.299$] & 59{,}923 \\
\midrule
\textbf{Mean across models} & $-0.174$ & [$-0.205$, $-0.142$] &  & $-0.085$ & [$-0.131$, $-0.039$] &  & $-0.252$ & [$-0.295$, $-0.210$] &  \\
\bottomrule
\end{tabular}\endgroup}

\newcommand{\TabFigOneLancaster}{%
\begingroup\scriptsize\setlength{\tabcolsep}{1.9pt}%
\begin{tabular}{@{}lrcrrcrrcr@{}}
\toprule
 & \multicolumn{3}{c}{Words (Fig.~1c)} & \multicolumn{3}{c}{Phrases (averaged)} & \multicolumn{3}{c}{Words and phrases} \\
\cmidrule(lr){2-4}\cmidrule(lr){5-7}\cmidrule(lr){8-10}
Model & $r$ & 95\% CI & $n$ & $r$ & 95\% CI & $n$ & $r$ & 95\% CI & $n$ \\
\midrule
FastText & $-0.051$ & [$-0.062$, $-0.041$] & 33{,}938 & $-0.116$ & [$-0.124$, $-0.108$] & 58{,}821 & $-0.056$ & [$-0.063$, $-0.050$] & 92{,}759 \\
word2vec & $-0.080$ & [$-0.090$, $-0.069$] & 33{,}789 & $-0.163$ & [$-0.172$, $-0.155$] & 50{,}214 & $-0.079$ & [$-0.085$, $-0.072$] & 84{,}003 \\
Wikipedia GloVe & $-0.043$ & [$-0.054$, $-0.032$] & 31{,}446 & $-0.190$ & [$-0.197$, $-0.182$] & 57{,}487 & $-0.065$ & [$-0.072$, $-0.059$] & 88{,}933 \\
ConceptNet & $-0.023$ & [$-0.033$, $-0.013$] & 38{,}126 & $-0.073$ & [$-0.081$, $-0.064$] & 58{,}052 & $-0.013$ & [$-0.019$, $-0.006$] & 96{,}178 \\
BERT & $-0.083$ & [$-0.092$, $-0.073$] & 39{,}706 & $-0.071$ & [$-0.079$, $-0.063$] & 62{,}321 & $-0.002$$^{\dagger}$ & [$-0.008$, $0.004$] & 102{,}027 \\
distilBERT & $-0.028$ & [$-0.038$, $-0.019$] & 39{,}706 & $-0.128$ & [$-0.136$, $-0.121$] & 62{,}321 & $-0.112$ & [$-0.118$, $-0.106$] & 102{,}027 \\
RoBERTa & $-0.037$ & [$-0.047$, $-0.027$] & 39{,}706 & $-0.018$ & [$-0.026$, $-0.010$] & 62{,}321 & $-0.013$ & [$-0.019$, $-0.007$] & 102{,}027 \\
MPNet & $-0.132$ & [$-0.142$, $-0.123$] & 39{,}706 & $-0.140$ & [$-0.148$, $-0.133$] & 62{,}321 & $-0.107$ & [$-0.113$, $-0.101$] & 102{,}027 \\
GPT4 & $-0.084$ & [$-0.094$, $-0.074$] & 39{,}706 & $-0.103$ & [$-0.111$, $-0.095$] & 62{,}321 & $-0.081$ & [$-0.087$, $-0.075$] & 102{,}027 \\
Qwen3 & $-0.088$ & [$-0.098$, $-0.078$] & 39{,}706 & $-0.132$ & [$-0.140$, $-0.125$] & 62{,}321 & $-0.121$ & [$-0.127$, $-0.115$] & 102{,}027 \\
Nomic v2 MoE & $-0.151$ & [$-0.161$, $-0.142$] & 39{,}706 & $-0.146$ & [$-0.154$, $-0.139$] & 62{,}321 & $-0.138$ & [$-0.144$, $-0.132$] & 102{,}027 \\
GPT2 & $-0.005$$^{\dagger}$ & [$-0.015$, $0.005$] & 39{,}706 & $-0.147$ & [$-0.155$, $-0.139$] & 62{,}321 & $-0.084$ & [$-0.091$, $-0.078$] & 102{,}027 \\
\midrule
\textbf{Mean across models} & $-0.067$ & [$-0.095$, $-0.039$] &  & $-0.119$ & [$-0.149$, $-0.089$] &  & $-0.073$ & [$-0.101$, $-0.044$] &  \\
\bottomrule
\end{tabular}\endgroup}

%% file: discussion.tex
%
%

\section*{Discussion}

Across the space of everyday concepts, more abstract terms and phrases are more strongly associated with men and more concrete analogs with women. This pattern appears whether abstractness is judged by people or by AI, whether it is measured as rated concreteness or as sensorimotor strength, and whether gender is measured in the distributional structure of billions of words, in the faces that populate millions of online images, or in the free associations that humans and large language models produce in controlled online tasks. Prior work has located gender differences in abstraction in how women and men communicate and construe events\cite{joshi2020,dodson2025}, or in how perceivers categorize them in different social roles\cite{bailey2026,bailey2022}, based consistently in modest-scale and context-specific observations (e.g., occupations in the workplace). Our results shift the level of analysis to a much broader scope: the male abstraction bias pervades conceptual structure not just as a bias for differentially describing men and women, but as a bias that warps the meanings of most conceptual domains including everyday objects and events. 

Moreover, we show that the male abstraction bias maps tightly onto grammatical structure: within and across a wide variety of domains, action-oriented terms---verbs and adverbs---are both more abstract and more male-coded, whereas object-oriented nouns are more concrete and more female-coded, suggesting as a broader pattern that men are afforded dynamic representations as \textit{verbs} whereas women are represented more statically and passively as \textit{nouns}. As a theoretical advance, this suggests that gender stereotypes relating to agency are not simply a macro bias that emerges in descriptions of men and women across different social roles and activities (e.g., relating to norms of leadership in the workplace); instead, they appear to be woven into the function of some of the most basic and central concepts organizing grammar. This suggests that interventions relating to changing role dynamics among women and men (e.g., by promoting women to leadership positions), while undoubtedly important and impactful \cite{lawson2022hiring}, need to be complemented with broader cultural shifts relating not just to the conceptualization of people, but also to basic, systemic patterns in the representation of concepts across topic domains. 

Generally, by showing how the male abstraction bias pervades the cultural record in both online textual and image data, this raises concern for how this bias is poised to be inherited and reproduced by any AI system trained on that record. The free-association results indicate that this male abstraction bias is not merely an aggregate regularity of corpora but is verifiably internalized in implicit concept activation patterns that individuals reproduce without any explicit gender prime, a pattern that AI models reproduce as strongly as people do. This reveals how the male abstraction bias is vulnerable to algorithmic amplification via a cultural evolutionary process: through AI systems, stereotypes increasingly shape decisions and judgments about what is real, and thereby become real in their consequences\cite{guilbeault2025,guilbeault2026review}. Indeed, the male abstraction bias is already visible in consequential settings. In more than half a million LinkedIn recommendations spanning industries and occupations, men are more likely to be recommended in abstract terms and women in concrete ones\cite{dodson2025}, and exposure to gender-biased online representations measurably shifts people's beliefs and hiring preferences, which search and generative algorithms then learn and reproduce\cite{guilbeault2024,guilbeault2025,vlasceanu2022}.

These findings also provide mechanistic insights that establish a link between literatures in cognitive science and sociology that are rarely brought together. A key premise of embodied cognition is that humans leverage concrete referents in bodily, perceptual experience to help ground and metaphorically represent less perceivable, abstract concepts. Our findings indicate that people play an outsized role as concrete referents for representing abstract concepts, raising the question of which kinds of people are central as concrete anchors throughout conceptual systems \cite{lakoff1980,varela1991,thelensmith1994,barsalou2008, guilbeault2020, nadler2025}. Our evidence indicates a bias toward male representation when people are used as concrete anchors in abstraction; our image data shows not only that abstract concepts are more likely to be depicted by people, but also in particular by men. This association is shaped by systemic inequalities in human cultural evolution. Historically, men have been privileged in their ability to cultivate abstract reasoning and to specialize in the high-status knowledge work of science, art, technology and politics, and the association of reason with masculinity is a long-standing cultural construction \cite{lloyd1984,schiebinger1989,ortner1974,bourdieu2001}. This has given men unequal opportunity to gain recognition as symbols and role models in domains of abstract reasoning and prestigious knowledge work, on which many foundational cultural narratives are constructed. If men are sociologically favored---through stereotypes, norms, and privileged access to cultivation and resources---to accumulate experience and symbolic recognition in abstract reasoning, then that experience may reinforce the associative bias that licenses it via a self-fulfilling prophecy \cite{merton1948selffulfilling}. As this bias becomes widespread and amplified via AI, it may further entrench the conceptual strategies (e.g., the use of embodied concrete anchors for representing abstraction) that warp the foundations of cognition along gender lines.

As a theoretical advance, our findings help identify an explanatory framework in which many distinct gender stereotypes are manifestations and context-specific expressions of a simpler, more systematic associative bias running through the entire conceptual system. Many stereotypes cast women as communal and warm---concrete, embodied qualities---and men as competent, rational, task-oriented and ``colder''\cite{fiske2002,ellemers2018,abele2014, bakan1966,eagly1984}; note, as well, that emotions themselves are often framed as ``hot'' cognition, in contrast to, in Lakoff's words, ``more than cool reason'' \cite{lakoff1989more}. In general, men are assumed to be better at mathematics and STEM despite no empirical basis for the belief\cite{nosek2002,hyde2008,leslie2015,bian2017}, with consequences for self-sorting into degree programs and occupations\cite{cheryan2017}. Each of these stereotype patterns may, in fact, be a subset or extension of a deeper, more systematic semantic association between men and abstraction that pervades human thinking. Such an association may have played a key role in the development of belief systems early in human cultural evolution, given the near-uniform association between men and the leading philosophers, scientists, political leaders and religious thinkers who set the abstract paradigms that have shaped human thought for millennia and continue to shape it today\cite{lloyd1984,schiebinger1989,lerner1986}.

For broader context on the implications of our findings, it is useful to consider how AI itself is gendered. AI understood as a highly abstract and rationally engineered model of intelligence appears to be male-coded. News imagery about AI is dominated by male faces\cite{chen2024}, people attribute a male gender to chatbots and to ChatGPT when no gender cue is given\cite{han2025,wongkim2023}, and the thinking machine and its makers have long been culturally coded as masculine\cite{adam1998,craiut2022,cave2023}, with a parallel argument made for its coding as white\cite{cavedihal2020}. However, commercial voice assistants such as Siri, Alexa and Cortana were launched with female voices and names by default in most markets, and most text-based chatbots are designed with female names and avatars\cite{unesco2019,feine2020}, a pattern feminist scholars have read as the automation of feminized service and domestic labor\cite{woods2018,strengers2020}. Experimental work also finds that bots labeled female are judged more human than those labeled male\cite{borau2021}, and that people exploit female-labeled AI partners more than female-labeled human ones\cite{bazazi2025}. Wong and Kim\cite{wongkim2023} find that the default perception of ChatGPT as male reverses when its capacity to provide emotional support is made salient, and Han et al.\cite{han2025} find that the male default weakens to roughly neutral when a chatbot is placed in a nursing role, which suggests AI is read as male when framed as competence and reasoning, and as female when framed as emotional care and service. This division sits closely alongside the gender--abstraction association we document. As AI becomes symbolic of what people believe ``superior'' intelligence looks like, it may reinforce the perception that intelligence is intrinsically a form of abstract and rationalized reasoning, and that by being so it is also implicitly male, while the assistive and relational work done by the same systems continues to be coded as female. Both tendencies arguably reinforce systemic gender inequalities.

We conclude by highlighting important limitations of our study and thus valuable directions for future research. Our evidence is correlational and cannot by itself establish the direction of causation between gendered experience and conceptual structure. Our aim in this study is to provide a comprehensive identification of the male abstraction bias spanning common concepts and communication modalities (online text and images), and to demonstrate the robustness of this pattern to a myriad of measurement techniques and controls. To ensure our findings are of psychological relevance at the individual level, and are not merely a statistical pattern that emerges in large-scale social data, we successfully conducted out-of-sample tests using existing experimental data to show that the male abstraction bias is measurable in the free associations generated by humans and AI. This provides a foundation for future research aimed at providing controlled experimental evidence of the male abstraction bias and its consequences for various psychological processes. Moreover, while we conducted a number of robustness tests demonstrating the replicability of the male abstraction bias across nine languages from different linguistic families (which vary in whether gender is morphologically grammaticalized), nevertheless, our largest and most comprehensive analyses are from English. Because English is the \textit{lingua franca} of the internet, findings from English are the most likely to capture patterns of content association and consumption that characterize internet activity around the globe, but there remains substantial cultural variation in gender norms and concept representation that cannot be captured from these analyses. Our cross-linguistic analyses are sufficient for demonstrating that the male abstraction bias generalizes beyond English, but identifying the extent of this generalization is a compelling direction for future studies \cite{boroditsky2003,prewitt2012,lewis2020}. Given that men have occupied positions of privilege in the pursuit and dissemination of high-status, abstract knowledge work across cultures and throughout recorded history\cite{ortner1974,wood2002,lloyd1984}, we expect the male abstraction bias to hold across cultures; discovering deviations from this pattern and linking such deviations to variation in conceptual, linguistic, and social structure, marks rich new terrain for computational studies of gender stereotypes and cultural evolution more broadly.

%% file: methods.tex
%

\section*{Acknowledgements}

We thank Stanford's DARC team for technical assistance with the classification of the image data, and T.~Hull for technical assistance with the collection of the image data. We thank A.~Martin, J.~Schroeder, A.~Goldberg, S.~Srivastava and R.~Fischli, and the members of the Computational Culture Lab, for constructive discussion and feedback in the preparation of this study.

\section*{Materials and Methods}

\subsection*{Textual Analysis (Embedding Methods)}

\paragraph{Concreteness Norms.}
Human judgements of concreteness were taken from the norms of Brysbaert et~al.\cite{brysbaert2014}, in which crowdsourced raters evaluated 39,953 generally known English words on a scale from 1 (abstract) to 5 (concrete), and from their extension by Muraki et~al.\cite{muraki2023} to 62,889 multiword expressions, collected with the same instructions and scale. GPT-generated concreteness ratings for the multiword expressions ($n = 62{,}890$; Fig.~\ref{fig:text}b) were taken from Martinez et~al.\cite{martinez2025}. Perceptual strength was taken from the Lancaster Sensorimotor Norms\cite{lynott2020}, which rate how strongly each word's referent is experienced through each of six perceptual modalities (0--5); we averaged the six ratings for the 39,706 words that also appear in the concreteness norms and extended the measure to phrases by averaging over each phrase's constituent words (Supplementary Section~\ref{si:lancaster}).

\paragraph{Language Models.}
Gender associations were measured in 12 language models that differ in architecture, training data and year of release (Supplementary Table~\ref{tab:embedding-models}; Supplementary Section~\ref{sec:embedding-models}): four fixed-vocabulary static word-embedding models (word2vec, FastText, Wikipedia GloVe and ConceptNet Numberbatch); four transformer language models used without a sentence-embedding head (BERT, distilBERT, RoBERTa and GPT2); three open-weight sentence-embedding models (MPNet, Qwen3 Embedding and Nomic Embed v2); and one closed, provider-managed embedding model (OpenAI's \code{text-embedding-3-large}, labelled GPT4). Each term is embedded as a complete string; the static models are queried for the phrase itself and otherwise fall back to the mean of the available token vectors. Eight of the models cover the whole vocabulary (or all but ten terms) and the four static models cover between 87,678 and 97,051 terms. The same 12 models are used wherever a text-based gender association appears in the paper (Figs.~\ref{fig:text}, \ref{fig:domains}, \ref{fig:many}c and \ref{fig:freeassoc}).

\paragraph{Measuring Gender Associations.}
We used the geometry-of-culture method\cite{kozlowski2019}, as applied to gender in our prior work\cite{guilbeault2024,guilbeault2025}, which represents gender association as a position along a dimension identified via contrasting poles in embedding space. In each model, a gender axis was built from five seed pairs (man/woman, him/her, he/she, male/female and boy/girl): each seed term was embedded and normalised to unit length, the male-minus-female difference was averaged across the five pairs, and the mean difference was normalised (Supplementary Section~\ref{sec:gender-scores}, Eq.~\ref{eq:gender-axis}). The gender association of a term is the cosine similarity between its unit-normalised embedding and this axis, with positive values indicating masculine associations; the zero point is fixed by the seed terms rather than by the mean of the vocabulary. Scores were computed for 103,428 terms (39,953 words and 63,475 multiword expressions) in each model. Because the models differ in scale, gender associations are compared within models: deciles (Fig.~\ref{fig:text}) were computed separately within each model, and where gender associations are combined across models (Figs.~\ref{fig:domains} and \ref{fig:many}c) they were first $z$-scored within each model and then averaged with equal weight, with confidence intervals computed across models. Our prior work validated this measure against human judgements of the gender associations of the same terms and showed that it is robust to the choice of seed pairs\cite{guilbeault2024,guilbeault2025}; seed-pair sensitivity analyses for the cross-language replications are reported in Supplementary Tables~\ref{tab:S-dutch-poles}, \ref{tab:S-estonian-poles}, \ref{tab:S-chinese-poles} and \ref{tab:S-persian-poles}.

\paragraph{Identifying Semantic Domains.}
To decompose the results by semantic domain (Fig.~\ref{fig:domains}), each word and phrase was matched to WordNet\cite{miller1995} and assigned the domain that WordNet records for its primary sense (for example \textit{noun.food} or \textit{verb.communication}). Words were matched within their dominant part of speech as recorded in the concreteness norms, which prevents nouns from entering verb domains through rare verb senses; phrases were matched through their own WordNet entry or, failing that, through their head noun. Function words, lexically gendered terms (for example `waitress') and nouns that denote persons were excluded, so that the analysis captures the gendering of concepts rather than of people, leaving 31,591 words in 35 domains and 56,879 phrases in 33 domains. Word forms (Fig.~\ref{fig:domains}c,d) follow the same mapping, with noun domains divided into those denoting events and states and those denoting entities. Supplementary Section~\ref{si:domain-sensitivity} and Supplementary Table~\ref{tab:S-domain-sensitivity} show that the within-domain results do not depend on the method of domain assignment.

\paragraph{Statistical Analysis of Textual Data and Robustness Tests.}
For each model we computed two-tailed Pearson's correlations between gender association and concreteness, with Fisher-transformed 95\% confidence intervals and \textit{P} values Bonferroni-corrected for the number of models. To summarize across models we treated models as the units of analysis: we report the unweighted mean of the per-model estimates, a 95\% confidence interval from the $t$ distribution across models and a two-tailed one-sample $t$-test against zero, so that no model is weighted more heavily than another. Within-domain contrasts (Fig.~\ref{fig:domains}a,b) use two-tailed binomial sign tests across domains and paired $t$-tests across models; word-form contrasts (Fig.~\ref{fig:domains}c,d) use two-tailed Wilcoxon signed-rank tests across models. Supplementary analyses show that the text-based results replicate for words and phrases analysed separately (Supplementary Section~\ref{si:words-phrases}); hold for perceptual strength extended to phrases (Supplementary Section~\ref{si:lancaster}); survive controls for word frequency, length, polysemy, age of acquisition, valence, arousal, dominance and part of speech, as well as the exclusion of person nouns, lexically gendered terms, function words and the most strongly gendered terms (Supplementary Section~\ref{si:text-covariates}); are visible at the level of individual words within domains (Supplementary Section~\ref{si:domain-examples}); and replicate in Dutch, Estonian, Mandarin Chinese and Persian, using native-speaker concreteness norms\cite{brysbaert2014dutch,proos2023,chan2024,song2021,nemati2026}, language-specific word vectors\cite{honnibal2020spacy,grave2018} and gender axes built from kinship and role seed pairs (Supplementary Sections \ref{si:dutch} to~\ref{si:persian}), as well as in smaller norms for five further languages (Supplementary Section~\ref{si:forest}).

\subsection*{Image Analysis}

\paragraph{Image Data.}
For each of the 39,956 terms in the single-word concreteness norms\cite{brysbaert2014}, we searched Google Images and retained the top results, up to 300 images per term (median 265), yielding 11,346,925 images. Images were downloaded into one directory per search term and named by the MD5 hash of their contents, so that byte-identical duplicates within a term collapse to a single file (Supplementary Section~\ref{sec:image-data}).

\paragraph{Image Classification.}
Every image was classified by \VLM\cite{meta2024llama32,grattafiori2024llama3}, an open-weight vision--language model, run on Stanford's high-performance computing infrastructure behind an OpenAI-compatible inference endpoint. The model was prompted with a single fixed instruction, identical for every image and containing no reference to the search term, so that classification is conditioned on the image alone. From its responses we obtained, for each image, a free-text list of the objects depicted, the number of human faces, and the gender (male or female) of each face together with the model's confidence in that classification. Object labels are open-vocabulary rather than drawn from a fixed taxonomy; they were lowercased, stripped of formatting, digits and punctuation, and matched to the lexicon of the concreteness norms, and labels identical to the search term were discarded. This yields 9,977,592 images with at least one lexicon-matched object across 39,918 search terms, and 1,211,815 detected faces (67.3\% classified as male) across the 35,913 terms whose images contained at least one face.

\paragraph{Image Analyses.}
We derived three term-level measures. The probability that a term's images depict a person is the fraction of its images in which at least one face was detected (Fig.~\ref{fig:images}a). The male association of a term is the fraction of male faces among all faces detected in its images (Fig.~\ref{fig:images}b,c); the 4,043 terms whose images contained no faces enter Fig.~\ref{fig:images} with a value of zero, and Supplementary Section~\ref{si:image-robustness} shows that the association with concreteness holds among terms with at least one face, after controlling for the number of faces, word frequency, classifier confidence and part of speech, and under alternative treatments of terms without faces (Supplementary Tables~\ref{tab:S-image-agg} to~\ref{tab:S-image-face-controls}). The visual diversity of a term is the number of unique object labels detected across its images (Fig.~\ref{fig:many}); because this count rises with the number of images retrieved, which varied across terms, all estimates in Fig.~\ref{fig:many} come from regressions that include the log-transformed number of images as a covariate (Supplementary Section~\ref{si:many-robustness}). The text-based gender association of a term in Fig.~\ref{fig:many}c is the equal-weight mean of its $z$-scored gender associations across the 12 language models. Associations were tested with two-tailed Pearson's (or partial Pearson's) correlations; Fig.~\ref{fig:images} displays means within 20 equal-count bins of the predictor with LOESS fits, and the comparison of concreteness quartiles in Fig.~\ref{fig:images}a uses Welch's $t$-test.

\subsection*{Free Association Tests}

\paragraph{Free Association Data.}
We analysed two existing free-association datasets collected by other teams for purposes unrelated to gender or concreteness. The Small World of Words English norms (SWOW)\cite{dedeyne2019} were collected by continued free association: each participant was shown a cue word and typed the first three words that came to mind, in order, with roughly 100 participants per cue and more than 12,000 cues; we used the released strength file, which aggregates the three response slots into cue--response counts. The LLM World of Words (LWOW)\cite{abramski2025} administered the same task to three large language models---Mistral (\code{mistral-7b}), Llama~3 (\code{llama3.1-8b}) and Claude~3.5 Haiku (\code{claude-3-5-haiku-latest})---alongside human responses to the same cues: each model was given a cue and instructed to write the first three words it associates with it, separated by commas, and the prompt was repeated 100 times per cue, yielding three response slots per repetition, which we aggregated into cue--response counts. In both datasets, cues and responses were lowercased and trimmed, non-responses and responses identical to the cue were removed, and cues were retained where a human concreteness rating\cite{brysbaert2014} was available, leaving 11,179 SWOW cues and 10,495 to 10,527 LWOW cues per source of responses (Supplementary Section~\ref{sec:free-association}). For each cue and each of the 12 language models, the gender association of the responses is the mean gender association of the responded words, weighted by how often each response was produced; it was correlated with the cue's concreteness using two-tailed Pearson's correlations, giving 60 correlations (five sources of responses $\times$ 12 scoring models) that were Bonferroni-corrected and summarized across scoring models as above.

\subsection*{Data Availability}
All data supporting the findings of this study are publicly available. The concreteness norms for English words\cite{brysbaert2014} and multiword expressions\cite{muraki2023}, the GPT-generated concreteness ratings\cite{martinez2025}, the Lancaster Sensorimotor Norms\cite{lynott2020}, the Small World of Words\cite{dedeyne2019} and LLM World of Words\cite{abramski2025} free-association norms, the Dutch\cite{brysbaert2014dutch}, Estonian\cite{proos2023}, Chinese\cite{chan2024,song2021} and Persian\cite{nemati2026} concreteness norms and the norms for the five further languages\cite{goz2017,montefinese2014,bonin2018,soares2017,sianipar2016} were obtained from the sources cited. WordNet\cite{miller1995}, the lexical covariates used in the supplementary analyses\cite{kuperman2012,warriner2013} and the pre-trained word vectors and language models (Supplementary Table~\ref{tab:embedding-models}; spaCy\cite{honnibal2020spacy} and fastText\cite{grave2018}) are likewise publicly available. The derived data generated for this study---the gender association of every term under each of the 12 language models and under the language-specific vectors, the per-term image data (number of images retrieved, number of detected faces, the gender classification of each face with its confidence, and the object labels detected in each image) and the cue-level free-association measures---are available through the project repository at \url{https://github.com/bhargavvader/gender-abstraction}, which documents each data file together with its provenance and licensing terms. The images themselves were retrieved from Google Images and are subject to third-party copyright; we therefore release the search terms and the complete per-image classification output rather than the image files. Source data for Figs.~\ref{fig:text}--\ref{fig:freeassoc} are provided with this paper.

\subsection*{Code Availability}
All code used to score gender associations, construct the WordNet domain mapping, aggregate the image classifications, run the free-association and cross-language analyses and produce every main figure and supplementary analysis is available at \url{https://github.com/bhargavvader/gender-abstraction}, together with documentation of the file structure and of where each data file must be placed for the scripts to run. Analyses were conducted in Python and R, using the most recent stable release of each at the time of analysis; the software environment is recorded in the repository. Image classification used \VLM, whose weights are publicly released\cite{meta2024llama32}; the prompt and inference scripts are included in the repository. Gender associations under OpenAI's \code{text-embedding-3-large} were obtained through the OpenAI API in July 2026; all other language models are open weights and are specified by name and version in Supplementary Table~\ref{tab:embedding-models}.

%% file: supplementary.tex
\clearpage
\setcounter{page}{1}
\ifdefined\hypersetup\hypersetup{pageanchor=false}\fi

\begin{center}
  {\Large\bfseries Supplementary Materials}\\[0.6em]
  {\large The Male Abstraction Bias in Internet Culture and Large Language Models}
\end{center}
\vspace{1em}

\section{Data and methods}\label{sec:methods}

\subsection{Image dataset construction}\label{sec:image-data}

Our lexicon is the concreteness norms of Brysbaert et~al.\cite{brysbaert2014}, which contain 39,954 entries covering generally known English words together with a set of two-word expressions, each rated on a five-point scale from abstract to concrete. We searched Google Images for every term and retained up to 300 images per term. Images were downloaded in advance into one directory per search term, each file named by the MD5 hash of its contents, so that byte-identical duplicates within a term collapse to a single file.

Each image was then labelled by Llama-3.2-11B-Vision-Instruct \cite{meta2024llama32, grattafiori2024llama3}, an open-weight vision--language model built on the LLaMA family, run on university high-performance computing in collaboration with the DARC team. The model was served behind an OpenAI-compatible inference endpoint and prompted with a single fixed instruction, identical for every image and containing no reference to the search term, so that labelling is conditioned on the image alone. The prompt was 41 tokens on every call. The model returns a short comma-separated list of the objects it identifies, in free text, so labelling is generative and open-vocabulary rather than detection against a fixed taxonomy. This produced 11,346,925 labelled images across 39,929 search terms.

Labels were parsed from the model's response and normalised: lowercased, stripped of bullet formatting, digits and punctuation, and collapsed to single spacing. Multiword expressions appearing in the concreteness norms were joined into single labels, labels identical to the search query were removed, and the remainder were retained where they matched the lexicon. After parsing and lexical filtering the analysis set retains 9,977,592 images across 39,918 search terms, with a median of 265 images per term and a mean of 250.

Labels were aggregated across all images for a search term and saved to disk in two forms: a dictionary mapping each label to its occurrence count, and the list of distinct labels. Repeated labels within an image are counted rather than collapsed, so an object token is a label occurrence. No frequency threshold was applied at aggregation. The dataset contains 77,170,514 object tokens across 63,540 distinct object categories, with a mean of 7.73 objects per image and a median of 5. Every one of the 39,918 search terms has at least one recorded object. Per term, the median is 1,952 object tokens and 800 distinct categories.

\subsection{Gender scores}\label{sec:gender-scores}

We follow the geometric approach of Kozlowski et~al.\cite{kozlowski2019}, applied to images and text in Guilbeault et~al.\cite{guilbeault2024}, which represents a cultural dimension as a direction in embedding space. The gender axis is built from five seed pairs: \emph{man}/\emph{woman}, \emph{him}/\emph{her}, \emph{he}/\emph{she}, \emph{male}/\emph{female}, and \emph{boy}/\emph{girl}. Each seed term is embedded and its vector normalised to unit length, the male-minus-female difference is averaged across the five pairs, and that mean difference is itself normalised to give the axis:
\begin{equation}\label{eq:gender-axis}
\vec{a} = \mathrm{unit}\!\left(\frac{1}{5}\sum_{i=1}^{5}\left[\hat{u}(m_i) - \hat{u}(f_i)\right]\right).
\end{equation}
A term's gender score is the cosine similarity between its unit-normalised embedding and this axis, $g(t) = \mathrm{unit}(\vec{v}_t)^{\top}\vec{a}$, so a positive score falls on the male side. The zero point is fixed by the anchors and is independent of the mean of any corpus.

Scores were generated for 103,428 terms: 39,953 from the concreteness norms above and a further 63,475 multiword expressions from Muraki et~al.\cite{muraki2023}, deduplicated case-insensitively with the first source taking precedence.

Gender at the object level, for a given search term, is the token-weighted mean of the gender scores of its labels,
\begin{equation}\label{eq:object-gender}
G(w) = \frac{\sum_{o \in S} c_o\, g(o)}{\sum_{o \in S} c_o},
\end{equation}
where $c_o$ is the occurrence count of label $o$ and $S$ is the set of labels with a finite score under the model in question. Labels without a score are excluded from both numerator and denominator. The measure is weighted by occurrence, not computed over distinct labels.

The text-based gender association of a search term in Fig.~4c is the equal-weight mean of its gender scores across the 12 language models of Supplementary Table~\ref{tab:embedding-models}, each first $z$-scored within the 39,953-word vocabulary; terms scored by fewer than eight models would be excluded, but every term with a concreteness rating is scored by at least eight.

Because the number of unique objects detected for a term rises with the number of images in its image set, which varied across terms (median 265; mean 256; s.d.\ 39; range 3--300 among the terms analysed), every estimate in Fig.~4 is from a regression of the number of unique objects on the predictor and the log-transformed number of images, and the figure displays each term's count adjusted to the mean number of images. The unadjusted estimates and alternative specifications of the control are reported in Supplementary Section~\ref{si:many-robustness}.

\subsection{Embedding models}\label{sec:embedding-models}

Gender scores were computed under 12 language models, which fall into four groups according to how a term is turned into a vector. The same 12 models are used wherever a text-based gender association appears in the paper: Figs.~1 and 2, the text-based gender association of search terms in Fig.~4c, the scoring of free-association responses in Fig.~5, and the text-based supplementary analyses.

Four fixed-vocabulary word-embedding models---\code{word2vec-google-news-300}, \code{fasttext-wiki-news-subwords-300}, 300-dimensional GloVe vectors trained on Wikipedia and Gigaword (`Wikipedia GloVe') and ConceptNet Numberbatch---are queried for the exact phrase first, then for lowercase and underscore variants, and otherwise fall back to the mean of the available token vectors. 
OpenAI's \code{text-embedding-3-large} (`GPT4') is called through the API and returns a 3,072-dimensional embedding of the complete string. Three sentence-transformer models use their own native pooling: \code{all-mpnet-base-v2} (`MPNet'), \code{Qwen3-Embedding-0.6B} (`Qwen3') and \code{nomic-embed-text-v2-moe} (`Nomic v2 MoE'); Nomic is given the prefix \code{search_document:} on both anchors and vocabulary items, as its model card requires. Four transformer language models are used without a sentence-embedding head: \code{gpt2} as a static embedding table, in which a term is tokenised and its vector is the mean of the input embeddings of its subword tokens, with special and padding tokens excluded; and \code{bert-base-uncased}, \code{distilbert-base-uncased} and \code{roberta-base}. 

Eight of the 12 models cover the full vocabulary of 103,429 distinct terms (the three BERT-family models all terms, and \code{text-embedding-3-large}, MPNet, Qwen3, Nomic and GPT2 all but ten); the fixed-vocabulary models cover fewer, the shortfall arising from terms absent from their vocabularies (Supplementary Table~\ref{tab:embedding-models}). Scoring ran at a batch size of 128, or 512 for the OpenAI batch jobs, with automatic device selection and no cap on vocabulary size.

\ifdefined\embmodelsbox\else\newsavebox{\embmodelsbox}\fi
\begin{table}[htbp]
\centering
\caption{Language models used for gender scoring in Figs.~1, 2, 4c and 5 and in the text-based supplementary analyses. Terms scored are distinct terms with a gender score; words and phrases give the number of these among the 39,953 single words and the multiword phrases of the concreteness norms.}\label{tab:embedding-models}
\vspace{4pt}
\sbox{\embmodelsbox}{\small\setlength{\tabcolsep}{4pt}%
\begin{tabular}{@{}lllrrr@{}}
\toprule
Model & \begin{tabular}[b]{@{}l@{}}Label in\\ figures\end{tabular} & Pooling & \begin{tabular}[b]{@{}r@{}}Terms\\ scored\end{tabular} & Words & Phrases \\
\midrule
\code{fasttext-wiki-news-subwords-300} & FastText & \begin{tabular}[t]{@{}l@{}}phrase lookup,\\ else token mean\end{tabular} & 96,657 & 34,132 & 62,626 \\\addlinespace[2pt]
\code{word2vec-google-news-300} & word2vec & \begin{tabular}[t]{@{}l@{}}phrase lookup,\\ else token mean\end{tabular} & 87,678 & 33,963 & 53,814 \\\addlinespace[2pt]
\begin{tabular}[t]{@{}l@{}}GloVe 300D,\\ Wikipedia + Gigaword\end{tabular} & Wikipedia GloVe & \begin{tabular}[t]{@{}l@{}}phrase lookup,\\ else token mean\end{tabular} & 92,674 & 31,609 & 61,163 \\\addlinespace[2pt]
ConceptNet Numberbatch & ConceptNet & \begin{tabular}[t]{@{}l@{}}phrase lookup,\\ else token mean\end{tabular} & 97,051 & 38,341 & 61,616 \\\addlinespace[2pt]
\code{bert-base-uncased} & BERT & see text & 103,429 & 39,953 & 66,432 \\
\code{distilbert-base-uncased} & distilBERT & see text & 103,429 & 39,953 & 66,432 \\
\code{roberta-base} & RoBERTa & see text & 103,429 & 39,953 & 66,432 \\
\code{all-mpnet-base-v2} & MPNet & model-native & 103,419 & 39,953 & 63,466 \\
\code{text-embedding-3-large} & GPT4 & provider-managed & 103,419 & 39,953 & 63,466 \\
\code{Qwen3-Embedding-0.6B} & Qwen3 & model-native & 103,419 & 39,953 & 63,466 \\
\code{nomic-embed-text-v2-moe} & Nomic v2 MoE & model-native & 103,419 & 39,953 & 63,466 \\
\code{gpt2} & GPT2 & \begin{tabular}[t]{@{}l@{}}mean static\\ subword\end{tabular} & 103,419 & 39,953 & 63,466 \\
\bottomrule
\end{tabular}}%
\ifdim\wd\embmodelsbox>\textwidth
  \resizebox{\textwidth}{!}{\usebox{\embmodelsbox}}%
\else
  \usebox{\embmodelsbox}%
\fi
\end{table}

\subsection{Free association datasets}\label{sec:free-association}

We draw on two free association datasets, one generated by language models and one collected from human participants.

The LLM World of Words\cite{abramski2025} provides free association norms produced by three generators: \code{mistral-7b}, \code{llama3.1-8b} and \code{claude-3-5-haiku-latest}. Each generator is given a cue with the instruction to write the first three words it associates with that cue, separated by commas, and the prompt is repeated one hundred times per cue, yielding three response slots on each repetition. We load the processed release, melt the three response columns into a single response column, normalise case and whitespace, drop blank and missing responses, and sum occurrences for every cue--response pair. Our association file retains 10,527 cues for each generator, of which 10,495 to 10,527 (depending on the generator) have a cue concreteness rating and enter the analysis.

The Small World of Words English release\cite{dedeyne2019} provides the human comparison. It was collected by continued free association: a participant is shown a cue and gives three responses in order, with roughly one hundred participants per cue. We use the strength release, \code{strength.SWOW-EN.R123.20180827.csv}, which aggregates the three response slots (R123) into cue--response counts: 1,389,711 cue--response pairs across 12,216 cues. Cues and responses are lower-cased and trimmed; non-response strings (\texttt{no more responses}, \texttt{unknown word}, \code{nan}, \code{none}, \code{na} and the empty string) and responses identical to the cue are excluded. Cues are retained where a cue concreteness rating is available in the concreteness norms, which leaves 11,179 cues for every scoring model; the number of responses scored varies by model with its vocabulary coverage (Supplementary Table~\ref{tab:embedding-models}).

In both datasets the outcome measure is the mean gender score of a cue's responses, weighted by how often each response was produced, with the weights normalised to sum to one.

\ifdefined\FigOneAAllMeanR\else\input{stats_fig1}\fi

\input{stats_si_text}

\section{Replication of the text-based result in Dutch}
\label{si:dutch}

To test whether the association between gender and concreteness in Fig.~1a is specific to English, we repeated the single-word analysis in Dutch, the only language other than English for which crowdsourced concreteness ratings exist at a comparable scale. In a different paper, Brysbaert et~al.\cite{brysbaert2014dutch} provide ratings of concreteness (1, abstract; 5, concrete) and of age of word acquisition for 30,000 Dutch lemmas, collected with the same instructions and rating scale as the English norms; we used the 25,881 distinct words with a concreteness rating in the distributed file (mean 2.93; s.d.\ 0.99). Dutch is also informative because its nouns carry no masculine or feminine grammatical gender (the two Dutch genders are common and neuter), so a replication cannot be attributed to grammatical gender marking.

Gender associations were measured in the Dutch-trained static word vectors distributed with spaCy's \code{nl_core_news_lg} model (version 3.8.0; 500,000 vocabulary entries, 300 dimensions)\cite{honnibal2020spacy}, a language-specific model rather than one of the English or multilingual models of Supplementary Table~\ref{tab:embedding-models}. The gender axis was constructed exactly as in Eq.~\ref{eq:gender-axis}, from five seed pairs chosen before any result was examined: \emph{man}/\emph{vrouw}, \emph{jongen}/\emph{meisje} (boy/girl), \emph{mannelijk}/\emph{vrouwelijk} (male/female), \emph{vader}/\emph{moeder} (father/mother) and \emph{broer}/\emph{zus} (brother/sister). The two pronoun pairs of the English axis were not transferred, because their Dutch counterparts are lexically ambiguous: \emph{zij} is both \emph{she} and \emph{they}, and \emph{haar} is both \emph{her} and \emph{hair}. They were retained as a separate specification so that the consequence of this ambiguity could be reported (Supplementary Table~\ref{tab:S-dutch-poles}, family C). An extended axis added five further kinship and role pairs (\emph{zoon}/\emph{dochter}, \emph{oom}/\emph{tante}, \emph{koning}/\emph{koningin}, \emph{heer}/\emph{dame}, \emph{opa}/\emph{oma}), and a third variant took the first principal component of the ten unit-normalised pair differences\cite{bolukbasi2016}, sign-aligned with their mean. Each rated word was scored as the cosine between its unit-normalised vector and the axis, positive values indicating masculine associations; words absent from the vector table (8.2\%) were not scored, leaving 23,758 words. As a check on the axis that does not involve concreteness, we verified that the primary axis orders the male above the female member in all 27 gendered word pairs that were not used as seeds (for example \emph{koning}/\emph{koningin}, \emph{acteur}/\emph{actrice}, \emph{oom}/\emph{tante}, \emph{meneer}/\emph{mevrouw}).

The statistics mirror Supplementary Sections~\ref{si:words-phrases} and~\ref{si:text-covariates}: two-tailed Pearson's correlations with Fisher-transformed 95\% confidence intervals, Spearman's rank correlations, mean concreteness within deciles of the gender score, and ordinary least squares regressions of the $z$-scored gender score on $z$-scored concreteness with heteroskedasticity-robust (HC1) standard errors, whose coefficient is a standardized $\beta$. Covariates were word frequency (the Zipf-scaled frequency of the \code{wordfreq} library\cite{speer2022wordfreq}, which combines subtitle, Wikipedia and web corpora; 23,407 words have an entry), length in characters, and age of acquisition from the same norms. We also report the association within each quintile of word frequency, after excluding words that contain a gendered morpheme or a kinship or role stem (for example \emph{-man}, \emph{vrouw-}, \emph{-ster}, \emph{vader}, \emph{koning}) together with the seed words themselves, and after excluding the 5\% or 10\% of words with the most extreme gender scores. Finally, we recomputed the correlation for each axis obtained by dropping one seed pair from the primary set, for each of the ten kinship and role pairs on its own, and for the pronoun pairs.

\begin{figure}[!htb]
  \centering
  \includegraphics[width=0.55\textwidth]{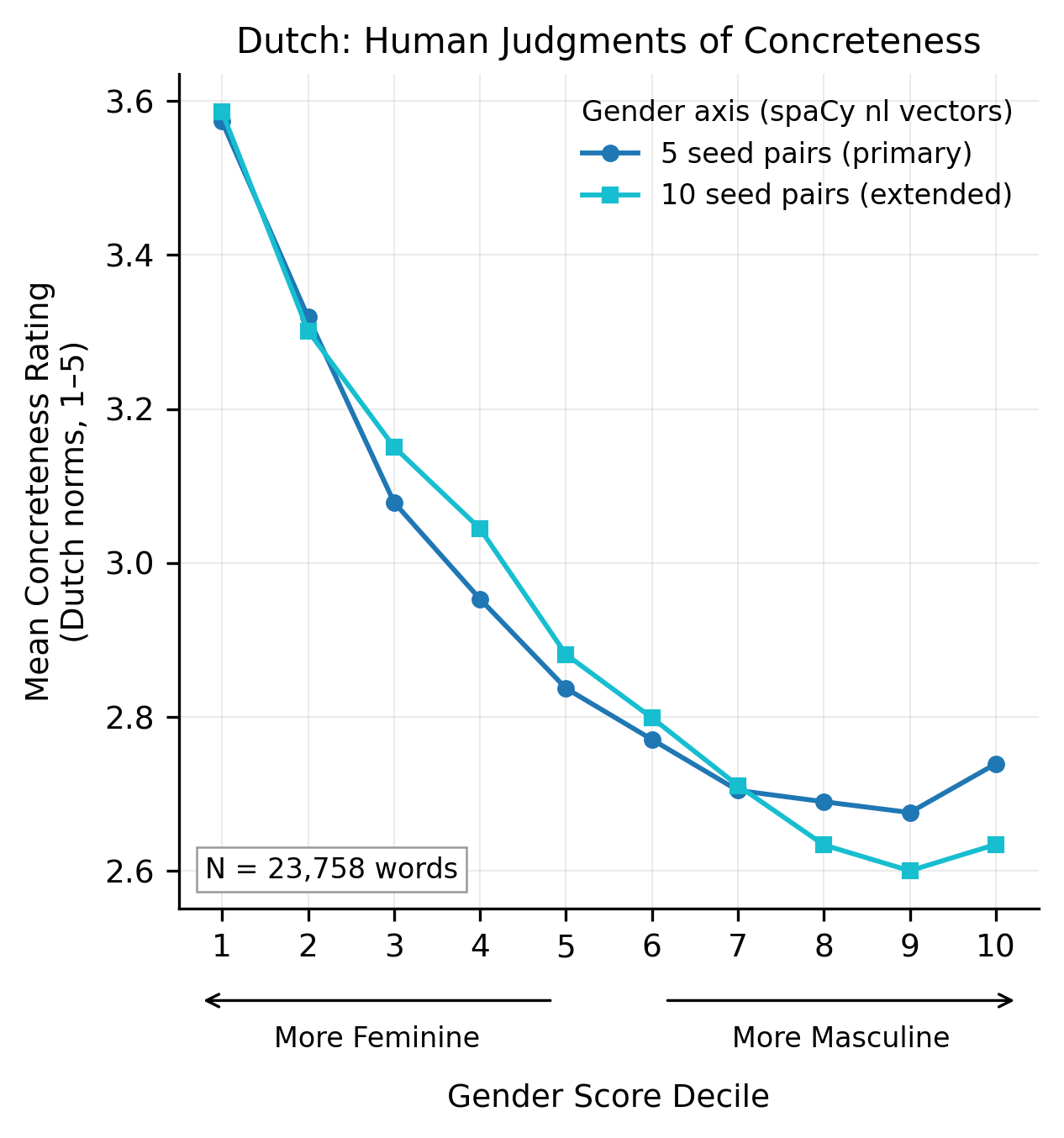}
  \caption{\textbf{More masculine Dutch words are more abstract.} Mean human-rated concreteness (1, abstract; 5, concrete)\cite{brysbaert2014dutch} of the $n = 23{,}758$ Dutch words within each decile of gender association (1, most feminine; 10, most masculine), as in Fig.~1a. Gender associations were measured in Dutch-trained word vectors (spaCy \protect\code{nl_core_news_lg}) along an axis built from five kinship and role seed pairs (\emph{man}/\emph{vrouw}, \emph{jongen}/\emph{meisje}, \emph{mannelijk}/\emph{vrouwelijk}, \emph{vader}/\emph{moeder}, \emph{broer}/\emph{zus}; circles) or from these and five further pairs (squares); deciles were computed separately for each axis. Pearson's $r = -0.268$ (95\% CI $-0.279$ to $-0.256$) and $-0.292$ ($-0.304$ to $-0.280$), respectively; Supplementary Tables~\ref{tab:S-dutch-spec} and~\ref{tab:S-dutch-poles} report the robustness of these estimates.}
  \label{fig:S-dutch}
\end{figure}

Supplementary Fig.~\ref{fig:S-dutch} shows that the pattern of Fig.~1a replicates. Across the 23,758 rated words with a vector, more masculine words are more abstract ($r = -0.268$; 95\% CI $= -0.279$ to $-0.256$; $P < 2.2 \times 10^{-16}$; Spearman's $\rho = -0.249$; two-tailed). Mean concreteness falls from 3.57 in the most feminine decile to 2.74 in the most masculine decile (minimum 2.68 in the ninth decile), a gap of 0.83 scale points between the extreme deciles, compared with 0.59 averaged across the 12 English models in Fig.~1a. The extended ten-pair axis gives $r = -0.292$ ($-0.304$ to $-0.280$) and a gap of 0.95 between the extreme deciles (3.59 versus 2.63).

The association is robust to the lexical properties examined for English in Supplementary Section~\ref{si:text-covariates} (Supplementary Table~\ref{tab:S-dutch-spec}). Controlling for word frequency left the estimate unchanged ($\beta = -0.269$), controlling for length reduced it slightly ($\beta = -0.235$), and controlling for frequency, length and age of acquisition together gave $\beta = -0.245$ ($-0.258$ to $-0.231$; $n = 23{,}398$). The correlation was negative in every quintile of word frequency ($r = -0.233$ to $-0.304$), and it was unchanged or slightly stronger after excluding words containing a gendered morpheme or kinship stem together with the seed words ($r = -0.268$; $n = 23{,}062$) and after excluding the 5\% or 10\% of words with the most extreme gender scores ($r = -0.281$ and $-0.284$), so it is not carried by a small set of explicitly gendered words. The same holds for the extended axis in every specification ($\beta$ or $r$ between $-0.221$ and $-0.335$).

The Dutch result is, however, more sensitive to the choice of seed pairs than the English results, and Supplementary Table~\ref{tab:S-dutch-poles} reports this sensitivity in full, together with the five most feminine and five most masculine words under each axis. Every axis built from several unambiguous kinship or role pairs is negative: the primary set ($-0.268$), the extended set ($-0.292$), each of the five axes obtained by dropping one pair from the primary set ($-0.148$ to $-0.339$), and the first principal component of the ten pair differences ($-0.215$), which captures 43\% of their variance and on which all ten pairs load positively (0.38 to 0.84). The feminine extreme of these axes is populated by words for girls, brides and feminine appearance (\emph{bruidsmeisje}, \emph{tienermeisje}, \emph{brunette}, \emph{lingerie}) and the masculine extreme by words for leaders and military or feudal roles (\emph{aanvoerder}, \emph{voorman}, \emph{bevelhebber}, \emph{veldheer}, \emph{leenheer}), as expected of a gender axis. Single pairs are noisier: six of the ten kinship and role pairs taken alone are negative (from \emph{heer}/\emph{dame}, $-0.375$, to \emph{opa}/\emph{oma}, $-0.136$), whereas \emph{oom}/\emph{tante} and \emph{mannelijk}/\emph{vrouwelijk} are close to zero and \emph{zoon}/\emph{dochter} and \emph{man}/\emph{vrouw} are positive ($0.122$ and $0.239$). The \emph{man}/\emph{vrouw} axis illustrates why single pairs are unreliable in a static vector table: its masculine extreme consists of colloquial terms for a foolish or idle man (\emph{prutser}, \emph{malloot}, \emph{mafkees}, \emph{hooligan}), reflecting the use of \emph{man} as an informal term of address, and its feminine extreme of abstract nouns for sexuality and motherhood (\emph{moederschap}, \emph{seksualiteit}, \emph{zwangerschap}), so that it conflates gender with register and with the concreteness of the seed words themselves. Averaging across several pairs removes this pair-specific variance, which is the rationale for the multi-pair construction of Eq.~\ref{eq:gender-axis}.

The pronoun pairs of the English axis cannot be used in Dutch, and including them reverses the sign (pronoun set, $r = 0.217$; \emph{hij}/\emph{zij} alone, $0.365$; \emph{hem}/\emph{haar} alone, $0.210$). These axes are visibly not gender axes: the ``feminine'' extreme of the \emph{hij}/\emph{zij} axis is \emph{wij}, \emph{we}, \emph{hun}, \emph{zij} and \emph{gezamenlijk} (we, they, their, jointly), because \emph{zij} also means \emph{they}, and that of the \emph{hem}/\emph{haar} axis is \emph{van}, \emph{hun}, \emph{alsook} and \emph{rondom} (of, their, as well as, around), because \emph{haar} also means \emph{hair} and is a possessive. Seed pairs for other languages should therefore be checked for this kind of ambiguity before an axis is built, and the extremes of the resulting axis inspected, as done here.

In sum, with seed pairs that refer unambiguously to women and men, the association between masculine connotation and abstractness replicates in Dutch with a magnitude at the upper end of the English estimates, in a language whose nouns carry no grammatical gender. Extending the test to languages with grammatical gender, and to languages further from English, requires concreteness norms of comparable scale, which currently exist at the scale of ten thousand words or more only for Estonian and Mandarin\cite{proos2023, xu2020}. The Estonian, Chinese and Persian tests are reported in Supplementary Sections~\ref{si:estonian} to~\ref{si:persian}.

\begin{table}[!htbp]
  \centering
  \caption{\textbf{The male abstraction bias in Dutch is robust to lexical covariates and to the composition of the vocabulary.} Each row gives the association between the gender score of a Dutch word (positive, masculine; spaCy \protect\code{nl_core_news_lg} vectors) and its human-rated concreteness\cite{brysbaert2014dutch}, for the primary five-pair axis (left) and the extended ten-pair axis (right). Rows labelled $\beta$ are standardized coefficients on concreteness from regressions of the $z$-scored gender score on $z$-scored concreteness and the covariates listed, with heteroskedasticity-robust (HC1) 95\% confidence intervals; all other rows are two-tailed Pearson's correlations with Fisher-transformed intervals. Word frequency is the Zipf-scaled frequency of \protect\code{wordfreq}\cite{speer2022wordfreq}; age of acquisition is from the same norms. $P$ values are two-tailed; $n$ is the number of words.}
  \label{tab:S-dutch-spec}
  \begingroup\footnotesize\setlength{\tabcolsep}{4pt}
  \resizebox{\textwidth}{!}{%
  \begin{tabular}{@{}>{\raggedright\arraybackslash}p{6.2cm}lrrlrr@{}}
    \toprule
    & \multicolumn{3}{c}{Primary axis (5 pairs)} & \multicolumn{3}{c}{Extended axis (10 pairs)} \\
    \cmidrule(lr){2-4}\cmidrule(lr){5-7}
    Specification & Estimate [95\% CI] & $P$ & $n$ & Estimate [95\% CI] & $P$ & $n$ \\
    \midrule
    No controls (Supplementary Fig.~\ref{fig:S-dutch}) & $-0.268$ [$-0.279$, $-0.256$] & $<2.2\times10^{-16}$ & 23{,}758 & $-0.292$ [$-0.304$, $-0.280$] & $<2.2\times10^{-16}$ & 23{,}758 \\
    Excluding lexically gendered terms and seed words & $-0.268$ [$-0.280$, $-0.256$] & $<2.2\times10^{-16}$ & 23{,}062 & $-0.296$ [$-0.307$, $-0.284$] & $<2.2\times10^{-16}$ & 23{,}062 \\
    Excluding the 5\% most strongly gendered terms & $-0.281$ [$-0.293$, $-0.269$] & $<2.2\times10^{-16}$ & 22{,}570 & $-0.283$ [$-0.295$, $-0.271$] & $<2.2\times10^{-16}$ & 22{,}570 \\
    Excluding the 10\% most strongly gendered terms & $-0.284$ [$-0.296$, $-0.271$] & $<2.2\times10^{-16}$ & 21{,}382 & $-0.270$ [$-0.283$, $-0.258$] & $<2.2\times10^{-16}$ & 21{,}382 \\
    $\beta$, controlling for word frequency & $-0.269$ [$-0.281$, $-0.256$] & $<2.2\times10^{-16}$ & 23{,}407 & $-0.293$ [$-0.305$, $-0.281$] & $<2.2\times10^{-16}$ & 23{,}407 \\
    $\beta$, controlling for length & $-0.235$ [$-0.247$, $-0.223$] & $<2.2\times10^{-16}$ & 23{,}758 & $-0.281$ [$-0.293$, $-0.269$] & $<2.2\times10^{-16}$ & 23{,}758 \\
    $\beta$, controlling for frequency and length & $-0.231$ [$-0.243$, $-0.218$] & $<2.2\times10^{-16}$ & 23{,}407 & $-0.279$ [$-0.292$, $-0.267$] & $<2.2\times10^{-16}$ & 23{,}407 \\
    $\beta$, controlling for frequency, length and age of acquisition & $-0.245$ [$-0.258$, $-0.231$] & $<2.2\times10^{-16}$ & 23{,}398 & $-0.221$ [$-0.235$, $-0.207$] & $<2.2\times10^{-16}$ & 23{,}398 \\
    Frequency quintile 1 (least frequent) & $-0.236$ [$-0.262$, $-0.208$] & $<2.2\times10^{-16}$ & 4{,}682 & $-0.281$ [$-0.308$, $-0.255$] & $<2.2\times10^{-16}$ & 4{,}682 \\
    Frequency quintile 2 & $-0.284$ [$-0.310$, $-0.257$] & $<2.2\times10^{-16}$ & 4{,}681 & $-0.310$ [$-0.336$, $-0.284$] & $<2.2\times10^{-16}$ & 4{,}681 \\
    Frequency quintile 3 & $-0.277$ [$-0.303$, $-0.250$] & $<2.2\times10^{-16}$ & 4{,}681 & $-0.298$ [$-0.323$, $-0.271$] & $<2.2\times10^{-16}$ & 4{,}681 \\
    Frequency quintile 4 & $-0.304$ [$-0.330$, $-0.278$] & $<2.2\times10^{-16}$ & 4{,}681 & $-0.335$ [$-0.361$, $-0.310$] & $<2.2\times10^{-16}$ & 4{,}681 \\
    Frequency quintile 5 (most frequent) & $-0.233$ [$-0.260$, $-0.206$] & $<2.2\times10^{-16}$ & 4{,}682 & $-0.232$ [$-0.259$, $-0.205$] & $<2.2\times10^{-16}$ & 4{,}682 \\
    \bottomrule
  \end{tabular}}\endgroup
\end{table}

\begin{table}[!htbp]
  \centering
  \caption{\textbf{Sensitivity of the Dutch result to the seed pairs used to build the gender axis.} Two-tailed Pearson's correlations between the gender score and human-rated concreteness across the 23,758 rated words with a vector (positive values, more masculine words are more concrete), with Fisher-transformed 95\% confidence intervals, for axes built from the specifications listed; the last two columns give the five words with the lowest (most feminine) and highest (most masculine) scores under each axis. The first principal component is computed from the ten unit-normalised male-minus-female difference vectors of the extended set without centring\cite{bolukbasi2016} and sign-aligned with their mean. Family C uses the Dutch counterparts of the English pronoun pairs, which are lexically ambiguous (\emph{zij}, she/they; \emph{haar}, her/hair) and yield axes whose extremes are plural pronouns and function words rather than gendered words.}
  \label{tab:S-dutch-poles}
  \begingroup\scriptsize\setlength{\tabcolsep}{3pt}
  \resizebox{\textwidth}{!}{%
  \begin{tabular}{@{}>{\raggedright\arraybackslash}p{5.2cm}rlr>{\raggedright\arraybackslash}p{4.1cm}>{\raggedright\arraybackslash}p{4.1cm}@{}}
    \toprule
    Axis & Pairs & $r$ [95\% CI] & $P$ & Most feminine words & Most masculine words \\
    \midrule
    \multicolumn{6}{@{}l}{\textit{A. Axes built from several unambiguous kinship/role pairs}} \\
    Primary set (5 pairs) & 5 & $-0.268$ [$-0.279$, $-0.256$] & $<2.2\times10^{-16}$ & \emph{babydoll}, \emph{bruidsmeisje}, \emph{lesbienne}, \emph{brunette}, \emph{topless} & \emph{aanvoerder}, \emph{voorman}, \emph{voorganger}, \emph{manschappen}, \emph{bevelhebber} \\
    Extended set (10 pairs) & 10 & $-0.292$ [$-0.304$, $-0.280$] & $<2.2\times10^{-16}$ & \emph{babydoll}, \emph{brunette}, \emph{blondine}, \emph{lingerie}, \emph{tienermeisje} & \emph{leenheer}, \emph{voorman}, \emph{aanvoerder}, \emph{veldheer}, \emph{hopman} \\
    Primary set without \emph{man}/\emph{vrouw} & 4 & $-0.339$ [$-0.350$, $-0.327$] & $<2.2\times10^{-16}$ & \emph{babydoll}, \emph{brunette}, \emph{topless}, \emph{bruidsmeisje}, \emph{pinup} & \emph{aanvoerder}, \emph{voorman}, \emph{voorganger}, \emph{gebroeders}, \emph{meesterzet} \\
    Primary set without \emph{jongen}/\emph{meisje} & 4 & $-0.148$ [$-0.160$, $-0.135$] & $<2.2\times10^{-16}$ & \emph{babydoll}, \emph{lingerie}, \emph{seksbom}, \emph{brunette}, \emph{lesbienne} & \emph{voorman}, \emph{veldheer}, \emph{kompaan}, \emph{aanvoerder}, \emph{opportunist} \\
    Primary set without \emph{mannelijk}/\emph{vrouwelijk} & 4 & $-0.288$ [$-0.299$, $-0.276$] & $<2.2\times10^{-16}$ & \emph{lesbienne}, \emph{babydoll}, \emph{topless}, \emph{brunette}, \emph{bruidsmeisje} & \emph{aanvoerder}, \emph{voorman}, \emph{voorganger}, \emph{veedrijver}, \emph{oudgediende} \\
    Primary set without \emph{vader}/\emph{moeder} & 4 & $-0.229$ [$-0.241$, $-0.216$] & $<2.2\times10^{-16}$ & \emph{bruidsmeisje}, \emph{topless}, \emph{babydoll}, \emph{lesbienne}, \emph{brunette} & \emph{manschappen}, \emph{aanvoerder}, \emph{tegenstander}, \emph{loopjongen}, \emph{verschalken} \\
    Primary set without \emph{broer}/\emph{zus} & 4 & $-0.226$ [$-0.238$, $-0.214$] & $<2.2\times10^{-16}$ & \emph{bruidsmeisje}, \emph{babydoll}, \emph{lesbienne}, \emph{brunette}, \emph{tienermeisje} & \emph{aanvoerder}, \emph{manschappen}, \emph{ploegbaas}, \emph{opportunist}, \emph{voorman} \\
    First principal component of the 10 pair differences (0.43 of variance) & 10 & $-0.215$ [$-0.227$, $-0.202$] & $<2.2\times10^{-16}$ & \emph{babydoll}, \emph{brunette}, \emph{blondine}, \emph{lingerie}, \emph{seksbom} & \emph{leenheer}, \emph{veldheer}, \emph{voorman}, \emph{hopman}, \emph{kompaan} \\
    \multicolumn{6}{@{}l}{\textit{B. Single kinship/role pairs}} \\
    \emph{man}/\emph{vrouw} & 1 & $0.239$ [$0.227$, $0.251$] & $<2.2\times10^{-16}$ & \emph{moederschap}, \emph{seksualiteit}, \emph{masturbatie}, \emph{zwangerschap}, \emph{echtgenote} & \emph{prutser}, \emph{malloot}, \emph{mafkees}, \emph{hooligan}, \emph{pakweg} \\
    \emph{jongen}/\emph{meisje} & 1 & $-0.288$ [$-0.300$, $-0.277$] & $<2.2\times10^{-16}$ & \emph{meisje}, \emph{bruidsmeisje}, \emph{tienermeisje}, \emph{poesje}, \emph{blondje} & \emph{koeren}, \emph{zieden}, \emph{voortrekken}, \emph{venten}, \emph{poeieren} \\
    \emph{mannelijk}/\emph{vrouwelijk} & 1 & $0.042$ [$0.029$, $0.054$] & $1.3\times10^{-10}$ & \emph{ontwerpster}, \emph{kunstenares}, \emph{styliste}, \emph{schrijfster}, \emph{presentatrice} & \emph{kapotslaan}, \emph{stompen}, \emph{blaffer}, \emph{hoorndol}, \emph{hooivork} \\
    \emph{vader}/\emph{moeder} & 1 & $-0.210$ [$-0.222$, $-0.198$] & $<2.2\times10^{-16}$ & \emph{babydoll}, \emph{borsten}, \emph{zwangerschap}, \emph{borstvoeding}, \emph{masturberen} & \emph{voorman}, \emph{toenmalig}, \emph{historicus}, \emph{hopman}, \emph{tacticus} \\
    \emph{broer}/\emph{zus} & 1 & $-0.302$ [$-0.313$, $-0.290$] & $<2.2\times10^{-16}$ & \emph{mama}, \emph{beauty}, \emph{babydoll}, \emph{pinup}, \emph{mam} & \emph{voorman}, \emph{aanvoerder}, \emph{meesterzet}, \emph{voorganger}, \emph{luitenant} \\
    \emph{zoon}/\emph{dochter} & 1 & $0.122$ [$0.110$, $0.135$] & $<2.2\times10^{-16}$ & \emph{zomeren}, \emph{blonderen}, \emph{verleidelijk}, \emph{voorbijkomen}, \emph{vlechten} & \emph{nazaat}, \emph{cynicus}, \emph{oudoom}, \emph{schlemiel}, \emph{fanaticus} \\
    \emph{oom}/\emph{tante} & 1 & $0.034$ [$0.022$, $0.047$] & $1.1\times10^{-7}$ & \emph{babydoll}, \emph{lingerie}, \emph{prachtvrouw}, \emph{brunette}, \emph{stoeipoes} & \emph{koopman}, \emph{oudoom}, \emph{hopman}, \emph{kompaan}, \emph{leenheer} \\
    \emph{koning}/\emph{koningin} & 1 & $-0.332$ [$-0.343$, $-0.321$] & $<2.2\times10^{-16}$ & \emph{actrice}, \emph{prinses}, \emph{zwerfster}, \emph{bruidsmeisje}, \emph{koningin} & \emph{leenheer}, \emph{belegering}, \emph{veldheer}, \emph{opstandeling}, \emph{onderhandeling} \\
    \emph{heer}/\emph{dame} & 1 & $-0.375$ [$-0.386$, $-0.364$] & $<2.2\times10^{-16}$ & \emph{stoeipoes}, \emph{jongedame}, \emph{blondine}, \emph{tienermeisje}, \emph{meisje} & \emph{voorzitter}, \emph{secretaris}, \emph{griffier}, \emph{raadsheer}, \emph{aldus} \\
    \emph{opa}/\emph{oma} & 1 & $-0.136$ [$-0.149$, $-0.124$] & $<2.2\times10^{-16}$ & \emph{brunette}, \emph{seksbom}, \emph{tiener}, \emph{lesbienne}, \emph{masturbatie} & \emph{oudoom}, \emph{toenmalig}, \emph{kompaan}, \emph{hopman}, \emph{voorman} \\
    \multicolumn{6}{@{}l}{\textit{C. Pronoun-based seeds (\emph{zij} = \emph{they}; \emph{haar} = \emph{hair})}} \\
    Pronoun set (\emph{man}/\emph{vrouw}, \emph{hij}/\emph{zij}, \emph{hem}/\emph{haar}, \emph{mannelijk}/\emph{vrouwelijk}, \emph{jongen}/\emph{meisje}) & 5 & $0.217$ [$0.205$, $0.229$] & $<2.2\times10^{-16}$ & \emph{styliste}, \emph{lingerie}, \emph{hun}, \emph{moederschap}, \emph{bruidsmeisje} & \emph{prutser}, \emph{ie}, \emph{klaploper}, \emph{wegrenner}, \emph{idioot} \\
    \emph{hij}/\emph{zij} & 1 & $0.365$ [$0.353$, $0.376$] & $<2.2\times10^{-16}$ & \emph{wij}, \emph{we}, \emph{hun}, \emph{zij}, \emph{gezamenlijk} & \emph{hij}, \emph{ie}, \emph{bibber}, \emph{prutser}, \emph{zwabber} \\
    \emph{hem}/\emph{haar} & 1 & $0.210$ [$0.198$, $0.222$] & $<2.2\times10^{-16}$ & \emph{van}, \emph{hun}, \emph{alsook}, \emph{weerspiegelen}, \emph{rondom} & \emph{hem}, \emph{ie}, \emph{mij}, \emph{idioot}, \emph{geintje} \\
    \bottomrule
  \end{tabular}}\endgroup
\end{table}

\section{Replication of the text-based result in Estonian}
\label{si:estonian}

Estonian offers a second, and typologically more distant, test of the male abstraction bias. It is a Finno-Ugric rather than an Indo-European language, and it is a genderless language: nouns, adjectives and pronouns carry no grammatical gender, and a single third-person pronoun (\emph{tema}/\emph{ta}) covers both \emph{he} and \emph{she}. The concreteness norms of Proos and Aigro\cite{proos2023} provide ratings for 35,979 Estonian lemmas on a continuous scale from 0 (abstract) to 10 (concrete), collected from 2,278 native speakers; we used all of them (mean 6.39; s.d.\ 1.88). Gender associations were measured in fastText's Estonian word vectors\cite{grave2018}, which were trained on the Estonian portions of Common Crawl and Wikipedia (2 million words, 300 dimensions) and are, like the Dutch spaCy vectors, a model trained on the language itself rather than an English or multilingual model. The 33,031 rated lemmas (91.8\%) that occur in the fastText vocabulary were scored; words absent from it were not reconstructed from character $n$-grams.

The procedure follows the Dutch replication shown above in every respect but two. First, because Estonian has no gendered pronouns, both seed sets consist entirely of kinship and role nouns and the adjective pair for \emph{masculine}/\emph{feminine}: the primary axis uses \emph{mees}/\emph{naine} (man/woman), \emph{poiss}/\emph{tüdruk} (boy/girl), \emph{mehelik}/\emph{naiselik} (masculine/feminine), \emph{isa}/\emph{ema} (father/mother) and \emph{vend}/\emph{õde} (brother/sister), which are the Estonian counterparts of the five pairs of the Dutch primary set, and the extended axis adds \emph{poeg}/\emph{tütar} (son/daughter), \emph{onu}/\emph{tädi} (uncle/aunt), \emph{kuningas}/\emph{kuninganna} (king/queen), \emph{härra}/\emph{proua} (Mr/Mrs) and \emph{vanaisa}/\emph{vanaema} (grandfather/grandmother). Both axes order the male above the female member in all 20 gendered word pairs that were not used as seeds (for example \emph{peigmees}/\emph{pruut}, \emph{näitleja}/\emph{näitlejanna}, \emph{noormees}/\emph{neiu}, \emph{mehed}/\emph{naised}). Second, word frequency was taken from the order of the fastText vocabulary, which lists words by their frequency in the training corpus, and the Estonian norms provide no age-of-acquisition ratings; in their place we use the rater standard deviation distributed with the norms to restrict the analysis to reliably rated words.

\begin{figure}[!htb]
  \centering
  \includegraphics[width=0.55\textwidth]{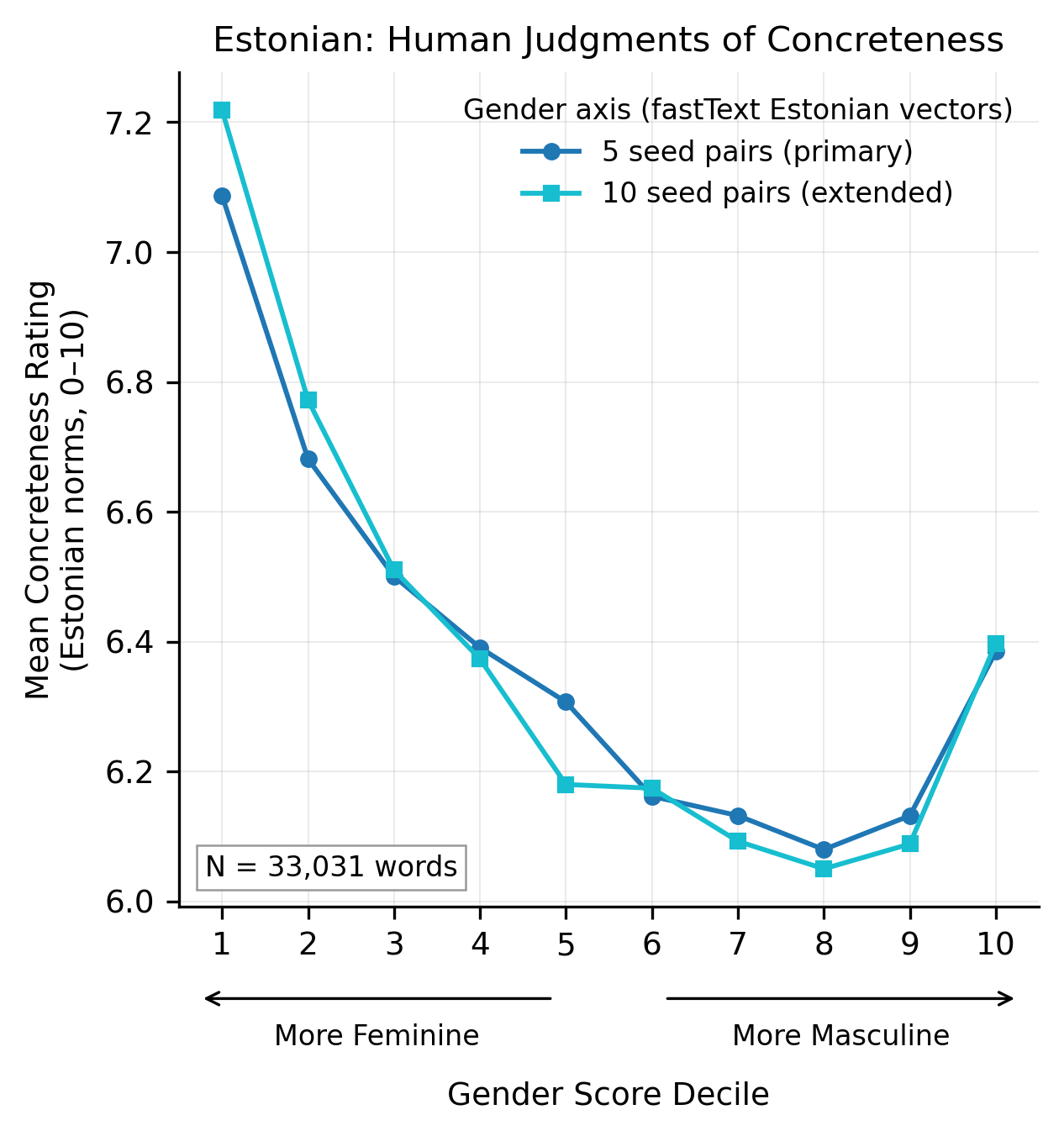}
  \caption{\textbf{More masculine Estonian words are more abstract.} Mean human-rated concreteness (0, abstract; 10, concrete)\cite{proos2023} of the $n = 33{,}031$ Estonian words within each decile of gender association (1, most feminine; 10, most masculine), as in Fig.~1a. Gender associations were measured in fastText's Estonian word vectors\cite{grave2018} along an axis built from five kinship and role seed pairs (\emph{mees}/\emph{naine}, \emph{poiss}/\emph{tüdruk}, \emph{mehelik}/\emph{naiselik}, \emph{isa}/\emph{ema}, \emph{vend}/\emph{õde}; circles) or from these and five further pairs (squares); deciles were computed separately for each axis. Pearson's $r = -0.124$ (95\% CI $-0.135$ to $-0.114$) and $-0.141$ ($-0.151$ to $-0.130$), respectively. The most masculine decile is populated by nouns for men in military, sporting and occupational roles, whose referents are people and therefore concrete (Supplementary Section~\ref{si:estonian}); Supplementary Tables~\ref{tab:S-estonian-spec} and~\ref{tab:S-estonian-poles} report the robustness of these estimates.}
  \label{fig:S-estonian}
\end{figure}

Supplementary Fig.~\ref{fig:S-estonian} shows that the pattern of Fig.~1a replicates in Estonian. Across the 33,031 words, more masculine words are more abstract ($r = -0.124$; 95\% CI $= -0.135$ to $-0.114$; $P < 2.2 \times 10^{-16}$; Spearman's $\rho = -0.124$; two-tailed). Mean concreteness declines from 7.09 in the most feminine decile to 6.08 in the eighth decile, and is 6.39 in the most masculine decile. The difference between the extreme deciles is 0.70 points on the 10-point scale, or 0.37 s.d.\ of the ratings; between the first and eighth deciles it is 1.01 points, or 0.54 s.d. For comparison, the corresponding difference between the extreme deciles is 0.28 s.d.\ for English single words (Supplementary Section~\ref{si:words-phrases}) and 0.84 s.d.\ for Dutch (Supplementary Section~\ref{si:dutch}), and the Estonian correlation lies between the mean of the twelve English single-word estimates ($-0.085$; Supplementary Table~\ref{tab:S-fig1-human}) and the Dutch estimate ($-0.268$). The extended ten-pair axis gives $r = -0.141$ ($-0.151$ to $-0.130$), a difference of 0.82 points (0.44 s.d.) between the extreme deciles and of 1.17 points (0.62 s.d.) between the first and eighth.

Supplementary Table~\ref{tab:S-estonian-spec} shows that the association is robust in the same ways as in Dutch. Controlling for word frequency, for length, or for both left it unchanged ($\beta = -0.126$, $-0.126$ and $-0.127$), it was negative in every quintile of word frequency ($r = -0.107$ to $-0.162$), and it was slightly stronger among reliably rated words ($r = -0.138$; $n = 16{,}609$), after excluding words that contain a gendered morpheme or a kinship or role stem together with the seed words ($r = -0.124$; $n = 32{,}263$), and after excluding the 5\% or 10\% of words with the most extreme gender scores ($r = -0.143$ and $-0.148$). The extended axis gives $r$ or $\beta$ between $-0.131$ and $-0.161$ across the same specifications.

The upturn in the most masculine decile has an identifiable source. The words with the highest gender scores are nouns for men in roles---military ranks (\emph{nooremseersant}, \emph{kapral}, \emph{kolonel}, \emph{major}), sports positions and occupations formed with the compound head \emph{-mees} `man' (\emph{roolimees} `helmsman', \emph{vilemees} `referee', \emph{jahimees} `hunter', \emph{ärimees} `businessman'), and kin (\emph{vend}, \emph{poeg}, \emph{pojapoeg})---together with weapons and military equipment (\emph{padrun} `cartridge', \emph{relvastus} `armament', \emph{sihik} `gunsight'). People and objects are perceivable referents, so this decile is more concrete than the two below it, just as the most masculine decile turns upward for some of the English models in Fig.~1a and as the person-noun exclusion of Fig.~2 anticipates. Removing the 5\% or 10\% of words with the most extreme gender scores removes most of these words and, as noted, strengthens the correlation; after the 10\% exclusion the decile profile is nearly flat at the masculine end (6.08 in the ninth decile and 6.13 in the tenth). A systematic exclusion of person-denoting nouns, as performed for English with WordNet, would require the Estonian WordNet and is a natural extension.

Supplementary Table~\ref{tab:S-estonian-poles} reports the sensitivity of the result to the seed pairs. As in Dutch, every axis built from several pairs is negative: the primary set ($-0.124$), the extended set ($-0.141$), each axis obtained by dropping one pair from the primary set ($-0.052$ to $-0.165$) and the first principal component of the ten pair differences ($-0.150$; 34\% of their variance; all ten pairs loading positively, 0.29 to 0.73). The feminine extremes of these axes are words for women, women's clothing and cosmetics (\emph{sõbranna} `female friend', \emph{õde} `sister', \emph{kleit} `dress', \emph{pulmakleit} `wedding dress', \emph{meik} `make-up', \emph{ämmaemand} `midwife') and the masculine extremes the role nouns listed above, as expected of a gender axis. Single pairs are again noisier: seven of the ten are negative on their own (from \emph{isa}/\emph{ema}, $-0.295$, to \emph{onu}/\emph{tädi}, $-0.052$), \emph{poiss}/\emph{tüdruk} and \emph{kuningas}/\emph{kuninganna} are indistinguishable from zero, and \emph{mees}/\emph{naine} is positive ($0.054$). The last is the same pair that reversed in Dutch (\emph{man}/\emph{vrouw}, $0.239$) and for a parallel reason: \emph{mees} is the productive compound head for male occupational nouns, so its neighbourhood consists of concrete role nouns (\emph{raudmees} `strongman', \emph{varumees} `substitute player', \emph{tsiklimees} `motorcyclist'), whereas the neighbourhood of \emph{naine} is dominated by nouns of pregnancy and motherhood (\emph{emadus} `motherhood', \emph{rasedus} `pregnancy', \emph{raseduskuu} `month of pregnancy'), so that the bare pair conflates gender with the concreteness of the seed words' own derivatives. The adjective pair \emph{mehelik}/\emph{naiselik} shows a further, fastText-specific weakness: because fastText represents words through character $n$-grams, its masculine extreme includes unrelated words that merely share the suffix string (\emph{ahelik}, \emph{võimeline}, \emph{liikmeline}), which is why that pair contributes little to the axis (loading 0.29) and why its omission leaves the primary estimate unchanged ($-0.124$). Averaging over several unambiguous pairs removes this pair-specific variance, as it did in Dutch.

In sum, the male abstraction bias replicates in a genderless Finno-Ugric language, in a second embedding family, with a magnitude between the English and Dutch estimates. Across the two replications, the same methodological regularities appear: axes averaged over several kinship and role pairs agree in sign and are validated by held-out gendered pairs, whereas the bare man/woman pair on its own is unreliable in both languages; and the words at the masculine extreme are predominantly nouns for men in roles, whose concreteness as person nouns attenuates rather than produces the association.

\begin{table}[!htbp]
  \centering
  \caption{\textbf{The male abstraction bias in Estonian is robust to lexical covariates and to the composition of the vocabulary.} Each row gives the association between the gender score of an Estonian word (positive, masculine; fastText \protect\code{cc.et.300} vectors\cite{grave2018}) and its human-rated concreteness\cite{proos2023}, for the primary five-pair axis (left) and the extended ten-pair axis (right). Rows labelled $\beta$ are standardized coefficients on concreteness from regressions of the $z$-scored gender score on $z$-scored concreteness and the covariates listed, with heteroskedasticity-robust (HC1) 95\% confidence intervals; all other rows are two-tailed Pearson's correlations with Fisher-transformed intervals. Word frequency is the log rank of the word in the fastText vocabulary, which is ordered by corpus frequency; the rater s.d.\ is distributed with the norms. $P$ values are two-tailed; $n$ is the number of words.}
  \label{tab:S-estonian-spec}
  \begingroup\footnotesize\setlength{\tabcolsep}{4pt}
  \resizebox{\textwidth}{!}{%
  \begin{tabular}{@{}>{\raggedright\arraybackslash}p{6.2cm}lrrlrr@{}}
    \toprule
    & \multicolumn{3}{c}{Primary axis (5 pairs)} & \multicolumn{3}{c}{Extended axis (10 pairs)} \\
    \cmidrule(lr){2-4}\cmidrule(lr){5-7}
    Specification & Estimate [95\% CI] & $P$ & $n$ & Estimate [95\% CI] & $P$ & $n$ \\
    \midrule
    No controls (Supplementary Fig.~\ref{fig:S-estonian}) & $-0.124$ [$-0.135$, $-0.114$] & $<2.2\times10^{-16}$ & 33{,}031 & $-0.141$ [$-0.151$, $-0.130$] & $<2.2\times10^{-16}$ & 33{,}031 \\
    Excluding lexically gendered terms and seed words & $-0.124$ [$-0.135$, $-0.113$] & $<2.2\times10^{-16}$ & 32{,}263 & $-0.141$ [$-0.152$, $-0.130$] & $<2.2\times10^{-16}$ & 32{,}263 \\
    Excluding the 5\% most strongly gendered terms & $-0.143$ [$-0.154$, $-0.133$] & $<2.2\times10^{-16}$ & 31{,}379 & $-0.161$ [$-0.172$, $-0.151$] & $<2.2\times10^{-16}$ & 31{,}379 \\
    Excluding the 10\% most strongly gendered terms & $-0.148$ [$-0.159$, $-0.137$] & $<2.2\times10^{-16}$ & 29{,}727 & $-0.161$ [$-0.172$, $-0.150$] & $<2.2\times10^{-16}$ & 29{,}727 \\
    Rater s.d.\ of concreteness $\leq$ median (reliably rated words) & $-0.138$ [$-0.153$, $-0.123$] & $<2.2\times10^{-16}$ & 16{,}609 & $-0.150$ [$-0.165$, $-0.135$] & $<2.2\times10^{-16}$ & 16{,}609 \\
    $\beta$, controlling for word frequency & $-0.126$ [$-0.137$, $-0.115$] & $<2.2\times10^{-16}$ & 33{,}031 & $-0.140$ [$-0.151$, $-0.129$] & $<2.2\times10^{-16}$ & 33{,}031 \\
    $\beta$, controlling for length & $-0.126$ [$-0.137$, $-0.115$] & $<2.2\times10^{-16}$ & 33{,}031 & $-0.142$ [$-0.153$, $-0.131$] & $<2.2\times10^{-16}$ & 33{,}031 \\
    $\beta$, controlling for frequency and length & $-0.127$ [$-0.138$, $-0.116$] & $<2.2\times10^{-16}$ & 33{,}031 & $-0.141$ [$-0.152$, $-0.130$] & $<2.2\times10^{-16}$ & 33{,}031 \\
    Frequency quintile 1 (least frequent) & $-0.107$ [$-0.131$, $-0.083$] & $<2.2\times10^{-16}$ & 6{,}607 & $-0.136$ [$-0.159$, $-0.112$] & $<2.2\times10^{-16}$ & 6{,}607 \\
    Frequency quintile 2 & $-0.116$ [$-0.140$, $-0.092$] & $<2.2\times10^{-16}$ & 6{,}606 & $-0.149$ [$-0.173$, $-0.126$] & $<2.2\times10^{-16}$ & 6{,}606 \\
    Frequency quintile 3 & $-0.119$ [$-0.142$, $-0.095$] & $<2.2\times10^{-16}$ & 6{,}606 & $-0.139$ [$-0.162$, $-0.115$] & $<2.2\times10^{-16}$ & 6{,}606 \\
    Frequency quintile 4 & $-0.120$ [$-0.144$, $-0.096$] & $<2.2\times10^{-16}$ & 6{,}606 & $-0.131$ [$-0.155$, $-0.107$] & $<2.2\times10^{-16}$ & 6{,}606 \\
    Frequency quintile 5 (most frequent) & $-0.162$ [$-0.186$, $-0.139$] & $<2.2\times10^{-16}$ & 6{,}606 & $-0.148$ [$-0.171$, $-0.124$] & $<2.2\times10^{-16}$ & 6{,}606 \\
    \bottomrule
  \end{tabular}}\endgroup
\end{table}

\begin{table}[!htbp]
  \centering
  \caption{\textbf{Sensitivity of the Estonian result to the seed pairs used to build the gender axis.} Two-tailed Pearson's correlations between the gender score and human-rated concreteness across the 33,031 rated words with a fastText vector (positive values, more masculine words are more concrete), with Fisher-transformed 95\% confidence intervals, for axes built from the specifications listed; the last two columns give the five words with the lowest (most feminine) and highest (most masculine) scores under each axis. The first principal component is computed from the ten unit-normalised male-minus-female difference vectors of the extended set without centring\cite{bolukbasi2016} and sign-aligned with their mean.}
  \label{tab:S-estonian-poles}
  \begingroup\scriptsize\setlength{\tabcolsep}{3pt}
  \resizebox{\textwidth}{!}{%
  \begin{tabular}{@{}>{\raggedright\arraybackslash}p{5.2cm}rlr>{\raggedright\arraybackslash}p{4.1cm}>{\raggedright\arraybackslash}p{4.1cm}@{}}
    \toprule
    Axis & Pairs & $r$ [95\% CI] & $P$ & Most feminine words & Most masculine words \\
    \midrule
    \multicolumn{6}{@{}l}{\textit{A. Axes built from several kinship/role pairs}} \\
    Primary set (5 pairs) & 5 & $-0.124$ [$-0.135$, $-0.114$] & $<2.2\times10^{-16}$ & \emph{sõbranna}, \emph{õde}, \emph{kleit}, \emph{pulmakleit}, \emph{rootslanna} & \emph{vend}, \emph{roolimees}, \emph{raudmees}, \emph{ülem}, \emph{pealik} \\
    Extended set (10 pairs) & 10 & $-0.141$ [$-0.151$, $-0.130$] & $<2.2\times10^{-16}$ & \emph{sõbranna}, \emph{tädi}, \emph{rootslanna}, \emph{kaaslanna}, \emph{eksnaine} & \emph{vend}, \emph{onu}, \emph{pojapoeg}, \emph{vilemees}, \emph{poeg} \\
    Primary set without \emph{mees}/\emph{naine} & 4 & $-0.165$ [$-0.175$, $-0.154$] & $<2.2\times10^{-16}$ & \emph{kleit}, \emph{sõbranna}, \emph{õde}, \emph{meik}, \emph{pulmakleit} & \emph{vend}, \emph{poeg}, \emph{pojapoeg}, \emph{pealik}, \emph{isa} \\
    Primary set without \emph{poiss}/\emph{tüdruk} & 4 & $-0.148$ [$-0.158$, $-0.137$] & $<2.2\times10^{-16}$ & \emph{õde}, \emph{ämmaemand}, \emph{sõbranna}, \emph{kleit}, \emph{eksnaine} & \emph{vend}, \emph{raudmees}, \emph{roolimees}, \emph{mängumees}, \emph{vilemees} \\
    Primary set without \emph{mehelik}/\emph{naiselik} & 4 & $-0.124$ [$-0.135$, $-0.114$] & $<2.2\times10^{-16}$ & \emph{õde}, \emph{sõbranna}, \emph{meditsiiniõde}, \emph{kleit}, \emph{ämmaemand} & \emph{vend}, \emph{roolimees}, \emph{pealik}, \emph{vilemees}, \emph{tehnikadirektor} \\
    Primary set without \emph{isa}/\emph{ema} & 4 & $-0.052$ [$-0.062$, $-0.041$] & $<2.2\times10^{-16}$ & \emph{õde}, \emph{pulmakleit}, \emph{kleit}, \emph{meik}, \emph{sõbranna} & \emph{vend}, \emph{raudmees}, \emph{roolimees}, \emph{mängumees}, \emph{varumees} \\
    Primary set without \emph{vend}/\emph{õde} & 4 & $-0.112$ [$-0.123$, $-0.101$] & $<2.2\times10^{-16}$ & \emph{naiselik}, \emph{sõbranna}, \emph{kleit}, \emph{parfüüm}, \emph{pulmakleit} & \emph{ülem}, \emph{roolimees}, \emph{vanemveebel}, \emph{julgeolekumees}, \emph{nooremseersant} \\
    First principal component of the 10 pair differences (0.34 of variance) & 10 & $-0.150$ [$-0.160$, $-0.139$] & $<2.2\times10^{-16}$ & \emph{sõbranna}, \emph{tädi}, \emph{eksnaine}, \emph{kaaslanna}, \emph{rootslanna} & \emph{vend}, \emph{vilemees}, \emph{poeg}, \emph{pojapoeg}, \emph{ülem} \\
    \multicolumn{6}{@{}l}{\textit{B. Single kinship/role pairs}} \\
    \emph{mees}/\emph{naine} & 1 & $0.054$ [$0.043$, $0.064$] & $<2.2\times10^{-16}$ & \emph{naine}, \emph{ema}, \emph{emadus}, \emph{eksnaine}, \emph{krahvinna} & \emph{raudmees}, \emph{varumees}, \emph{vilemees}, \emph{mees}, \emph{ekskavaator} \\
    \emph{poiss}/\emph{tüdruk} & 1 & $-0.007$ [$-0.018$, $0.004$] & $0.20$ & \emph{tüdruk}, \emph{meik}, \emph{kangelanna}, \emph{brünett}, \emph{seksikas} & \emph{poiss}, \emph{paipoiss}, \emph{pojapoeg}, \emph{seenior}, \emph{karjapoiss} \\
    \emph{mehelik}/\emph{naiselik} & 1 & $-0.055$ [$-0.066$, $-0.044$] & $<2.2\times10^{-16}$ & \emph{naiselik}, \emph{naiselikkus}, \emph{parfüüm}, \emph{meigikunstnik}, \emph{naiselikum} & \emph{mehelik}, \emph{riigimehelikult}, \emph{ahelik}, \emph{riigimehelik}, \emph{võimeline} \\
    \emph{isa}/\emph{ema} & 1 & $-0.295$ [$-0.305$, $-0.285$] & $<2.2\times10^{-16}$ & \emph{ema}, \emph{rinnapiim}, \emph{vastsündinu}, \emph{raseduskuu}, \emph{sõbranna} & \emph{isa}, \emph{ülem}, \emph{pojapoeg}, \emph{juhtfiguur}, \emph{etteotsa} \\
    \emph{vend}/\emph{õde} & 1 & $-0.116$ [$-0.126$, $-0.105$] & $<2.2\times10^{-16}$ & \emph{õde}, \emph{meditsiiniõde}, \emph{medõde}, \emph{viit}, \emph{ämmaemand} & \emph{vend}, \emph{poeg}, \emph{raudmees}, \emph{mängumees}, \emph{relvavend} \\
    \emph{poeg}/\emph{tütar} & 1 & $-0.101$ [$-0.111$, $-0.090$] & $<2.2\times10^{-16}$ & \emph{tütar}, \emph{õetütar}, \emph{paruness}, \emph{sõbranna}, \emph{pisitütar} & \emph{poeg}, \emph{munk}, \emph{vennapoeg}, \emph{jahimees}, \emph{nooremseersant} \\
    \emph{onu}/\emph{tädi} & 1 & $-0.052$ [$-0.063$, $-0.042$] & $<2.2\times10^{-16}$ & \emph{tädi}, \emph{perenaine}, \emph{ärinaine}, \emph{täditütar}, \emph{esmamulje} & \emph{onu}, \emph{kandevõime}, \emph{poolvend}, \emph{pojapoeg}, \emph{isa} \\
    \emph{kuningas}/\emph{kuninganna} & 1 & $0.010$ [$0.000$, $0.021$] & $0.06$ & \emph{kuninganna}, \emph{iluduskuninganna}, \emph{kaaslanna}, \emph{sõbranna}, \emph{rootslanna} & \emph{kuningas}, \emph{piiskop}, \emph{peapiiskop}, \emph{varitsus}, \emph{punavägi} \\
    \emph{härra}/\emph{proua} & 1 & $-0.170$ [$-0.180$, $-0.159$] & $<2.2\times10^{-16}$ & \emph{proua}, \emph{teenijanna}, \emph{vanaproua}, \emph{lastekodu}, \emph{haiglaravi} & \emph{härra}, \emph{paljukiidetud}, \emph{oponent}, \emph{härras}, \emph{legendaarne} \\
    \emph{vanaisa}/\emph{vanaema} & 1 & $-0.176$ [$-0.187$, $-0.166$] & $<2.2\times10^{-16}$ & \emph{vanaema}, \emph{sõbranna}, \emph{cappuccino}, \emph{saiake}, \emph{piss} & \emph{vanaisa}, \emph{vanavanaisa}, \emph{pojapoeg}, \emph{eesots}, \emph{kapten} \\
    \bottomrule
  \end{tabular}}\endgroup
\end{table}

\section{Replication of the text-based result in Chinese}
\label{si:chinese}
\makeatletter
\@ifundefined{zh}{%
  \@ifpackageloaded{CJKutf8}{\newcommand{\zh}[1]{\begin{CJK*}{UTF8}{gbsn}#1\end{CJK*}}}{%
  \@ifpackageloaded{xeCJK}{\newcommand{\zh}[1]{#1}}{%
  \PackageError{supplementary}{Chinese words need CJK support: add \string\usepackage{CJKutf8} (pdflatex) or \string\usepackage{xeCJK} (XeLaTeX) to the preamble}{}}}}{}
\makeatother

Chinese provides a third test, in a non-alphabetic script and in the language with the largest number of native speakers. Like Estonian, Mandarin Chinese has no grammatical gender: nouns and adjectives are unmarked, and the third-person pronoun is a single syllable, \emph{tā}, whose written forms \zh{他} (he) and \zh{她} (she) were only differentiated in the twentieth century. Two sets of human concreteness ratings for two-character words are available. The norms of Chan and Tse\cite{chan2024} cover 25,281 two-character words of standard written Chinese, rated on a 7-point scale by Cantonese-speaking participants in Hong Kong; because the distributed file gives each word in both traditional and simplified characters, the 25,044 distinct simplified forms could be scored directly (mean 4.60; s.d.\ 0.72). The norms of Song and Li\cite{song2021} cover 3,783 simplified two-character words rated on the same 7-point scale by Mainland Mandarin speakers (mean 4.77; s.d.\ 1.05), and serve to check that the result does not depend on the Hong Kong sample. Gender associations were measured in the static word vectors distributed with spaCy's \code{zh_core_web_lg} model (version 3.8.0; 500,000 entries, 300 dimensions)\cite{honnibal2020spacy}, which are fastText CBOW vectors trained on the Chinese portions of OSCAR Common Crawl and Wikipedia; 23,961 of the Chan and Tse words (95.7\%) and 3,782 of the Song and Li words are in its vocabulary.

The seed pairs are the Chinese counterparts of the Dutch and Estonian primary and extended sets: \zh{男人}/\zh{女人} (man/woman), \zh{男孩}/\zh{女孩} (boy/girl), \zh{男性}/\zh{女性} (male/female), \zh{父亲}/\zh{母亲} (father/mother) and \zh{哥哥}/\zh{姐姐} (elder brother/elder sister) for the primary axis, with \zh{儿子}/\zh{女儿} (son/daughter), \zh{叔叔}/\zh{阿姨} (uncle/aunt), \zh{国王}/\zh{女王} (king/queen), \zh{先生}/\zh{女士} (Mr/Ms) and \zh{爷爷}/\zh{奶奶} (grandfather/grandmother) added for the extended axis. Because the written pronoun distinction is unambiguous in Chinese, unlike in Dutch, a pronoun variant that replaces \zh{哥哥}/\zh{姐姐} with \zh{他}/\zh{她} is also reported. The primary axis orders the male above the female member in all 18 held-out gendered pairs (for example \zh{丈夫}/\zh{妻子}, \zh{新郎}/\zh{新娘}, \zh{王子}/\zh{公主}, \zh{皇帝}/\zh{皇后}), and the extended and pronoun axes in 17 of 18. Because every item is a two-character word, length cannot vary and is not controlled; word frequency is the log corpus frequency distributed with each set of norms, and the Chan and Tse rater standard deviations define the reliably rated subset. Words containing a gender-marking character (\zh{男}, \zh{女}, \zh{父}, \zh{母}, \zh{哥}, \zh{姐}, \zh{夫}, \zh{妻}, \zh{他}, \zh{她} and the like) are the lexically gendered terms for the exclusion check.

\begin{figure}[!htb]
  \centering
  \includegraphics[width=0.55\textwidth]{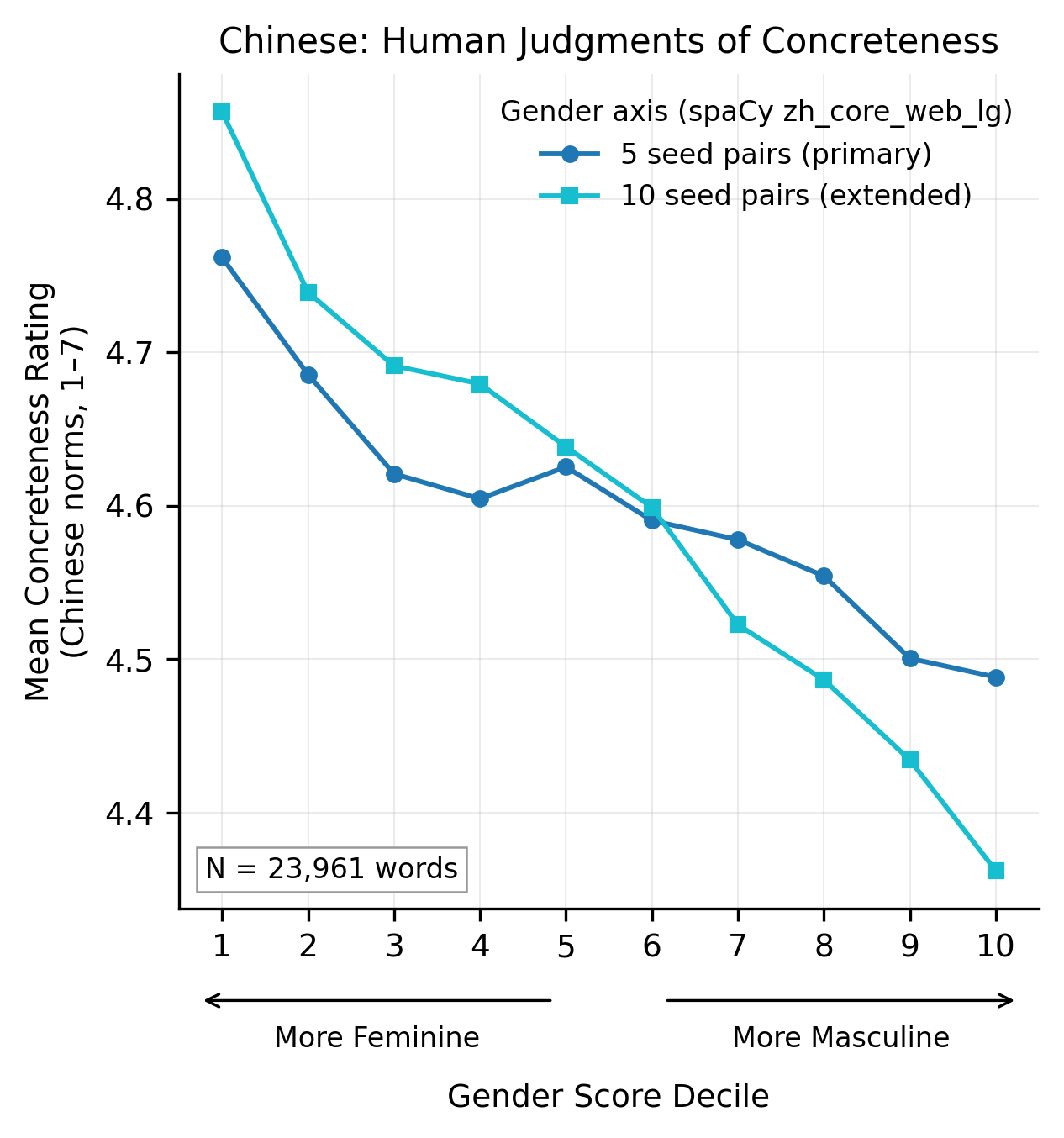}
  \caption{\textbf{More masculine Chinese words are more abstract.} Mean human-rated concreteness (1, abstract; 7, concrete)\cite{chan2024} of the $n = 23{,}961$ two-character Chinese words within each decile of gender association (1, most feminine; 10, most masculine), as in Fig.~1a. Gender associations were measured in Chinese-trained word vectors (spaCy \protect\code{zh_core_web_lg}) along an axis built from five kinship and role seed pairs (\zh{男人}/\zh{女人}, \zh{男孩}/\zh{女孩}, \zh{男性}/\zh{女性}, \zh{父亲}/\zh{母亲}, \zh{哥哥}/\zh{姐姐}; circles) or from these and five further pairs (squares); deciles were computed separately for each axis. Pearson's $r = -0.111$ (95\% CI $-0.123$ to $-0.098$) and $-0.193$ ($-0.205$ to $-0.181$), respectively; Supplementary Tables~\ref{tab:S-chinese-spec} and~\ref{tab:S-chinese-poles} report the robustness of these estimates, and Supplementary Section~\ref{si:chinese} the replication in the Mainland Mandarin norms of Song and Li\cite{song2021}.}
  \label{fig:S-chinese}
\end{figure}

Supplementary Fig.~\ref{fig:S-chinese} shows that the pattern of Fig.~1a replicates in Chinese. Across the 23,961 Chan and Tse words, more masculine words are more abstract ($r = -0.111$; 95\% CI $= -0.123$ to $-0.098$; $P < 2.2 \times 10^{-16}$; Spearman's $\rho = -0.101$), and mean concreteness falls from 4.76 in the most feminine decile to 4.49 in the most masculine, a difference of 0.27 points on the 7-point scale or 0.38 s.d.\ of the ratings. The extended axis gives $r = -0.193$ ($-0.205$ to $-0.181$; $\rho = -0.193$) with a strictly monotone decline across all ten deciles, from 4.86 to 4.36 (0.68 s.d.); the pronoun axis gives $r = -0.168$ ($-0.180$ to $-0.155$). In units of the ratings' standard deviation, the Chinese estimates are thus in the range of the Estonian estimates (0.37 and 0.44 s.d.\ for the two axes) and between the English single-word (0.28 s.d.) and Dutch (0.84 s.d.) estimates. Supplementary Table~\ref{tab:S-chinese-spec} shows that the association survives the available controls and exclusions: controlling for word frequency ($\beta = -0.112$ and $-0.197$ for the two axes), restricting the analysis to reliably rated words ($r = -0.129$ and $-0.200$), excluding words that contain a gender-marking character together with the seed words ($r = -0.095$ and $-0.187$), and excluding the 5\% or 10\% of words with the most extreme gender scores ($r = -0.088$ and $-0.082$ for the primary axis; $-0.184$ and $-0.168$ for the extended axis). Unlike in Estonian, the last exclusion weakens the primary estimate somewhat, because in Chinese it is the feminine extreme that is populated by lexically gendered words (see below).

The Mainland sample gives the same answer with larger effects. Across the 3,782 Song and Li words, $r = -0.189$ (95\% CI $= -0.219$ to $-0.158$) for the primary axis, $-0.226$ ($-0.256$ to $-0.196$) for the extended axis and $-0.288$ ($-0.316$ to $-0.258$) for the pronoun axis (all $P < 10^{-30}$); mean concreteness falls from 5.22 to 4.50 across the deciles of the primary axis (0.68 s.d.) and from 5.33 to 4.44 across those of the extended axis (0.84 s.d.), and the estimates remain between $-0.121$ and $-0.283$ after the frequency control and the three exclusions. The male abstraction bias in written Chinese is therefore present whether the ratings come from Cantonese-speaking or Mandarin-speaking raters, and is if anything stronger in the Mainland sample.

The words at the two extremes of the Chinese axes (Supplementary Table~\ref{tab:S-chinese-poles}) show the bias in an unusually direct form. The feminine extreme of the primary axis consists of words for women and for feminine appearance---\zh{少妇} `young married woman', \zh{美女} `beautiful woman', \zh{女人} `woman', \zh{少女} `girl', \zh{姐姐} `elder sister', \zh{嫂子} `sister-in-law', \zh{白嫩} `fair and delicate (of skin)'---whereas its masculine extreme consists not of words for men but of abstract words for chance, inference and strategy: \zh{运气} `luck', \zh{如何} `how', \zh{数值} `numerical value', \zh{设法} `to find a way', \zh{概率} `probability', \zh{祈祷} `to pray', with \zh{分析} `analysis' and \zh{推测} `to infer' at the masculine extreme of the pronoun axis. The extended axis, whose seeds include more kin and rank terms, adds the vocabulary of rank and command to the masculine extreme (\zh{长兄} `eldest brother', \zh{继位} `to succeed to the throne', \zh{带兵} `to lead troops', \zh{功劳} `merit'). The asymmetry---women represented by words for women, men by the vocabulary of reasoning and authority---is the lexical counterpart of the pattern documented in Figs.~2 and~4, and it is the reason the exclusion of lexically gendered words lowers the Chinese estimate: the words removed lie almost entirely at the concrete, feminine end.

Supplementary Table~\ref{tab:S-chinese-poles} also reports the sensitivity to the seed pairs, which repeats the pattern of the other two languages. Every multi-pair axis is negative: the primary set ($-0.111$), the extended set ($-0.193$), the pronoun set ($-0.168$), each axis obtained by dropping one pair from the primary set ($-0.040$ to $-0.168$) and the first principal component of the ten pair differences ($-0.191$; 29\% of their variance; all ten pairs loading positively, 0.30 to 0.68). Nine of the eleven single pairs are negative on their own, from \zh{先生}/\zh{女士} ($-0.332$) and \zh{他}/\zh{她} ($-0.266$) to \zh{爷爷}/\zh{奶奶} ($-0.058$); \zh{男孩}/\zh{女孩} is indistinguishable from zero; and \zh{男性}/\zh{女性} is positive ($0.139$). As with \emph{man}/\emph{vrouw} in Dutch and \emph{mees}/\emph{naine} in Estonian, the reversed pair is the one whose members have a dominant non-referential use: \zh{男性} and \zh{女性} are the terms of medical and sexual-health discourse, and the masculine extreme of that axis is accordingly \zh{早泄}, \zh{肾虚}, \zh{睾丸}, \zh{包皮}, \zh{阳痿} (premature ejaculation, kidney deficiency, testicle, foreskin, impotence), which are concrete. Conversely, the strongest single pair, \zh{先生}/\zh{女士}, owes part of its strength to the use of \zh{先生} for `teacher' or `master', so that its masculine extreme is populated by scholarly words (\zh{先师} `late teacher', \zh{著书} `to write books', \zh{遗著} `posthumous works'); this pair is not in the primary set, whose estimate is the conservative one. Averaging over several unambiguous pairs again removes these pair-specific effects.

In sum, the male abstraction bias replicates in written Chinese across two independently collected sets of human concreteness norms, two speaker populations and a logographic script, with a magnitude comparable to Estonian and, in the Mainland sample, to Dutch. With Dutch and Estonian, the text-based result now holds in three languages from three families (Germanic, Finno-Ugric and Sinitic), two of which have no grammatical gender at all, and in each the pattern appears whenever the gender axis is built from several unambiguous kinship and role pairs; Supplementary Section~\ref{si:persian} adds a fourth language, Persian.

\begin{table}[!htbp]
  \centering
  \caption{\textbf{The male abstraction bias in Chinese is robust to word frequency, rater reliability and the composition of the vocabulary.} Each row gives the association between the gender score of a Chinese two-character word (positive, masculine; spaCy \protect\code{zh_core_web_lg} vectors) and its human-rated concreteness in the norms of Chan and Tse\cite{chan2024}, for the primary five-pair axis (left) and the extended ten-pair axis (right). The row labelled $\beta$ is the standardized coefficient on concreteness from a regression of the $z$-scored gender score on $z$-scored concreteness and log word frequency (distributed with the norms), with heteroskedasticity-robust (HC1) 95\% confidence intervals; all other rows are two-tailed Pearson's correlations with Fisher-transformed intervals. Length is constant (two characters) and is not controlled. $P$ values are two-tailed; $n$ is the number of words.}
  \label{tab:S-chinese-spec}
  \begingroup\footnotesize\setlength{\tabcolsep}{4pt}
  \resizebox{\textwidth}{!}{%
  \begin{tabular}{@{}>{\raggedright\arraybackslash}p{6.2cm}lrrlrr@{}}
    \toprule
    & \multicolumn{3}{c}{Primary axis (5 pairs)} & \multicolumn{3}{c}{Extended axis (10 pairs)} \\
    \cmidrule(lr){2-4}\cmidrule(lr){5-7}
    Specification & Estimate [95\% CI] & $P$ & $n$ & Estimate [95\% CI] & $P$ & $n$ \\
    \midrule
    No controls (Supplementary Fig.~\ref{fig:S-chinese}) & $-0.111$ [$-0.123$, $-0.098$] & $<2.2\times10^{-16}$ & 23{,}961 & $-0.193$ [$-0.205$, $-0.181$] & $<2.2\times10^{-16}$ & 23{,}961 \\
    Excluding lexically gendered terms and seed words & $-0.095$ [$-0.108$, $-0.083$] & $<2.2\times10^{-16}$ & 23{,}559 & $-0.187$ [$-0.200$, $-0.175$] & $<2.2\times10^{-16}$ & 23{,}556 \\
    Excluding the 5\% most strongly gendered terms & $-0.088$ [$-0.101$, $-0.076$] & $<2.2\times10^{-16}$ & 22{,}762 & $-0.184$ [$-0.197$, $-0.172$] & $<2.2\times10^{-16}$ & 22{,}762 \\
    Excluding the 10\% most strongly gendered terms & $-0.082$ [$-0.095$, $-0.069$] & $<2.2\times10^{-16}$ & 21{,}564 & $-0.168$ [$-0.181$, $-0.155$] & $<2.2\times10^{-16}$ & 21{,}564 \\
    Rater s.d.\ of concreteness $\leq$ median (reliably rated words) & $-0.129$ [$-0.146$, $-0.112$] & $<2.2\times10^{-16}$ & 12{,}152 & $-0.200$ [$-0.217$, $-0.183$] & $<2.2\times10^{-16}$ & 12{,}152 \\
    $\beta$, controlling for word frequency & $-0.112$ [$-0.126$, $-0.097$] & $<2.2\times10^{-16}$ & 19{,}832 & $-0.197$ [$-0.211$, $-0.183$] & $<2.2\times10^{-16}$ & 19{,}832 \\
    \bottomrule
  \end{tabular}}\endgroup
\end{table}

\begin{table}[!htbp]
  \centering
  \caption{\textbf{Sensitivity of the Chinese result to the seed pairs used to build the gender axis.} Two-tailed Pearson's correlations between the gender score and human-rated concreteness\cite{chan2024} across the 23,961 rated words with a vector (positive values, more masculine words are more concrete), with Fisher-transformed 95\% confidence intervals, for axes built from the specifications listed; the last two columns give the five words with the lowest (most feminine) and highest (most masculine) scores under each axis. The first principal component is computed from the ten unit-normalised male-minus-female difference vectors of the extended set without centring\cite{bolukbasi2016} and sign-aligned with their mean. Family C uses the written pronoun pair \zh{他}/\zh{她}, which is unambiguous in Chinese.}
  \label{tab:S-chinese-poles}
  \begingroup\scriptsize\setlength{\tabcolsep}{3pt}
  \resizebox{\textwidth}{!}{%
  \begin{tabular}{@{}>{\raggedright\arraybackslash}p{5.2cm}rlr>{\raggedright\arraybackslash}p{4.1cm}>{\raggedright\arraybackslash}p{4.1cm}@{}}
    \toprule
    Axis & Pairs & $r$ [95\% CI] & $P$ & Most feminine words & Most masculine words \\
    \midrule
    \multicolumn{6}{@{}l}{\textit{A. Axes built from several kinship/role pairs}} \\
    Primary set (5 pairs) & 5 & $-0.111$ [$-0.123$, $-0.098$] & $<2.2\times10^{-16}$ & \zh{少妇}, \zh{美女}, \zh{女人}, \zh{少女}, \zh{女性} & \zh{运气}, \zh{长兄}, \zh{如何}, \zh{数值}, \zh{设法} \\
    Extended set (10 pairs) & 10 & $-0.193$ [$-0.205$, $-0.181$] & $<2.2\times10^{-16}$ & \zh{美女}, \zh{少妇}, \zh{女人}, \zh{女性}, \zh{少女} & \zh{长兄}, \zh{兄长}, \zh{他们}, \zh{继位}, \zh{晋见} \\
    Primary set without \zh{男人}/\zh{女人} & 4 & $-0.088$ [$-0.101$, $-0.075$] & $<2.2\times10^{-16}$ & \zh{少妇}, \zh{姐姐}, \zh{美女}, \zh{少女}, \zh{嫂子} & \zh{次子}, \zh{出生}, \zh{肾虚}, \zh{继任}, \zh{长兄} \\
    Primary set without \zh{男孩}/\zh{女孩} & 4 & $-0.132$ [$-0.144$, $-0.119$] & $<2.2\times10^{-16}$ & \zh{少妇}, \zh{女人}, \zh{女性}, \zh{美女}, \zh{少女} & \zh{运气}, \zh{外围}, \zh{长兄}, \zh{离婚}, \zh{破解} \\
    Primary set without \zh{男性}/\zh{女性} & 4 & $-0.168$ [$-0.180$, $-0.156$] & $<2.2\times10^{-16}$ & \zh{少妇}, \zh{美女}, \zh{女人}, \zh{少女}, \zh{姐姐} & \zh{运气}, \zh{长兄}, \zh{如何}, \zh{次子}, \zh{反而} \\
    Primary set without \zh{父亲}/\zh{母亲} & 4 & $-0.040$ [$-0.052$, $-0.027$] & $9.0\times10^{-10}$ & \zh{女人}, \zh{少妇}, \zh{美女}, \zh{少女}, \zh{女性} & \zh{运气}, \zh{男人}, \zh{如何}, \zh{数值}, \zh{破解} \\
    Primary set without \zh{哥哥}/\zh{姐姐} & 4 & $-0.096$ [$-0.109$, $-0.084$] & $<2.2\times10^{-16}$ & \zh{少妇}, \zh{美女}, \zh{女人}, \zh{少女}, \zh{女性} & \zh{如何}, \zh{运气}, \zh{离婚}, \zh{祈祷}, \zh{设法} \\
    First principal component of the 10 pair differences (0.29 of variance) & 10 & $-0.191$ [$-0.203$, $-0.179$] & $<2.2\times10^{-16}$ & \zh{少妇}, \zh{美女}, \zh{女人}, \zh{少女}, \zh{女性} & \zh{长兄}, \zh{兄长}, \zh{继位}, \zh{次子}, \zh{带兵} \\
    \multicolumn{6}{@{}l}{\textit{B. Single kinship/role pairs}} \\
    \zh{男人}/\zh{女人} & 1 & $-0.108$ [$-0.120$, $-0.095$] & $<2.2\times10^{-16}$ & \zh{女人}, \zh{美女}, \zh{少妇}, \zh{女性}, \zh{少女} & \zh{男人}, \zh{争霸}, \zh{暗号}, \zh{祈祷}, \zh{可否} \\
    \zh{男孩}/\zh{女孩} & 1 & $0.004$ [$-0.009$, $0.016$] & $0.58$ & \zh{美女}, \zh{少妇}, \zh{少女}, \zh{白嫩}, \zh{妹妹} & \zh{男孩}, \zh{咳嗽}, \zh{拳击}, \zh{补钙}, \zh{食谱} \\
    \zh{男性}/\zh{女性} & 1 & $0.139$ [$0.127$, $0.152$] & $<2.2\times10^{-16}$ & \zh{女色}, \zh{女性}, \zh{村姑}, \zh{妇人}, \zh{艺术} & \zh{早泄}, \zh{肾虚}, \zh{男性}, \zh{睾丸}, \zh{包皮} \\
    \zh{父亲}/\zh{母亲} & 1 & $-0.225$ [$-0.237$, $-0.213$] & $<2.2\times10^{-16}$ & \zh{少妇}, \zh{美女}, \zh{岳母}, \zh{妈妈}, \zh{黄色} & \zh{当时}, \zh{祖父}, \zh{当年}, \zh{出身}, \zh{后来} \\
    \zh{哥哥}/\zh{姐姐} & 1 & $-0.105$ [$-0.117$, $-0.092$] & $<2.2\times10^{-16}$ & \zh{姐姐}, \zh{姨妈}, \zh{表嫂}, \zh{后母}, \zh{阿姨} & \zh{哥哥}, \zh{降职}, \zh{属意}, \zh{靶子}, \zh{迁怒} \\
    \zh{儿子}/\zh{女儿} & 1 & $-0.103$ [$-0.116$, $-0.090$] & $<2.2\times10^{-16}$ & \zh{粉色}, \zh{女儿}, \zh{粉红}, \zh{校花}, \zh{桃红} & \zh{儿子}, \zh{欠账}, \zh{继子}, \zh{部下}, \zh{造反} \\
    \zh{叔叔}/\zh{阿姨} & 1 & $-0.094$ [$-0.106$, $-0.081$] & $<2.2\times10^{-16}$ & \zh{阿姨}, \zh{少妇}, \zh{表嫂}, \zh{后母}, \zh{岳母} & \zh{叔叔}, \zh{上校}, \zh{二叔}, \zh{写信}, \zh{副官} \\
    \zh{国王}/\zh{女王} & 1 & $-0.137$ [$-0.150$, $-0.125$] & $<2.2\times10^{-16}$ & \zh{女王}, \zh{美女}, \zh{交友}, \zh{女星}, \zh{按摩} & \zh{国王}, \zh{酋长}, \zh{继位}, \zh{他们}, \zh{亲王} \\
    \zh{先生}/\zh{女士} & 1 & $-0.332$ [$-0.343$, $-0.320$] & $<2.2\times10^{-16}$ & \zh{男士}, \zh{女士}, \zh{女性}, \zh{女孩}, \zh{妇女} & \zh{先师}, \zh{先生}, \zh{不是}, \zh{著书}, \zh{遗著} \\
    \zh{爷爷}/\zh{奶奶} & 1 & $-0.058$ [$-0.070$, $-0.045$] & $<2.2\times10^{-16}$ & \zh{同行}, \zh{转盘}, \zh{私吞}, \zh{抽水}, \zh{暗号} & \zh{爷爷}, \zh{父亲}, \zh{他们}, \zh{你们}, \zh{当年} \\
    \multicolumn{6}{@{}l}{\textit{C. Pronoun-based seeds}} \\
    Pronoun set & 5 & $-0.168$ [$-0.180$, $-0.155$] & $<2.2\times10^{-16}$ & \zh{少妇}, \zh{美女}, \zh{女人}, \zh{少女}, \zh{女性} & \zh{运气}, \zh{分析}, \zh{如何}, \zh{设法}, \zh{推测} \\
    \zh{他}/\zh{她} & 1 & $-0.266$ [$-0.277$, $-0.254$] & $<2.2\times10^{-16}$ & \zh{姐姐}, \zh{妈妈}, \zh{阿姨}, \zh{妹妹}, \zh{女孩} & \zh{战术}, \zh{功绩}, \zh{将帅}, \zh{效力}, \zh{防守} \\
    \bottomrule
  \end{tabular}}\endgroup
\end{table}

\section{Replication of the text-based result in Persian}
\label{si:persian}

Persian provides a fourth test, in a language of the Indo-Iranian branch of Indo-European written in a right-to-left script. Like Estonian and Chinese, Persian is genderless: nouns, adjectives and articles carry no grammatical gender, and the third-person pronoun \emph{u} covers both \emph{he} and \emph{she}. The concreteness norms are those of the Extrapolated Persian Lexical Affect Norms (E-PLAN) of Nemati et~al.\cite{nemati2026}, which differ from the other norms used in this paper in how they were collected: rather than rating each word on a scale, 1,071 Persian speakers judged small sets of words, choosing the most and the least concrete word in each set (best--worst scaling), and each word's score is derived from how often it was chosen as most versus least concrete and expressed on a scale from 0 (abstract) to 1 (concrete). The human-rated set covers 2,924 words (mean 0.51; s.d.\ 0.13), and the authors validate it against earlier Persian rating-scale norms. They also provide extrapolated scores for 85,716 words, obtained by fitting generalized additive models on principal components of Persian skip-gram embeddings to the human judgements ($R^2 = 0.68$ for concreteness); because these scores are themselves derived from word embeddings, we use them only as a secondary check and report the human-rated words as the result. Gender associations were measured in fastText's Persian word vectors\cite{grave2018} (Common Crawl and Wikipedia; 2 million words, 300 dimensions); every one of the 2,924 human-rated words has a vector.

The seed pairs are the Persian counterparts of the primary and extended sets of the other languages, all kinship and role nouns and one adjective pair, since Persian has no gendered pronouns: \emph{mard}/\emph{zan} (man/woman), \emph{pesar}/\emph{dokhtar} (boy/girl), \emph{mard\={a}ne}/\emph{zan\={a}ne} (masculine/feminine), \emph{pedar}/\emph{m\={a}dar} (father/mother) and \emph{bar\={a}dar}/\emph{kh\={a}har} (brother/sister) for the primary axis, with \emph{amu}/\emph{amme} (uncle/aunt), \emph{sh\={a}h}/\emph{malake} (king/queen), \emph{\={a}q\={a}}/\emph{kh\={a}nom} (Mr/Ms), \emph{pedarbozorg}/\emph{m\={a}darbozorg} (grandfather/grandmother) and \emph{d\={a}m\={a}d}/\emph{arus} (groom/bride) added for the extended axis. Both axes order the male above the female member in 10 of 11 held-out gendered pairs (for example \emph{mard\={a}n}/\emph{zan\={a}n} `men/women', \emph{pirmard}/\emph{pirzan} `old man/old woman', \emph{sh\={a}hz\={a}de}/\emph{sh\={a}hdokht} `prince/princess'). The procedure otherwise follows the Dutch replication: word frequency is the rank of the word in the fastText vocabulary (the norms carry no frequency counts or rater standard deviations), length is the number of characters, and words containing a gender-marking stem (\emph{mard}, \emph{zan}, \emph{pesar}, \emph{dokhtar}, \emph{pedar}, \emph{m\={a}dar} and the like) are the lexically gendered terms for the exclusion check.

\begin{figure}[!htb]
  \centering
  \includegraphics[width=0.55\textwidth]{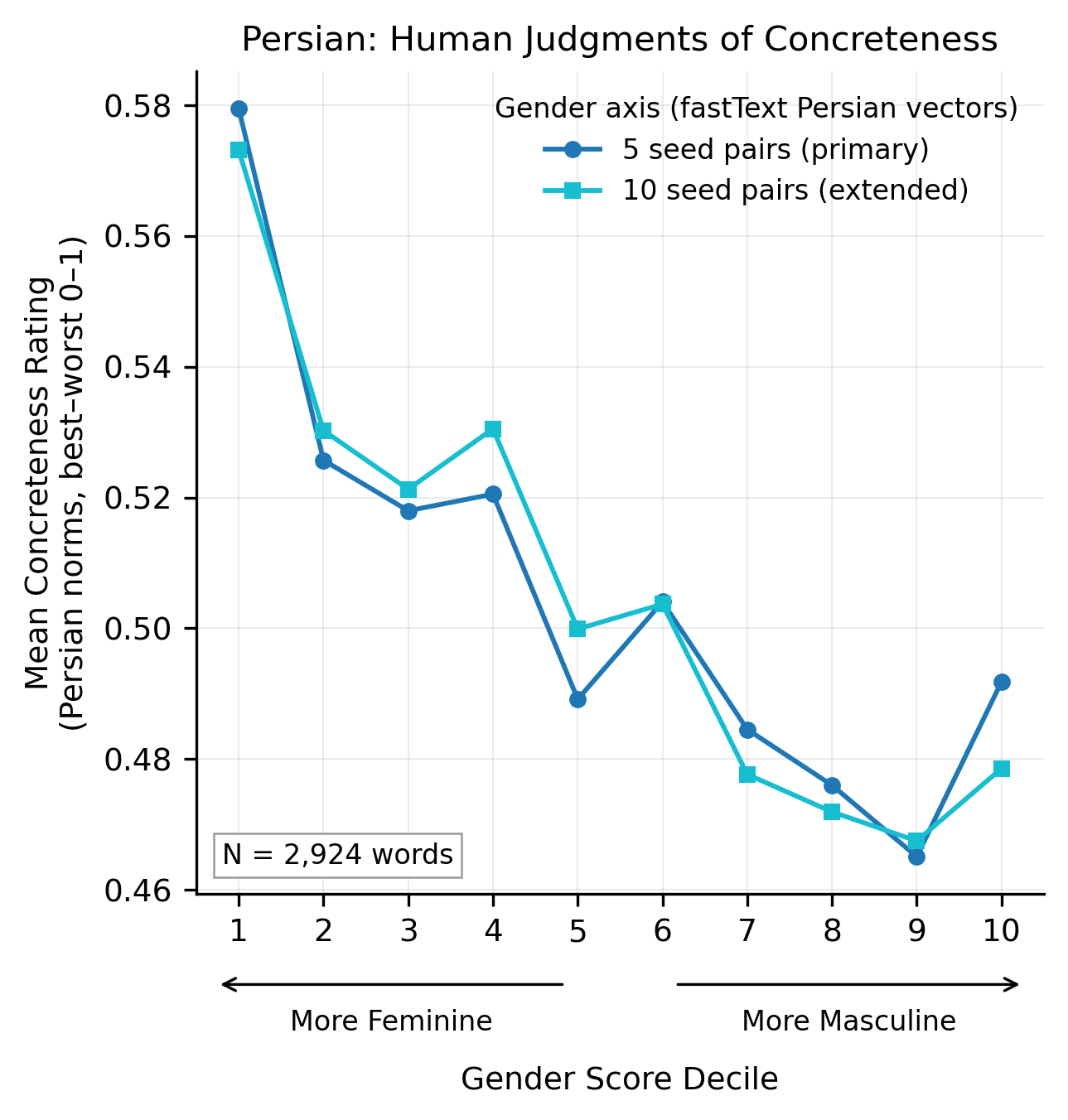}
  \caption{\textbf{More masculine Persian words are more abstract.} Mean human best--worst concreteness score (0, abstract; 1, concrete)\cite{nemati2026} of the $n = 2{,}924$ Persian words within each decile of gender association (1, most feminine; 10, most masculine), as in Fig.~1a. Gender associations were measured in fastText's Persian word vectors\cite{grave2018} along an axis built from five kinship and role seed pairs (\emph{mard}/\emph{zan}, \emph{pesar}/\emph{dokhtar}, \emph{mard\={a}ne}/\emph{zan\={a}ne}, \emph{pedar}/\emph{m\={a}dar}, \emph{bar\={a}dar}/\emph{kh\={a}har}; circles) or from these and five further pairs (squares); deciles were computed separately for each axis. Pearson's $r = -0.194$ (95\% CI $-0.228$ to $-0.158$) and $-0.208$ ($-0.243$ to $-0.173$), respectively. The most masculine decile is populated by titles of rank and male given names, whose referents are people and therefore concrete (Supplementary Section~\ref{si:persian}); Supplementary Tables~\ref{tab:S-persian-spec} and~\ref{tab:S-persian-poles} report the robustness of these estimates.}
  \label{fig:S-persian}
\end{figure}

Supplementary Fig.~\ref{fig:S-persian} shows that the pattern of Fig.~1a replicates in Persian. Across the 2,924 human-rated words, more masculine words are more abstract ($r = -0.194$; 95\% CI $= -0.228$ to $-0.158$; $P < 2.2 \times 10^{-16}$; Spearman's $\rho = -0.191$), and mean concreteness falls from 0.58 in the most feminine decile to 0.47 in the ninth decile, with 0.49 in the most masculine. The difference between the extreme deciles is 0.087 on the 0--1 scale, or 0.65 s.d.\ of the ratings; between the first and ninth deciles it is 0.86 s.d. The extended axis gives $r = -0.208$ ($-0.243$ to $-0.173$; $\rho = -0.216$) and a difference of 0.71 s.d.\ between the extreme deciles. In units of the ratings' standard deviation the Persian effect is thus the largest after Dutch (0.84 s.d.) and above the Estonian (0.37 s.d.) and Chinese (0.38 s.d.) estimates, and the correlation lies between the Chinese and Dutch ones; its confidence interval is wider than in the other languages because the human-rated set is smaller.

Supplementary Table~\ref{tab:S-persian-spec} shows that the association is robust. Controlling for word frequency left it unchanged ($\beta = -0.194$), controlling for length or for frequency and length together reduced it marginally ($\beta = -0.187$), and it was negative in every quintile of word frequency ($r = -0.155$ to $-0.246$; each $P < 0.001$). Excluding words that contain a gender-marking stem together with the seed words left it unchanged ($r = -0.194$; $n = 2{,}882$), and excluding the 5\% or 10\% of words with the most extreme gender scores strengthened it ($r = -0.227$ and $-0.224$), for the same reason as in Estonian: the words at the two extremes are concrete nouns for people and bodies, so their removal leaves the association among the remaining vocabulary intact and the most masculine decile of the primary axis no longer turns upward. The extended axis behaves in the same way ($r$ or $\beta$ between $-0.147$ and $-0.259$). The extrapolated set gives the same answer at lower magnitude, as expected from scores that are regression predictions with a compressed range (s.d.\ 0.07): across 85,007 words, $r = -0.129$ (95\% CI $= -0.136$ to $-0.122$) for the primary axis and $-0.147$ ($-0.154$ to $-0.140$) for the extended axis, mean concreteness declines monotonically from the first to the ninth decile (0.50 to 0.47), and the estimates remain between $-0.096$ and $-0.197$ across the same controls and exclusions.

Supplementary Table~\ref{tab:S-persian-poles} reports the sensitivity to the seed pairs, which is the smallest of the four languages. Every multi-pair axis is negative, within a narrow range: the primary set ($-0.194$), the extended set ($-0.208$), each axis obtained by dropping one pair from the primary set ($-0.170$ to $-0.188$) and the first principal component of the ten pair differences ($-0.211$; 26\% of their variance; all ten pairs loading positively, 0.24 to 0.64). Nine of the ten single pairs are negative on their own, from \emph{d\={a}m\={a}d}/\emph{arus} ($-0.299$) to \emph{bar\={a}dar}/\emph{kh\={a}har} ($-0.093$); unlike in Dutch, Estonian and Chinese, the bare man/woman pair \emph{mard}/\emph{zan} does not reverse ($-0.102$), although its masculine extreme does show the pattern that made that pair unreliable elsewhere, namely the generic words for a person (\emph{fard}, \emph{shakhs}, \emph{afr\={a}d} `individuals')---the lexical form of the androcentric default in which the generic person is a man\cite{bailey2022}. The one reversed pair is \emph{amu}/\emph{amme} ($0.131$), and its extremes show that it is not a gender axis at all: they are everyday nouns unrelated to gender (\emph{entesh\={a}r} `publication', \emph{ot\={a}q} `room', \emph{kh\={a}b} `sleep' at one end; \emph{sabz} `green', \emph{barg} `leaf', \emph{k\={a}rtun} `cartoon' at the other), consistent with \emph{amu} doubling as a familiar term of address for any adult man in children's speech. Averaging over several pairs removes this pair-specific variance, as in the other languages.

The words at the two extremes again show the bias in a direct form, with a different emphasis from Chinese. The feminine extreme of the primary axis consists of words for women (\emph{zan} `woman', \emph{b\={a}nu} `lady', \emph{kh\={a}nom} `Ms', \emph{malake} `queen') and, prominently, of words for the female body, sexuality and maternity (\emph{pest\={a}n} `breast', \emph{seksi} `sexy', \emph{lokht} `naked', \emph{h\={a}mele} `pregnant', \emph{zan\={a}ne} `feminine'); the masculine extreme consists of kinship terms (\emph{bar\={a}dar}, \emph{pedar}, \emph{pesar}) followed by titles of military, religious and social rank and by male given names (\emph{sard\={a}r} `commander', \emph{kh\={a}n}, \emph{jen\={a}b} `his excellency', \emph{moll\={a}}, \emph{h\={a}ji}, \emph{seyyed}, \emph{\={a}q\={a}} `sir', \emph{Mohammad}, \emph{Ali}). Women are thus represented at the feminine pole through the body and through marital and maternal states (\emph{arus} `bride', \emph{h\={a}mele} `pregnant'), and men at the masculine pole through titles of authority and status. This is the Persian counterpart of the asymmetries seen in Estonian (men as role nouns) and Chinese (men as the vocabulary of reasoning and command, women as women), and it speaks directly to the objectification argument of the Discussion: in each language the feminine pole is anchored in concrete, embodied referents and the masculine pole in roles, rank and abstraction.

In sum, the male abstraction bias replicates in Persian, a fourth language from a third family and a third script, with human concreteness judgements collected by a different method (best--worst scaling) and with the most stable seed-pair behaviour of the four languages. With Dutch, Estonian and Chinese, the text-based result now holds in four languages, three of them without grammatical gender, and in every case the sign is the same whenever the gender axis is built from several unambiguous kinship and role pairs.

\begin{table}[!htbp]
  \centering
  \caption{\textbf{The male abstraction bias in Persian is robust to word frequency, length and the composition of the vocabulary.} Each row gives the association between the gender score of a Persian word (positive, masculine; fastText \protect\code{cc.fa.300} vectors\cite{grave2018}) and its human best--worst concreteness score in the E-PLAN norms\cite{nemati2026}, for the primary five-pair axis (left) and the extended ten-pair axis (right). Rows labelled $\beta$ are standardized coefficients on concreteness from regressions of the $z$-scored gender score on $z$-scored concreteness and the covariates listed, with heteroskedasticity-robust (HC1) 95\% confidence intervals; all other rows are two-tailed Pearson's correlations with Fisher-transformed intervals. Word frequency is the log rank of the word in the fastText vocabulary, which is ordered by corpus frequency; length is the number of characters. $P$ values are two-tailed; $n$ is the number of words.}
  \label{tab:S-persian-spec}
  \begingroup\footnotesize\setlength{\tabcolsep}{4pt}
  \resizebox{\textwidth}{!}{%
  \begin{tabular}{@{}>{\raggedright\arraybackslash}p{6.2cm}lrrlrr@{}}
    \toprule
    & \multicolumn{3}{c}{Primary axis (5 pairs)} & \multicolumn{3}{c}{Extended axis (10 pairs)} \\
    \cmidrule(lr){2-4}\cmidrule(lr){5-7}
    Specification & Estimate [95\% CI] & $P$ & $n$ & Estimate [95\% CI] & $P$ & $n$ \\
    \midrule
    No controls (Supplementary Fig.~\ref{fig:S-persian}) & $-0.194$ [$-0.228$, $-0.158$] & $<2.2\times10^{-16}$ & 2{,}924 & $-0.208$ [$-0.243$, $-0.173$] & $<2.2\times10^{-16}$ & 2{,}924 \\
    Excluding lexically gendered terms and seed words & $-0.194$ [$-0.229$, $-0.159$] & $<2.2\times10^{-16}$ & 2{,}882 & $-0.208$ [$-0.243$, $-0.173$] & $<2.2\times10^{-16}$ & 2{,}882 \\
    Excluding the 5\% most strongly gendered terms & $-0.227$ [$-0.262$, $-0.192$] & $<2.2\times10^{-16}$ & 2{,}777 & $-0.227$ [$-0.262$, $-0.192$] & $<2.2\times10^{-16}$ & 2{,}777 \\
    Excluding the 10\% most strongly gendered terms & $-0.224$ [$-0.260$, $-0.188$] & $<2.2\times10^{-16}$ & 2{,}631 & $-0.205$ [$-0.241$, $-0.168$] & $<2.2\times10^{-16}$ & 2{,}631 \\
    $\beta$, controlling for word frequency & $-0.194$ [$-0.229$, $-0.158$] & $<2.2\times10^{-16}$ & 2{,}924 & $-0.206$ [$-0.240$, $-0.171$] & $<2.2\times10^{-16}$ & 2{,}924 \\
    $\beta$, controlling for length & $-0.187$ [$-0.222$, $-0.151$] & $<2.2\times10^{-16}$ & 2{,}924 & $-0.208$ [$-0.242$, $-0.174$] & $<2.2\times10^{-16}$ & 2{,}924 \\
    $\beta$, controlling for frequency and length & $-0.187$ [$-0.222$, $-0.152$] & $<2.2\times10^{-16}$ & 2{,}924 & $-0.207$ [$-0.241$, $-0.173$] & $<2.2\times10^{-16}$ & 2{,}924 \\
    Frequency quintile 1 (least frequent) & $-0.169$ [$-0.247$, $-0.089$] & $3.8\times10^{-5}$ & 585 & $-0.218$ [$-0.294$, $-0.140$] & $9.8\times10^{-8}$ & 585 \\
    Frequency quintile 2 & $-0.221$ [$-0.297$, $-0.143$] & $6.6\times10^{-8}$ & 585 & $-0.259$ [$-0.333$, $-0.182$] & $2.0\times10^{-10}$ & 585 \\
    Frequency quintile 3 & $-0.246$ [$-0.321$, $-0.168$] & $1.7\times10^{-9}$ & 584 & $-0.249$ [$-0.324$, $-0.171$] & $1.1\times10^{-9}$ & 584 \\
    Frequency quintile 4 & $-0.179$ [$-0.256$, $-0.099$] & $1.3\times10^{-5}$ & 585 & $-0.147$ [$-0.225$, $-0.067$] & $3.6\times10^{-4}$ & 585 \\
    Frequency quintile 5 (most frequent) & $-0.155$ [$-0.233$, $-0.075$] & $1.7\times10^{-4}$ & 585 & $-0.162$ [$-0.240$, $-0.082$] & $8.6\times10^{-5}$ & 585 \\
    \bottomrule
  \end{tabular}}\endgroup
\end{table}

\begin{table}[!htbp]
  \centering
  \caption{\textbf{Sensitivity of the Persian result to the seed pairs used to build the gender axis.} Two-tailed Pearson's correlations between the gender score and human best--worst concreteness\cite{nemati2026} across the 2,924 rated words (positive values, more masculine words are more concrete), with Fisher-transformed 95\% confidence intervals, for axes built from the specifications listed; the last two columns give the five words with the lowest (most feminine) and highest (most masculine) scores under each axis, in Latin transliteration (\={a}, long \emph{a}). The first principal component is computed from the ten unit-normalised male-minus-female difference vectors of the extended set without centring\cite{bolukbasi2016} and sign-aligned with their mean. $^{\dag}$Subword tokens of the vector vocabulary that are not words on their own.}
  \label{tab:S-persian-poles}
  \begingroup\scriptsize\setlength{\tabcolsep}{3pt}
  \resizebox{\textwidth}{!}{%
  \begin{tabular}{@{}>{\raggedright\arraybackslash}p{5.2cm}rlr>{\raggedright\arraybackslash}p{4.1cm}>{\raggedright\arraybackslash}p{4.1cm}@{}}
    \toprule
    Axis & Pairs & $r$ [95\% CI] & $P$ & Most feminine words & Most masculine words \\
    \midrule
    \multicolumn{6}{@{}l}{\textit{A. Axes built from several kinship/role pairs}} \\
    Primary set (5 pairs) & 5 & $-0.194$ [$-0.228$, $-0.158$] & $<2.2\times10^{-16}$ & \emph{zan}, \emph{seksi}, \emph{pest\={a}n}, \emph{b\={a}nu}, \emph{set\textsuperscript{\dag}} & \emph{bar\={a}dar}, \emph{pedar}, \emph{pesar}, \emph{sard\={a}r}, \emph{kh\={a}n} \\
    Extended set (10 pairs) & 10 & $-0.208$ [$-0.243$, $-0.173$] & $<2.2\times10^{-16}$ & \emph{malake}, \emph{zan}, \emph{seksi}, \emph{kh\={a}nom}, \emph{h\={a}mele} & \emph{moll\={a}}, \emph{Mohammad}, \emph{h\={a}ji}, \emph{\={a}q\={a}}, \emph{kh\={a}n} \\
    Primary set without \emph{mard}/\emph{zan} & 4 & $-0.170$ [$-0.205$, $-0.135$] & $<2.2\times10^{-16}$ & \emph{pest\={a}n}, \emph{b\={a}nu}, \emph{kh\={a}nom}, \emph{seksi}, \emph{sadaf} & \emph{bar\={a}dar}, \emph{pedar}, \emph{shurid}, \emph{farm\={a}ndehi}, \emph{ben\textsuperscript{\dag}} \\
    Primary set without \emph{pesar}/\emph{dokhtar} & 4 & $-0.174$ [$-0.209$, $-0.139$] & $<2.2\times10^{-16}$ & \emph{zan}, \emph{seksi}, \emph{ti\textsuperscript{\dag}}, \emph{gol}, \emph{set\textsuperscript{\dag}} & \emph{bar\={a}dar}, \emph{mardi}, \emph{sard\={a}r}, \emph{pedar}, \emph{jen\={a}b} \\
    Primary set without \emph{mard\={a}ne}/\emph{zan\={a}ne} & 4 & $-0.187$ [$-0.222$, $-0.152$] & $<2.2\times10^{-16}$ & \emph{zan}, \emph{mu}, \emph{set\textsuperscript{\dag}}, \emph{ti\textsuperscript{\dag}}, \emph{seksi} & \emph{bar\={a}dar}, \emph{pesar}, \emph{pedar}, \emph{shurid}, \emph{mardi} \\
    Primary set without \emph{pedar}/\emph{m\={a}dar} & 4 & $-0.188$ [$-0.222$, $-0.152$] & $<2.2\times10^{-16}$ & \emph{zan}, \emph{mu}, \emph{seksi}, \emph{dar}, \emph{ti\textsuperscript{\dag}} & \emph{bar\={a}dar}, \emph{sard\={a}r}, \emph{pesar}, \emph{jen\={a}b}, \emph{kh\={a}n} \\
    Primary set without \emph{bar\={a}dar}/\emph{kh\={a}har} & 4 & $-0.186$ [$-0.221$, $-0.151$] & $<2.2\times10^{-16}$ & \emph{zan}, \emph{mu}, \emph{ti\textsuperscript{\dag}}, \emph{gol}, \emph{set\textsuperscript{\dag}} & \emph{pedar}, \emph{pesar}, \emph{mardi}, \emph{sard\={a}r}, \emph{vaf\={a}d\={a}r} \\
    First principal component of the 10 pair differences (0.26 of variance) & 10 & $-0.211$ [$-0.245$, $-0.176$] & $<2.2\times10^{-16}$ & \emph{kh\={a}nom}, \emph{malake}, \emph{seksi}, \emph{zan\={a}ne}, \emph{pest\={a}n} & \emph{bar\={a}dar}, \emph{moll\={a}}, \emph{\={a}q\={a}}, \emph{kh\={a}n}, \emph{Mohammad} \\
    \multicolumn{6}{@{}l}{\textit{B. Single kinship/role pairs}} \\
    \emph{mard}/\emph{zan} & 1 & $-0.102$ [$-0.138$, $-0.066$] & $3.1\times10^{-8}$ & \emph{zan}, \emph{ti\textsuperscript{\dag}}, \emph{si\textsuperscript{\dag}}, \emph{mu}, \emph{pey\textsuperscript{\dag}} & \emph{mard}, \emph{mardi}, \emph{pesar}, \emph{ruzi}, \emph{afr\={a}d} \\
    \emph{pesar}/\emph{dokhtar} & 1 & $-0.157$ [$-0.192$, $-0.121$] & $<2.2\times10^{-16}$ & \emph{dokhtar}, \emph{kh\={a}nom}, \emph{b\={a}nu}, \emph{z\={a}er}, \emph{ir\={a}ni} & \emph{farzand}, \emph{pesar}, \emph{pedar}, \emph{bar\={a}dar}, \emph{ben\textsuperscript{\dag}} \\
    \emph{mard\={a}ne}/\emph{zan\={a}ne} & 1 & $-0.124$ [$-0.160$, $-0.088$] & $1.6\times10^{-11}$ & \emph{zan\={a}ne}, \emph{p\={a}shne}, \emph{s\={a}f}, \emph{seksi}, \emph{khoshgel} & \emph{mard\={a}ne}, \emph{sard\={a}r}, \emph{qabile}, \emph{doshman}, \emph{qor'e} \\
    \emph{pedar}/\emph{m\={a}dar} & 1 & $-0.117$ [$-0.152$, $-0.081$] & $2.5\times10^{-10}$ & \emph{m\={a}dar}, \emph{pest\={a}n}, \emph{janin}, \emph{b\={a}rd\={a}ri}, \emph{shir} & \emph{pedar}, \emph{t\={a}jer}, \emph{gorikht}, \emph{shurid}, \emph{bar\={a}dar} \\
    \emph{bar\={a}dar}/\emph{kh\={a}har} & 1 & $-0.093$ [$-0.129$, $-0.057$] & $4.8\times10^{-7}$ & \emph{kh\={a}har}, \emph{shohar}, \emph{ezdev\={a}j}, \emph{kh\={a}bg\={a}h}, \emph{mehriye} & \emph{bar\={a}dar}, \emph{dar\={o}ft\={a}d}, \emph{jen\={a}b}, \emph{gom\={a}sht}, \emph{shurid} \\
    \emph{amu}/\emph{amme} & 1 & $0.131$ [$0.095$, $0.166$] & $1.3\times10^{-12}$ & \emph{amme}, \emph{entesh\={a}r}, \emph{makhfi}, \emph{khosusi}, \emph{ot\={a}q} & \emph{h\={a}ji}, \emph{S\={a}m}, \emph{sabz}, \emph{k\={a}rtun}, \emph{Ali} \\
    \emph{sh\={a}h}/\emph{malake} & 1 & $-0.097$ [$-0.133$, $-0.061$] & $1.3\times10^{-7}$ & \emph{malake}, \emph{mor\={a}qebat}, \emph{s\={a}here}, \emph{festiv\={a}l}, \emph{mos\={a}beq\={a}t} & \emph{sh\={a}h}, \emph{h\={a}ji}, \emph{moll\={a}}, \emph{Rez\={a}}, \emph{kh\={a}n} \\
    \emph{\={a}q\={a}}/\emph{kh\={a}nom} & 1 & $-0.134$ [$-0.169$, $-0.098$] & $4.1\times10^{-13}$ & \emph{kh\={a}nom}, \emph{zoj}, \emph{zan\={a}ne}, \emph{raqsande}, \emph{b\={a}rd\={a}ri} & \emph{\={a}q\={a}}, \emph{d\={a}d\={a}sh}, \emph{em\={a}m}, \emph{b\={a}b\={a}}, \emph{moll\={a}} \\
    \emph{pedarbozorg}/\emph{m\={a}darbozorg} & 1 & $-0.189$ [$-0.224$, $-0.154$] & $<2.2\times10^{-16}$ & \emph{seksi}, \emph{ch\={a}q}, \emph{maq'ad}, \emph{khoshgel}, \emph{kir} & \emph{pedar}, \emph{ersi}, \emph{farzand}, \emph{pedari}, \emph{enhel\={a}l} \\
    \emph{d\={a}m\={a}d}/\emph{arus} & 1 & $-0.299$ [$-0.332$, $-0.266$] & $<2.2\times10^{-16}$ & \emph{arus}, \emph{leb\={a}s}, \emph{kafsh}, \emph{model}, \emph{bacheg\={a}ne} & \emph{mohaqqeq}, \emph{moll\={a}}, \emph{Abdoll\={a}h}, \emph{Mahmud}, \emph{rowh\={a}ni} \\
    \bottomrule
  \end{tabular}}\endgroup
\end{table}

\section{Five further languages with smaller concreteness norms, and the pattern across languages}
\label{si:forest}

Human concreteness norms also exist, at a smaller scale, for several further languages, and the same test can be run on them with the same language-specific fastText vectors\cite{grave2018} and the same construction of the gender axis from five kinship and role seed pairs. We did so for every such language whose norms are publicly redistributed and whose seed words are all present in the vectors: Italian (1,121 words; the Italian adaptation of the ANEW list\cite{montefinese2014}), French (1,659 words\cite{bonin2018}), Turkish (568 words\cite{goz2017}), Portuguese (3,800 words; the Minho Word Pool\cite{soares2017}) and Indonesian (1,487 words\cite{sianipar2016}). Two caveats apply to all five. With the exception of the Portuguese set, these word lists were assembled for affective or age-of-acquisition norming and so sample the vocabulary unevenly, with an emphasis on emotionally loaded words and words for people; and at a few hundred to a few thousand words they do not support the decile, covariate and within-domain analyses of the preceding sections. We therefore report for each language the correlation between the gender score and concreteness with its 95\% confidence interval, the same correlation for the extended ten-pair axis, the range of correlations obtained by dropping one seed pair from the primary set, and a check that the axis orders the male above the female member of ten held-out gendered word pairs (Supplementary Table~\ref{tab:S-forest}). The seed pairs are the counterparts of those used for the other languages (for example \emph{uomo}/\emph{donna}, \emph{ragazzo}/\emph{ragazza}, \emph{padre}/\emph{madre} in Italian; \emph{adam}/\emph{kadın}, \emph{oğlan}/\emph{kız}, \emph{baba}/\emph{anne} in Turkish; \emph{pria}/\emph{wanita}, \emph{ayah}/\emph{ibu}, \emph{saudara}/\emph{saudari} in Indonesian), and the axis passed the held-out check in all ten pairs in every language.

In all five languages the correlation is negative (Supplementary Table~\ref{tab:S-forest}): Turkish ($r = -0.244$; 95\% CI $= -0.320$ to $-0.165$), Italian ($-0.187$; $-0.243$ to $-0.129$), French ($-0.118$; $-0.166$ to $-0.070$), Portuguese ($-0.039$; $-0.071$ to $-0.008$; $P = 0.015$) and Indonesian ($-0.027$; $-0.079$ to $0.025$; $P = 0.31$). The extended axis gives the same sign in the first four ($-0.192$, $-0.203$, $-0.141$ and $-0.061$) and the leave-one-pair-out ranges lie entirely below zero for them; in Indonesian the estimate is indistinguishable from zero and its sign depends on the seed set (extended axis $0.054$; leave-one-out range $-0.118$ to $0.033$), so we treat Indonesian as inconclusive rather than as a replication. The words at the extremes of the axes are gendered in every language---\emph{donna}, \emph{ragazza}, \emph{regina}, \emph{suora}, \emph{madre} against \emph{fratello}, \emph{amico}, \emph{padre}, \emph{ragazzo} in Italian; \emph{mère}, \emph{reine}, \emph{soeur}, \emph{tante}, \emph{dame} against \emph{homme}, \emph{frère}, \emph{bougre} in French; \emph{hamile} `pregnant', \emph{kadın} `woman', \emph{makyaj} `make-up' against \emph{dede} `grandfather', \emph{baba} `father', \emph{berber} `barber' in Turkish; \emph{mulher}, \emph{rainha}, \emph{freira}, \emph{mãe}, \emph{parteira} `midwife' against \emph{velho} `old man', \emph{irmão}, \emph{pai}, \emph{homem}, \emph{sábio} `wise man' in Portuguese; \emph{anggun} `graceful', \emph{cantik} `beautiful', \emph{gaun} `gown', \emph{ibu}, \emph{wanita} against \emph{paman} `uncle', \emph{berkelahi} `to fight', \emph{ayah}, \emph{abang}, \emph{tampan} `handsome' in Indonesian---so that each axis is a gender axis in the sense of the preceding sections, with the feminine pole again anchored in appearance and maternity and the masculine pole in kin, age and roles.

\begin{figure}[!htbp]
  \centering
  \includegraphics[width=0.9\textwidth]{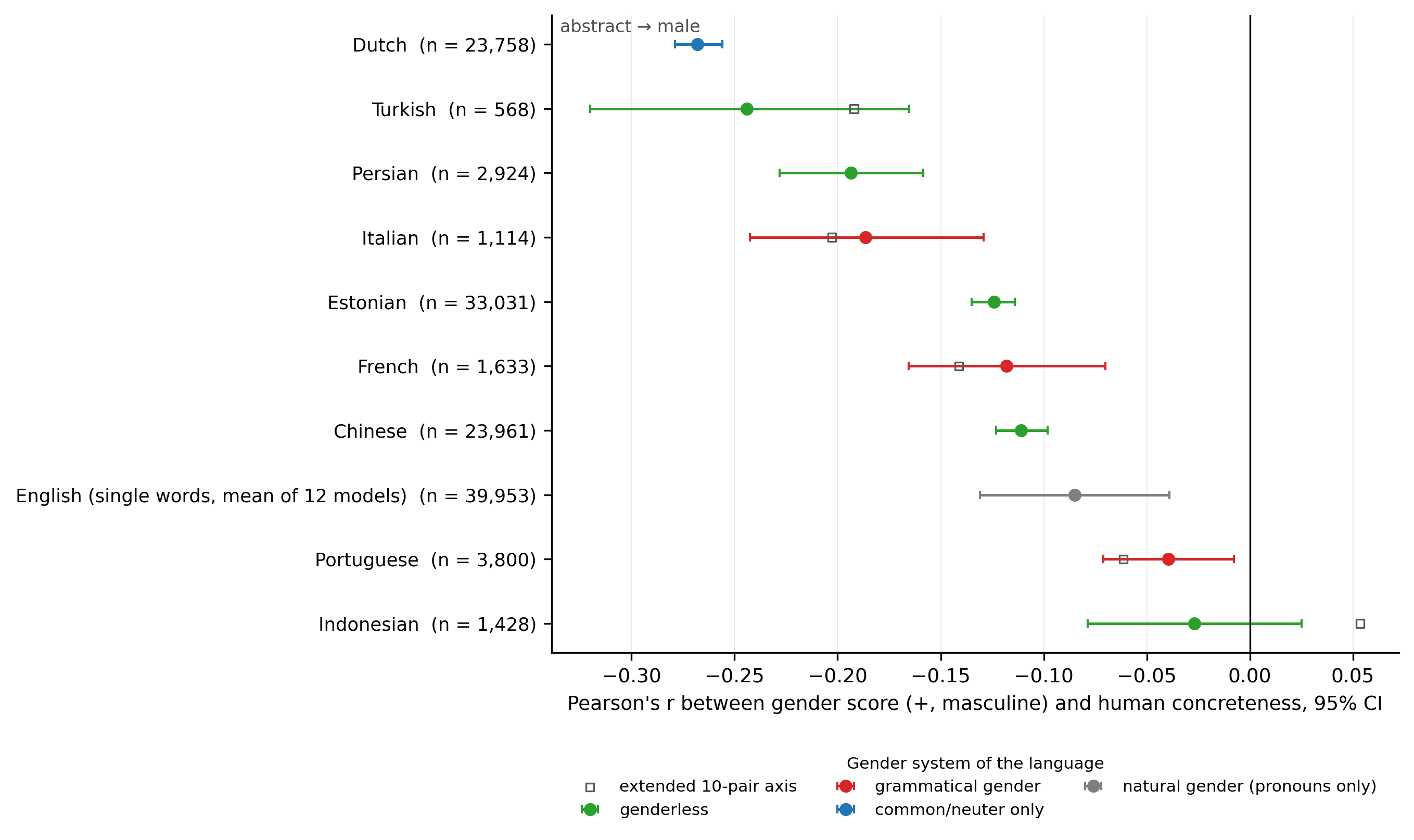}
  \caption{\textbf{The male abstraction bias across languages.} Two-tailed Pearson's correlation (with Fisher-transformed 95\% confidence interval) between the gender score of a word in language-specific word vectors (positive, masculine) and its human concreteness rating, for the nine languages other than English examined in Supplementary Sections~\ref{si:dutch} to~\ref{si:forest}, together with the mean across the 12 English models for single words from Fig.~1a (Supplementary Table~\ref{tab:S-fig1-human}; its interval is the 95\% confidence interval across models). Filled circles, primary five-pair axis; open squares, extended ten-pair axis for the five languages of Supplementary Table~\ref{tab:S-forest}. Colours give the gender system of the language. $n$ is the number of rated words with a vector. Negative values indicate that more masculine words are more abstract.}
  \label{fig:S-forest}
\end{figure}

Supplementary Fig.~\ref{fig:S-forest} places these five estimates alongside the four large-norm replications and the English single-word estimate of Fig.~1a. Across the nine languages other than English, the correlation is negative in all nine and its confidence interval excludes zero in eight (two-tailed binomial sign test on the direction, $P = 0.0039$); the median correlation is $-0.124$, and the English single-word mean across the 12 models ($-0.085$) lies in the middle of the cross-language distribution. The estimates do not sort by the gender system of the language: the three largest are Dutch (common/neuter gender only), Turkish and Persian (genderless), and the languages with masculine/feminine grammatical gender (Italian, French, Portuguese) span the range from the fourth-largest (Italian) to the smallest significant estimate (Portuguese). The Romance estimates are, on average, somewhat smaller than those of the genderless languages, which would be expected if the gender axis of a language with grammatical gender partly encodes the grammatical gender of nouns---a component unrelated to concreteness that adds noise to the gender score---but the present data cannot separate that explanation from differences in the norm sets, and the question is better addressed with norms that record grammatical gender directly. In sum, the available non-English norms, large and small, give the same answer: more masculine words are rated as more abstract, in nine languages from four families (Indo-European, in its Germanic, Romance and Iranian branches; Uralic; Sino-Tibetan; and Turkic, with Austronesian inconclusive) written in three scripts, with and without grammatical gender.

\begin{table}[!htbp]
  \centering
  \caption{\textbf{The male abstraction bias in five further languages with smaller concreteness norms.} For each language, the human concreteness norms used (with the number of rated words that have a fastText vector, and that number as a percentage of all rated words), the number of held-out gendered word pairs in which the primary gender axis orders the male above the female member, the two-tailed Pearson's correlation between the gender score (positive, masculine) and concreteness for the primary five-pair axis with its Fisher-transformed 95\% confidence interval and $P$ value, the same correlation for the extended ten-pair axis, and the range of correlations obtained by dropping each of the five primary seed pairs in turn. Gender associations were measured in fastText's vectors for each language\cite{grave2018}.}
  \label{tab:S-forest}
  \begingroup\footnotesize\setlength{\tabcolsep}{4pt}
  \resizebox{\textwidth}{!}{%
  \begin{tabular}{@{}llrcllll@{}}
    \toprule
    Language (gender system) & Norms & $n$ (coverage) & Held-out pairs & $r$ [95\% CI], primary axis & $P$ & $r$ [95\% CI], extended axis & Leave-one-pair-out range \\
    \midrule
    Turkish (genderless) & Göz et~al.\cite{goz2017} & 568 (100\%) & 10/10 & $-0.244$ [$-0.320$, $-0.165$] & $3.7\times10^{-9}$ & $-0.192$ [$-0.270$, $-0.112$] & $-0.324$ to $-0.150$ \\
    Italian (grammatical gender) & Montefinese et~al.\cite{montefinese2014} & 1{,}114 (99\%) & 10/10 & $-0.187$ [$-0.243$, $-0.129$] & $3.5\times10^{-10}$ & $-0.203$ [$-0.258$, $-0.146$] & $-0.205$ to $-0.162$ \\
    French (grammatical gender) & Bonin et~al.\cite{bonin2018} & 1{,}633 (98\%) & 10/10 & $-0.118$ [$-0.166$, $-0.070$] & $1.7\times10^{-6}$ & $-0.141$ [$-0.188$, $-0.093$] & $-0.145$ to $-0.091$ \\
    Portuguese (grammatical gender) & Soares et~al.\cite{soares2017} & 3{,}800 (100\%) & 10/10 & $-0.039$ [$-0.071$, $-0.008$] & $0.015$ & $-0.061$ [$-0.093$, $-0.030$] & $-0.063$ to $-0.019$ \\
    Indonesian (genderless) & Sianipar et~al.\cite{sianipar2016} & 1{,}428 (96\%) & 10/10 & $-0.027$ [$-0.079$, $0.025$] & $0.31$ & $0.054$ [$0.002$, $0.105$] & $-0.118$ to $0.033$ \\
    \bottomrule
  \end{tabular}}\endgroup
\end{table}

\section{Text-based results for words and phrases analysed separately}
\label{si:words-phrases}

The main text reports the relationship between the gender association of a term and its
human-rated concreteness while pooling single words\cite{brysbaert2014} and multiword
phrases\cite{muraki2023} (Fig.~1a). The two sets of norms were collected with the same instructions
and rating scale, but they differ in ways that could matter for our analyses: phrases are
predominantly nominal and they are less polysemous than single words. Here we confirm that the male
abstraction bias does not depend on pooling the two.

For each language model, we computed the two-tailed Pearson's correlation between the gender
association of a term (positive values indicate masculine associations) and its concreteness,
separately for (i) words and phrases pooled, (ii) words only and (iii) phrases only, with 95\%
confidence intervals obtained using the Fisher transformation. Within each of these analyses,
\textit{P} values were Bonferroni-corrected for the number of language models examined. To
summarize across models, we treated language models as the units of analysis: we report the
unweighted mean of the per-model correlations, with a 95\% confidence interval from the $t$
distribution across models, and a two-tailed one-sample $t$-test of the per-model correlations
against zero. No model is weighted more heavily than any other.

Supplementary Table~\ref{tab:S-fig1-human} shows that the results replicate in each dataset. For
words, the mean correlation across models was \FigOneAWrdMeanR\ (95\% CI $=$ \FigOneAWrdCIlo\ to
\FigOneAWrdCIhi; $t(\FigOneAWrdDf) =$~\FigOneAWrdT; \textit{P}~\FigOneAWrdP), and the correlation
was negative in \FigOneAWrdKNeg\ of \FigOneAWrdKTot\ models (significantly so after Bonferroni
correction in \FigOneAWrdKSigNeg). Terms in the most feminine decile had a mean concreteness of
\FigOneAWrdDecOne, compared with \FigOneAWrdDecTen\ for terms in the most masculine decile. For
phrases, the mean correlation was \FigOneAPhrMeanR\ (95\% CI $=$ \FigOneAPhrCIlo\ to
\FigOneAPhrCIhi; $t(\FigOneAPhrDf) =$~\FigOneAPhrT; \textit{P}~\FigOneAPhrP), and the correlation
was negative in \FigOneAPhrKNeg\ of \FigOneAPhrKTot\ models (significantly so in
\FigOneAPhrKSigNeg); mean concreteness was \FigOneAPhrDecOne\ in the most feminine decile and
\FigOneAPhrDecTen\ in the most masculine decile. The one exception among single words was
ConceptNet, for which the correlation was small and positive ($r = 0.04$; 95\% CI $= 0.03$ to
$0.05$; $n = 38{,}341$ words). The bias is thus present in both datasets but is markedly stronger
for phrases than for single words. Supplementary Section~\ref{si:text-covariates} shows that roughly
half of this difference is accounted for by the frequency, length and part of speech of the phrases,
and that the association in both datasets is robust to these and other lexical covariates.

\begin{table}[!htbp]
  \centering
  \caption{\textbf{The male abstraction bias replicates when words and phrases are analysed
  separately.} Two-tailed Pearson's correlations between the gender association of a term
  (positive, masculine) and human judgements of its concreteness, for each language model. $n$ is
  the number of rated terms scored by the model. The final row gives the unweighted mean across
  models with a 95\% confidence interval across models. $^{\dagger}$Not significant after Bonferroni
  correction for the number of models within the column.}
  \label{tab:S-fig1-human}
  \TabFigOneHuman
\end{table}

\section{Sensorimotor results extended to phrases}
\label{si:lancaster}

The Lancaster Sensorimotor Norms\cite{lynott2020} provide human ratings, for approximately 40,000
single words, of how strongly the referent of each word is experienced through each of six
perceptual modalities (auditory, gustatory, haptic, interoceptive, olfactory and visual), each on
a scale from 0 (not experienced at all) to 5 (experienced greatly). We measured the perceptual
strength of a word as its mean rating across the six modalities. Fig.~1c of the main text reports
the relationship between gender association and perceptual strength for the 39,706 words in the
concreteness norms that were rated in the Lancaster norms (Supplementary Table~\ref{tab:S-fig1-lanc},
left). Here we ask whether the same relationship holds for multiword phrases, which the Lancaster
norms do not cover.

We extended the norms to phrases by assigning each phrase the mean perceptual strength of its
constituent words that appear in the norms. Of the 62,889 phrases in the concreteness norms,
62,321 contained at least one such word; on average 91\% of a phrase's words were covered, and
80\% of phrases were covered completely. This extension rests on two simplifying assumptions:
that the perceptual strength of a phrase is compositional, which does not hold for idiomatic
phrases, and that function words, which are low in perceptual strength, contribute equally to the
average. Averaging also compresses the range of perceptual strength among phrases relative to
words. The extension should therefore be read as a conservative approximation.

We repeated the analysis described in Supplementary Section~\ref{si:words-phrases}, replacing
concreteness with perceptual strength, for (i) the words alone, as in Fig.~1c, (ii) the phrases
alone and (iii) words and phrases pooled. Supplementary Table~\ref{tab:S-fig1-lanc} shows that the
relationship replicates in the extended data. For the phrases alone, the mean correlation across
models between masculine association and perceptual strength was \FigOneCPhrMeanR\ (95\% CI $=$
\FigOneCPhrCIlo\ to \FigOneCPhrCIhi; $t(\FigOneCPhrDf) =$~\FigOneCPhrT; \textit{P}~\FigOneCPhrP),
and the correlation was negative and significant after Bonferroni correction in every model
(\FigOneCPhrKSigNeg\ of \FigOneCPhrKTot). Phrases in the most feminine decile had a mean perceptual
strength of \FigOneCPhrDecOne, compared with \FigOneCPhrDecTen\ for phrases in the most masculine
decile. For words and phrases pooled, the mean correlation was \FigOneCAllMeanR\ (95\% CI $=$
\FigOneCAllCIlo\ to \FigOneCAllCIhi; $t(\FigOneCAllDf) =$~\FigOneCAllT; \textit{P}~\FigOneCAllP;
negative in \FigOneCAllKNeg\ of \FigOneCAllKTot\ models and significantly so in \FigOneCAllKSigNeg,
the exception being BERT, $r = -0.002$, \textit{P}~$= 0.47$). For comparison, the corresponding
mean correlation for the words alone (Fig.~1c) was \FigOneCWrdMeanR\ (95\% CI $=$
\FigOneCWrdCIlo\ to \FigOneCWrdCIhi; negative in \FigOneCWrdKNeg\ of \FigOneCWrdKTot\ models and
significantly so in \FigOneCWrdKSigNeg; the exception was GPT2, $r = -0.005$,
\textit{P}~$= 0.31$). The association between masculine connotations and weaker perceptual
grounding is therefore not limited to the single words for which the norms were collected; if
anything it is stronger among phrases, as was the case for concreteness (Supplementary
Section~\ref{si:words-phrases}).

\begin{table}[!htbp]
  \centering
  \caption{\textbf{More masculine terms are lower in perceptual strength, for the single words in
  the original Lancaster Sensorimotor Norms (Fig.~1c) and for phrases when the norms are extended
  to them.} Two-tailed Pearson's correlations between the gender association of a term (positive,
  masculine) and its perceptual strength (mean rating across six perceptual modalities; 0--5), for
  each language model. For phrases, perceptual strength is the mean across the constituent words
  that appear in the norms. $n$ is the number of rated terms scored by the model. The final row
  gives the unweighted mean across models with a 95\% confidence interval across models.
  $^{\dagger}$Not significant after Bonferroni correction for the number of models within the
  column.}
  \label{tab:S-fig1-lanc}
  \TabFigOneLancaster
\end{table}

\section{Robustness of the text-based results to lexical covariates and to the composition of the vocabulary}
\label{si:text-covariates}

Concreteness is correlated with several other lexical properties that embedding-based gender
associations could, in principle, be tracking. Among the single words in the concreteness
norms\cite{brysbaert2014}, more concrete words are more frequent ($r = 0.17$ with log-transformed
SUBTLEX-US frequency), shorter ($r = -0.29$ with length in characters and $-0.39$ with the number of
syllables), acquired earlier ($r = -0.38$ with age of acquisition\cite{kuperman2012}), more positive
($r = 0.10$ with valence\cite{warriner2013}) and less arousing ($r = -0.17$). Because embedding-based
measures are known to be sensitive to word frequency and length, we tested whether the association
between gender and concreteness reported in Fig.~1a survives controlling for these properties, and
whether it depends on particular parts of the vocabulary.

For each language model, we regressed the gender association of a term ($z$-scored within the
model; positive values indicate masculine associations) on its human-rated concreteness
($z$-scored) and on a set of covariates, using ordinary least squares with heteroskedasticity-robust
(HC1) standard errors; the coefficient on concreteness is a standardized $\beta$, which equals
Pearson's $r$ when no covariates are included. For single words, the covariates were word frequency
(log-transformed SUBTLEX-US frequency, as distributed with the concreteness
norms\cite{brysbaert2014}); length, measured as the number of characters and the number of syllables
(from the CMU Pronouncing Dictionary, with an orthographic approximation for words it does not
contain); polysemy, measured as the log-transformed number of senses recorded for the word in
WordNet\cite{miller1995} ($n = 34{,}026$ words with an entry); age of acquisition\cite{kuperman2012}
($n = 23{,}860$); valence, arousal and dominance\cite{warriner2013} ($n = 13{,}385$); and the word's
dominant part of speech, as recorded in the concreteness norms or, for words without one, the part of
speech of the word's first WordNet sense, entered as fixed effects. Because
polysemy, age of acquisition and the affective ratings are available for subsets of the words, for
these covariates we also report the association without controls on the same subset, which separates
the effect of the control from the effect of the change in sample. For phrases, frequency, age of
acquisition and the affective ratings were assigned as the mean over the phrase's constituent words
that appear in the respective norms; length was measured as the number of characters and the number
of words; polysemy as the number of WordNet senses of the phrase itself, together with an indicator
for whether the phrase has a WordNet entry (25,202 of the 62,889 phrases do); and part of speech was
taken from the phrase's WordNet entry or, failing that, from its head word. Results were summarized
across models as in Supplementary Section~\ref{si:words-phrases}: an unweighted mean of the per-model
estimates, a 95\% confidence interval and a one-sample $t$-test across models, and a count of the
models in which the estimate was negative and significant after Bonferroni correction for the number
of models. We further (i) restricted the analysis to words known by at least 95\% of the raters who
judged their concreteness, and to words whose concreteness ratings had below-median standard
deviation across raters, and (ii) estimated the association without controls separately within each
quintile of word frequency.

Supplementary Table~\ref{tab:S-text-cov-words} reports the results for single words. Word frequency
did not account for the association: controlling for it left the mean estimate across models
essentially unchanged (mean $\beta = -0.091$; 95\% CI $= -0.143$ to $-0.039$; $t(11) = -3.88$;
\textit{P}~$= 0.003$; compared with $-0.085$ without controls), and the association was negative in
every quintile of word frequency (mean $\beta = -0.127$ among the least frequent fifth of words,
$-0.096$, $-0.071$ and $-0.065$ in the next three fifths, and $-0.089$ among the most frequent fifth).
Controlling for length strengthened the association (mean $\beta = -0.117$; 95\% CI $= -0.169$ to
$-0.064$), as did restricting the analysis to reliably rated words (mean $\beta = -0.115$ for words with
below-median rater disagreement and $-0.114$ with both restrictions; negative in 11 of 12 models in
each case). Polysemy and age of acquisition made no material difference once the change in sample is
taken into account. The one covariate that attenuated the association was affective content: among
the 13,385 words with affective ratings, controlling for valence, arousal and dominance reduced the
mean estimate from $-0.063$ to $-0.052$ (95\% CI $= -0.094$ to $-0.010$; negative in 9 of 12 models),
because more positive words are both more concrete and more feminine. Controlling for all covariates
simultaneously, together with part-of-speech fixed effects, nevertheless yielded a larger estimate
than the association without controls on the same words (mean $\beta = -0.099$; 95\% CI $= -0.143$ to
$-0.054$; $t(11) = -4.90$; \textit{P}~$= 4.7\times10^{-4}$; negative in 11 of 12 models and significantly so in 10,
compared with $-0.063$ without controls; $n = 13{,}069$ to $13{,}082$ words per model), and controlling
for frequency, length, polysemy and part of speech in the much larger set of words with a WordNet
entry gave a similar estimate (mean $\beta = -0.101$; 95\% CI $= -0.149$ to $-0.053$; negative in 11 of
12 models; $n = 28{,}958$ to $34{,}026$). Supplementary Table~\ref{tab:S-text-cov-permodel} reports the
estimates for each model. As in Fig.~1a, ConceptNet is the exception among single words, with a small
positive association that becomes negative only when all covariates are included
($\beta = -0.032$), and GPT2 and RoBERTa remain close to zero throughout.

For phrases (Supplementary Table~\ref{tab:S-text-cov-phrases}), the association was negative in all
12 models in every specification with fewer than the full set of controls, but its magnitude depended
on lexical covariates more than it did for single words. Controlling for the frequency of the
constituent words reduced the mean estimate from $-0.252$ (95\% CI $= -0.295$ to $-0.210$) to $-0.146$
($-0.185$ to $-0.108$), and controlling for length or for part of speech reduced it to $-0.181$ and
$-0.196$, respectively, whereas polysemy, age of acquisition and affective content had little effect.
With all covariates included, the mean estimate was $-0.106$ (95\% CI $= -0.150$ to $-0.062$;
$t(11) = -5.30$; \textit{P}~$= 2.5\times10^{-4}$; negative in 11 of 12 models and significantly so in
10; the exception was RoBERTa, $\beta = 0.041$), compared with $-0.247$ without controls on the same
phrases.
Roughly half of the stronger association observed for phrases than for words in Fig.~1a therefore
reflects the fact that more abstract phrases are composed of rarer words, are longer and are more
often verb phrases, properties that are correlated with gender association in their own right; the
remaining association, which is comparable in size to that observed for single words, is independent
of these properties.

Finally, we asked whether the association depends on particular parts of the vocabulary
(Supplementary Table~\ref{tab:S-text-composition}). First, we excluded nouns that denote persons,
lexically gendered terms, function words and proper names, using the criteria described under
`Identifying Semantic Domains' in Materials and Methods for Fig.~2, which removes terms whose gender association reflects the gender of
a referent rather than of a concept. The mean estimate was $-0.092$ for words (95\% CI $= -0.147$ to
$-0.037$; negative in 11 of 12 models) and $-0.269$ for phrases ($-0.317$ to $-0.222$; 12 of 12), and
adding controls for frequency, length and part of speech to this restricted vocabulary gave $-0.103$
and $-0.126$. Second, we excluded the 5\% or 10\% of terms with the most extreme gender associations in
each model, to check that the association is not carried by a small set of strongly gendered terms.
The estimates were slightly smaller for words ($-0.072$ and $-0.068$; negative in 10 of 12 models) and
essentially unchanged for phrases ($-0.258$ and $-0.246$), and Spearman's rank correlations, which are
insensitive to extreme values, gave $-0.081$ and $-0.273$. Third, for each model we recomputed the
association omitting one WordNet domain at a time (39 or 40 domains with at least 100 words and 36 or
37 domains with at least 100 phrases, depending on the model's coverage) and retained the least negative
of the resulting estimates. This worst case averaged $-0.067$ for words (negative in 11 of 12 models;
the most influential domain was the adverbs for seven of the models) and $-0.243$ for phrases (12 of
12; \textit{noun.food} for four models). No single domain, and no small set of strongly gendered terms,
accounts for the association.

\begin{table}[!htbp]
  \centering
  \caption{\textbf{The male abstraction bias in single words is robust to lexical covariates, to
  restricting the analysis to reliably rated words, and to word frequency.} Each row is a separate
  regression, estimated once for each of the 12 language models, of the gender
  association of a word ($z$-scored within the model; positive, masculine) on its human-rated
  concreteness ($z$-scored) and the covariates listed, with heteroskedasticity-robust (HC1) standard
  errors. Mean $\beta$ is the unweighted mean of the standardized coefficient on concreteness across
  models, with a 95\% confidence interval from the $t$ distribution across models and the range across
  models; Negative gives the number of models in which the coefficient was negative and Sig.\ the
  number in which it was negative and significant after Bonferroni correction for 12 models
  (two-tailed). Where a covariate is available for a subset of words, the association without controls
  is also reported for the same subset. $n$ per model, number of words in the regression (range across
  models).}
  \label{tab:S-text-cov-words}
  \TabTextCovWords
\end{table}

\begin{table}[!htbp]
  \centering
  \caption{\textbf{The male abstraction bias in phrases is robust to lexical covariates, although
  constituent-word frequency, length and part of speech account for part of its magnitude.} As
  Supplementary Table~\ref{tab:S-text-cov-words}, for the 62,889 human-rated phrases and the same 12
  language models. Frequency, age of acquisition and affective ratings are the means over each phrase's
  constituent words that appear in the respective norms; polysemy is the number of WordNet senses of
  the phrase itself, entered together with an indicator for whether the phrase has a WordNet entry;
  part of speech is that of the phrase's WordNet entry or, failing that, of its head word. Sig., number
  of models in which the coefficient was negative and significant after Bonferroni correction for 12
  models.}
  \label{tab:S-text-cov-phrases}
  \TabTextCovPhrases
\end{table}

\begin{table}[!htbp]
  \centering
  \caption{\textbf{Per-model estimates for selected specifications of Supplementary
  Tables~\ref{tab:S-text-cov-words} and~\ref{tab:S-text-cov-phrases}.} Standardized coefficients
  ($\beta$) on concreteness from regressions of the gender association of a term on its concreteness
  with no controls (None); controlling for word frequency (+ Freq.); for length (+ Length); for
  valence, arousal and dominance (+ VAD; words only, $n \approx 13{,}300$); for part of speech (+ POS;
  phrases only); and for all covariates and part-of-speech fixed effects (+ All; $n \approx 13{,}100$
  words and $48{,}000$--$57{,}500$ phrases per model). The final row gives the unweighted mean across
  models. $^{\dagger}$Not significant after Bonferroni correction for the number of models within the
  column (two-tailed).}
  \label{tab:S-text-cov-permodel}
  \TabTextCovPerModel
\end{table}

\begin{table}[!htbp]
  \centering
  \caption{\textbf{The male abstraction bias in text does not depend on particular parts of the
  vocabulary.} Standardized coefficients ($\beta$) on concreteness, summarized across models as in
  Supplementary Table~\ref{tab:S-text-cov-words} (Neg., number of models with a negative coefficient;
  in parentheses, the number that were also significant after Bonferroni correction). Rows exclude
  person nouns, lexically gendered terms, function words and proper names (for phrases, person nouns
  and lexically gendered terms), with and without controls for frequency, length and part of speech;
  exclude the 5\% or 10\% of terms with the most extreme gender association ($|z|$) within each model;
  report Spearman's rank correlation; and report, for each model, the least negative estimate obtained
  when omitting each WordNet domain (with at least 100 terms) in turn.}
  \label{tab:S-text-composition}
  \TabTextComposition
\end{table}

\section{Word-level illustrations of the bias within example domains}
\label{si:domain-examples}

\begin{figure}[!htbp]
  \centering
  \includegraphics[width=\textwidth]{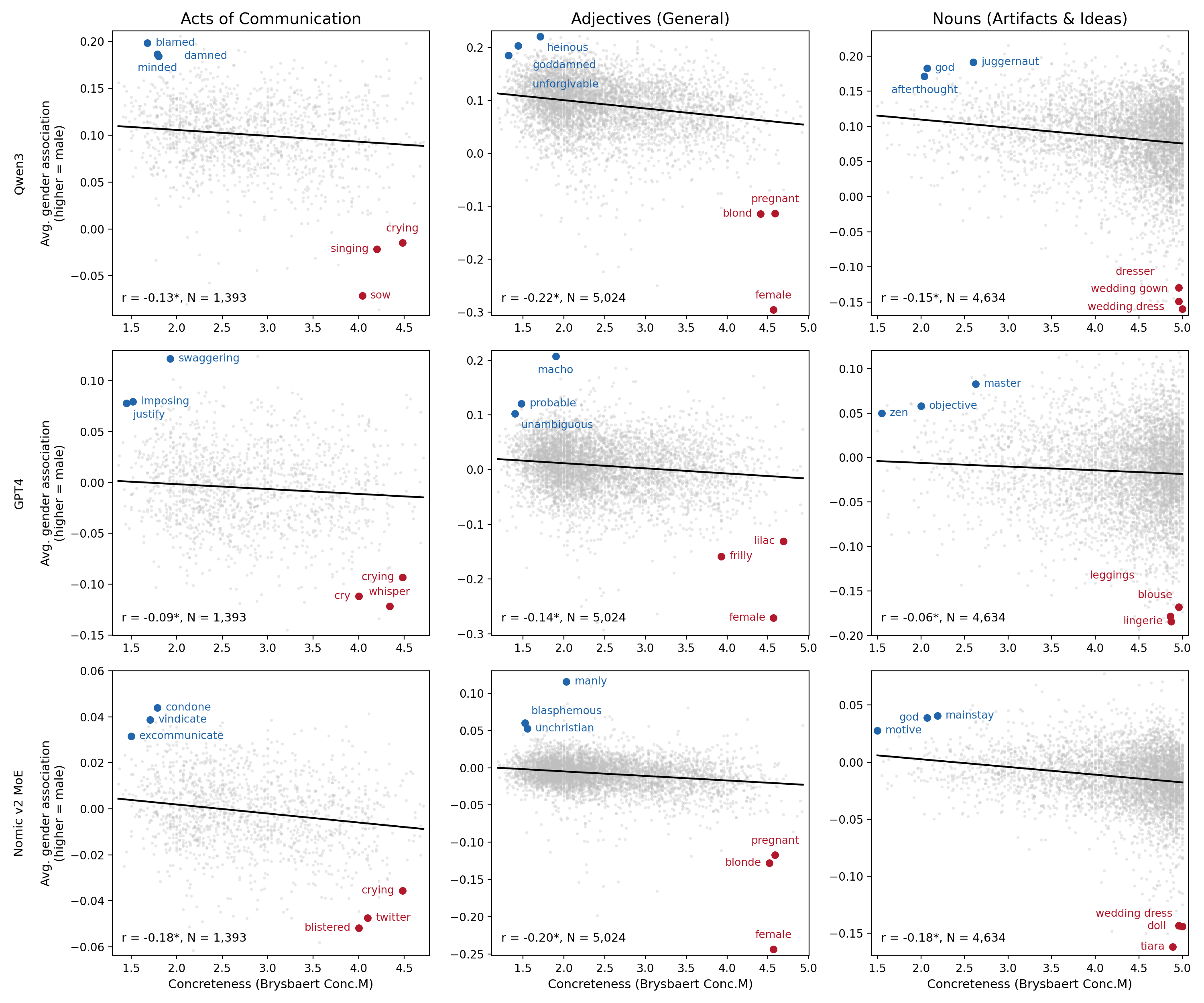}
  \caption{\textbf{Word-level illustrations of the male abstraction bias within three
  example WordNet domains.} Each panel plots, for one embedding model (rows:
  \protect\code{Qwen3-Embedding-0.6B}; \protect\code{text-embedding-3-large}, labelled GPT4 as in Figs.~1 and 2;
  \protect\code{nomic-embed-text-v2-moe}) and one WordNet domain (columns: acts of communication,
  \textit{verb.communication}; general adjectives, \textit{adj.all}; artifact nouns,
  \textit{noun.artifact}), the gender association of every word in the domain (vertical
  axis; higher values, more masculine) against its human-rated concreteness (horizontal
  axis; 1, abstract; 5, concrete)\cite{brysbaert2014}. Grey points show all words in the
  domain; the black line is the least-squares fit; the annotation gives the Pearson
  correlation (asterisk, $P_{\mathrm{FDR}} < 0.05$; Supplementary
  Table~\ref{tab:S-domain-examples}) and the number of words. In each panel the three
  most strongly male-associated abstract words (blue) and the three most strongly
  female-associated concrete words (red) are highlighted, ranked by
  $-z(\text{gender}) \times z(\text{concreteness})$ within the panel. For readability,
  the vertical axis is truncated at the male extreme in four panels, cropping 14, 4, 10
  and 0 of the panel's words (\protect\code{text-embedding-3-large} artifact nouns;
  \protect\code{nomic-embed-text-v2-moe} adjectives, artifact nouns and acts of communication,
  respectively; the cropped words are strongly male-associated terms such as \emph{man},
  \emph{king} and \emph{boss}); all correlations are computed on the full samples.}
  \label{fig:S-domain-examples}
\end{figure}

The main text (Fig.~2) establishes that the male abstraction bias holds within semantic
domains by contrasting, in every WordNet domain, the mean gender association of the most
abstract third of terms with that of the most concrete third, averaged across language
models. That analysis is deliberately aggregated: it collapses the concreteness continuum
into terciles, pools across models, and---to give the most conservative test---excludes
function words, lexically gendered terms and person-denoting nouns. Here we complement it
with a disaggregated view of the same phenomenon, illustrating what the within-domain
association looks like at the level of individual words. For three example domains and
three embedding models, Supplementary Fig.~\ref{fig:S-domain-examples} plots every word's
gender association against its human-rated concreteness and highlights, in each panel, the
individual words that most strongly anchor the two ends of the bias: the most strongly
male-associated abstract words and the most strongly female-associated concrete words.

\paragraph{Domains and vocabulary.}
The vocabulary is the 39,953 entries of the Brysbaert et~al.\cite{brysbaert2014}
concreteness norms scored by our embedding models (Supplementary
Section~\ref{sec:gender-scores}). We selected three WordNet\cite{miller1995} domains as
illustrative examples spanning the major word forms analysed in Fig.~2: acts of
communication (\textit{verb.communication}; for example, \emph{whisper}, \emph{condone}),
general adjectives (\textit{adj.all}; WordNet's participial adjective file
\textit{adj.ppl} is folded in), and artifact nouns (\textit{noun.artifact}; for example,
\emph{blouse}, \emph{doll}), a domain whose membership nonetheless spans the full
concreteness scale because WordNet assigns some abstract-rated words to it through
non-physical senses (for example, \emph{mainstay}, \emph{objective}). Because the purpose
of this analysis is illustrative rather than the aggregate estimation of Fig.~2, domain
membership is defined more inclusively than in the main text: rather than assigning each
word only the domain of its primary WordNet sense, a word is included in a domain if
\emph{any} of its WordNet senses falls in that domain, subject to a part-of-speech gate
that requires the word's dominant part of speech in the Brysbaert norms
(\code{Dom_Pos}) to match the part of speech of the domain (verbs for
\textit{verb.*}, nouns for \textit{noun.*}, and so on). The gate is what makes the
sense-based mapping safe: without it, common nouns leak into verb domains through rare
verb senses (WordNet lists \emph{butterfly} as a communication verb, ``to talk
amorously'', and \emph{skirt} as ``to evade an issue''), even though the gender score of
the word form is dominated by its nominal use. The small set of two-word entries in the
norms carries no dominant part of speech, and such entries are admitted only to
\textit{noun.artifact}, and only when WordNet lists them exclusively as nouns (for
example, \emph{wedding dress}). This yields 1,393 words for \textit{verb.communication},
5,024 for \textit{adj.all} and 4,634 for \textit{noun.artifact}. Unlike Fig.~2, the
vocabulary here is not filtered for lexically gendered terms: the aim is to show which
words carry the extremes of the association, and the fact that explicitly gendered
adjectives (\emph{female}, \emph{macho}, \emph{manly}) surface at those extremes is itself
informative; Fig.~2 and Supplementary Section~\ref{si:text-covariates} establish that the
aggregate bias survives their exclusion.

\paragraph{Models and statistics.}
The three models shown are \code{text-embedding-3-large} (GPT4),
\code{Qwen3-Embedding-0.6B} and \code{nomic-embed-text-v2-moe}, spanning a closed
provider-managed model and two open-weight sentence transformers (Supplementary
Table~\ref{tab:embedding-models}); gender scores are computed as in Supplementary
Section~\ref{sec:gender-scores}. For each model and domain we computed the two-tailed
Pearson's correlation between gender association and concreteness across all words in the
domain, with 95\% confidence intervals from the Fisher transformation.
\textit{P} values were corrected for multiple comparisons by the Benjamini--Hochberg
procedure across the full family of 48 model--domain cells: the 12 language models of
Figs.~1 and 2 (Supplementary Table~\ref{tab:embedding-models}) crossed with the three
domains above plus \textit{adv.all}, the fourth domain family examined in this analysis.

In all nine panels of Supplementary Fig.~\ref{fig:S-domain-examples} the association is
negative: more concrete words within a domain carry more feminine associations, and more
abstract words more masculine ones, with correlations ranging from $r = -0.06$
(\code{text-embedding-3-large}, artifact nouns) to $r = -0.22$
(\code{Qwen3-Embedding-0.6B}, adjectives), all $P_{\mathrm{FDR}} < 0.002$ (Supplementary
Table~\ref{tab:S-domain-examples}). The same holds in the fourth, unshown domain,
\textit{adv.all} ($n = 1{,}654$), for \code{Qwen3-Embedding-0.6B} ($r = -0.10$;
$P_{\mathrm{FDR}} = 8.2\times10^{-5}$) and \code{nomic-embed-text-v2-moe} ($r = -0.14$;
$P_{\mathrm{FDR}} = 1.9\times10^{-8}$), though not significantly for
\code{text-embedding-3-large} ($r = -0.03$; $P_{\mathrm{FDR}} = 0.32$). Across the full
48-cell family, the correlation was negative in 39 cells and significantly so after
correction in 33; the nine positive estimates were all small ($r \leq 0.03$), none was
significant after correction, and they were confined to the three BERT-family models
(\code{bert-base-uncased}, \code{distilbert-base-uncased} and \code{roberta-base}),
\code{gpt2} and ConceptNet Numberbatch.

The highlighted words make the two poles of the association concrete. Within each panel,
every word was scored for how strongly it exemplifies the bias, as the product
$-z(\text{gender}) \times z(\text{concreteness})$ of its standardized coordinates, and
the three highest-scoring words in each of the two bias-consistent quadrants are labelled:
abstract-and-male in blue, concrete-and-female in red. At the female--concrete pole the
domains fill with garments and domestic artifacts (\emph{wedding dress}, \emph{blouse},
\emph{lingerie}, \emph{doll}, \emph{tiara}), gendered descriptors of appearance
(\emph{blonde}, \emph{pregnant}) and expressive, embodied acts of communication
(\emph{crying}, \emph{singing}, \emph{whisper}); at the male--abstract pole sit evaluative
and normative terms (\emph{heinous}, \emph{unforgivable}, \emph{blasphemous}), verbs of
judgement and authority (\emph{justify}, \emph{condone}, \emph{vindicate},
\emph{excommunicate}) and abstract-rated nouns (\emph{juggernaut}, \emph{mainstay},
\emph{motive}). Where the top-ranked candidate for a label was redundant with another
highlighted word form (for example, \emph{goddamn} beside \emph{goddamned}) or opaque in
isolation (for example, abbreviations), the next-ranked candidate is shown instead;
highlighted words are always members of the plotted domain sample, and for the
\code{text-embedding-3-large} artifact panel labels are restricted to single-word entries.
Supplementary Section~\ref{si:domain-sensitivity} shows that neither the Fig.~2
results nor these example panels are sensitive to the domain-assignment rule.

\begin{table}[!htbp]
  \centering
  \caption{\textbf{Within-domain correlations between gender association and concreteness
  for the nine panels of Supplementary Fig.~\ref{fig:S-domain-examples}.} Two-tailed
  Pearson's correlations between the gender association of a word (positive, masculine)
  and its human-rated concreteness\cite{brysbaert2014}, computed across all words assigned
  to the domain. 95\% confidence intervals use the Fisher transformation.
  $P_{\mathrm{FDR}}$, Benjamini--Hochberg-corrected \textit{P} values across the full
  family of 48 model--domain cells (the 12 language models of Figs.~1 and 2 $\times$ four
  domains, including \textit{adv.all}, which is not displayed).}
  \label{tab:S-domain-examples}
  \vspace{4pt}
  \begingroup\footnotesize\setlength{\tabcolsep}{6pt}
  \begin{tabular}{@{}llrrr@{}}
  \toprule
  Model & WordNet domain & $n$ words & $r$ [95\% CI] & $P_{\mathrm{FDR}}$ \\
  \midrule
  \code{Qwen3-Embedding-0.6B} & \textit{verb.communication} & 1,393 & $-0.125$ [$-0.177$, $-0.073$] & $6.2\times10^{-6}$ \\
  \code{Qwen3-Embedding-0.6B} & \textit{adj.all} & 5,024 & $-0.224$ [$-0.250$, $-0.198$] & $1.4\times10^{-56}$ \\
  \code{Qwen3-Embedding-0.6B} & \textit{noun.artifact} & 4,634 & $-0.149$ [$-0.177$, $-0.121$] & $2.5\times10^{-23}$ \\
  \addlinespace[2pt]
  \code{text-embedding-3-large} & \textit{verb.communication} & 1,393 & $-0.092$ [$-0.143$, $-0.039$] & $1.1\times10^{-3}$ \\
  \code{text-embedding-3-large} & \textit{adj.all} & 5,024 & $-0.138$ [$-0.165$, $-0.111$] & $5.1\times10^{-22}$ \\
  \code{text-embedding-3-large} & \textit{noun.artifact} & 4,634 & $-0.058$ [$-0.086$, $-0.029$] & $1.6\times10^{-4}$ \\
  \addlinespace[2pt]
  \code{nomic-embed-text-v2-moe} & \textit{verb.communication} & 1,393 & $-0.182$ [$-0.233$, $-0.131$] & $2.3\times10^{-11}$ \\
  \code{nomic-embed-text-v2-moe} & \textit{adj.all} & 5,024 & $-0.203$ [$-0.229$, $-0.176$] & $1.9\times10^{-46}$ \\
  \code{nomic-embed-text-v2-moe} & \textit{noun.artifact} & 4,634 & $-0.179$ [$-0.207$, $-0.151$] & $1.3\times10^{-33}$ \\
  \bottomrule
  \end{tabular}\endgroup
\end{table}

\section{Sensitivity of the within-domain results to the WordNet domain assignment}
\label{si:domain-sensitivity}


The within-domain analyses of Fig.~2 and of Supplementary
Section~\ref{si:domain-examples} both depend on a mapping from words to WordNet
domains, and WordNet leaves that mapping underdetermined for polysemous words:
the main text assigns each word the domain of its primary (first-listed) sense,
whereas the preceding section admits a word to the domain of any of its senses,
in both cases within the word's dominant part of speech. Here we show that the
central within-domain result---that the most abstract third of a domain is more
male-associated than its most concrete third---does not depend on this choice,
or more generally on how individual words are assigned to domains.

\paragraph{Design.}
All comparisons hold the vocabulary and the gender scores fixed and vary only
domain membership, so that any difference between schemes isolates the
contribution of the assignment rule. The vocabulary comprises the 25,597
single-word entries of the concreteness norms\cite{brysbaert2014} whose
dominant part of speech is a noun, verb, adjective or adverb and that have at
least one WordNet\cite{miller1995} sense in that part of speech (two-word
entries carry no dominant part of speech and are excluded; WordNet's
participial adjective file \textit{adj.ppl} is folded into \textit{adj.all}).
Mirroring the main text's exclusion of person-denoting nouns, the
\textit{noun.person} domain is excluded; the main text's remaining lexical
filters (function words and lexically gendered terms) are not reapplied here,
because the manipulation targets the mapping alone and the robustness of the
aggregate association to those filters is established in Supplementary
Section~\ref{si:text-covariates}. Within each of the 12 language models of
Fig.~2, gender associations are
$z$-scored over this fixed vocabulary. We then recompute the Fig.~2 quantity
under each assignment scheme: within every domain containing at least 100
assigned words, words are split into terciles by their concreteness rating; the
domain's gap is the mean gender association of the most abstract third minus
that of the most concrete third, averaged with equal weight across the 12
models; each domain's gap is tested with a two-tailed one-sample $t$-test
across models ($\mathrm{df} = 11$), and the share of domains with a positive gap
is tested with a two-tailed binomial sign test.

Four assignment schemes are compared. (i) \emph{Primary sense}, the main-text
rule: each word receives the single domain of its first-listed WordNet sense
within its dominant part of speech. (ii) \emph{All senses}, the rule of
Supplementary Section~\ref{si:domain-examples}: a word joins every domain in
which it has a sense within its dominant part of speech, so a polysemous word
can contribute to several domains. (iii) \emph{Domain-unambiguous words only}:
the analysis is restricted to the 16,553 words whose senses all fall in a
single domain, for which assignment is beyond dispute and the primary/all-sense
distinction vanishes. (iv) \emph{Random sense}: each word receives the domain
of one uniformly sampled sense within its dominant part of speech, and the
entire analysis is repeated over 200 independent replicates, characterizing
the full distribution of results attainable by perturbing the assignment of
every polysemous word at once.

\paragraph{Results.}
Supplementary Table~\ref{tab:S-domain-sensitivity} summarizes the outcome
under every scheme. Under the main-text rule, the most abstract third was more
male-associated than the most concrete third in 32 of 37 domains (two-tailed
binomial \textit{P}~$= 7.4\times10^{-6}$; median gap across domains
$= 0.10$~s.d.; significantly positive in 17 domains and significantly negative
in none), closely matching Fig.~2 (33 of 35 domains; median gap
$= 0.10$~s.d.) despite the more lightly filtered vocabulary. The result was
essentially unchanged under
the all-senses rule (35 of 40 domains; \textit{P}~$= 1.4\times10^{-6}$; median
gap $= 0.12$~s.d.), and held among the domain-unambiguous words alone (19 of
22 domains; \textit{P}~$= 8.6\times10^{-4}$; median gap $= 0.12$~s.d.), showing that the
pattern is carried by words whose domain membership involves no analytic
choice at all. Across the 200 random-sense replicates, the share of domains
with a positive gap ranged from 83\% to 97\% (median 92\%), the sign test
rejected the null in every replicate (largest \textit{P}~$= 7.0\times10^{-5}$),
and the median gap ranged from 0.10 to 0.15~s.d. No domain showed a
significant reversal under any of the three deterministic schemes; across the
random replicates, 195 of 200 contained no nominally significant reversal and
none contained more than one.

Finally, we verified that the word-level example panels of Supplementary
Fig.~\ref{fig:S-domain-examples} are not a product of their inclusive
mapping. Recomputing the nine panel correlations with membership restricted to
each word's primary sense (on the common single-word vocabulary; the panel
sample sizes in Supplementary Table~\ref{tab:S-domain-examples} additionally
include two-word artifact entries), every correlation remained negative,
ranging from $r = -0.05$ (\code{text-embedding-3-large}, artifact nouns;
$n = 2{,}266$) to $r = -0.24$ (\code{Qwen3-Embedding-0.6B}, acts of
communication; $n = 630$), all two-tailed \textit{P}~$\leq 0.014$. If
anything, the primary-sense correlations were stronger---most notably for acts
of communication ($r = -0.13$, $-0.09$ and $-0.18$ under the all-senses rule
versus $-0.24$, $-0.13$ and $-0.23$ under the primary-sense rule for
\code{Qwen3-Embedding-0.6B}, \code{text-embedding-3-large} and
\code{nomic-embed-text-v2-moe}, respectively)---indicating that the inclusive
mapping used for illustration attenuates rather than inflates the within-domain
associations.

\begin{table}[!htbp]
  \centering
  \caption{\textbf{The within-domain male abstraction bias is robust to the
  WordNet domain-assignment rule.} Each row recomputes the Fig.~2 quantity---%
  within each domain, the mean $z$-scored gender association of the most
  abstract third of words minus that of the most concrete third, averaged with
  equal weight across the 12 language models of Fig.~2---under a different rule for
  assigning words to domains, on a fixed vocabulary of 25,597 single words
  (see text). Domains, WordNet domains with at least 100 assigned words
  (\textit{noun.person} excluded); Abstract $>$ concrete, domains with a
  positive gap, with the two-tailed binomial sign-test \textit{P}; Sig.\
  $+$/$-$, domains whose gap is significantly positive/negative at
  \textit{P}~$< 0.05$ (two-tailed one-sample $t$-tests across the 12
  models). The random-sense row reports the range over 200 replicates.}
  \label{tab:S-domain-sensitivity}
  \vspace{4pt}
  \resizebox{\textwidth}{!}{%
  \begin{tabular}{lrrlrrr}
  \toprule
  Assignment scheme & Words assigned & Domains & Abstract $>$ concrete (sign-test \textit{P}) & Median gap (s.d.) & Sig.\ $+$ & Sig.\ $-$ \\
  \midrule
  Primary sense (main-text rule) & 22,831 & 37 & 32 ($7.4\times10^{-6}$) & 0.10 & 17 & 0 \\
  All senses (Supplementary Fig.~\ref{fig:S-domain-examples} rule) & 39,430 & 40 & 35 ($1.4\times10^{-6}$) & 0.12 & 20 & 0 \\
  Domain-unambiguous words only & 13,481 & 22 & 19 ($8.6\times10^{-4}$) & 0.12 & 8 & 0 \\
  Random sense (200 replicates; min--max) & --- & 36--38 & 30--37 ($\leq 7.0\times10^{-5}$) & 0.10--0.15 & 13--23 & 0--1 \\
  \bottomrule
  \end{tabular}}
\end{table}

\section{Robustness of the image-based results}
\label{si:image-robustness}

\subsection{Robustness to the exclusion of terms without faces and to statistical controls}
\label{si:image-robustness-general}

In Fig.~3 of the main text, the male association of a search term is the fraction of male faces
among all faces detected in its Google Images, and all 39,956 search terms are shown. Of these,
4,043 terms (10.1\%) returned no images containing a detectable face, and these terms take a value
of zero. Terms without faces are not distributed evenly: more concrete terms, which more often
depict objects than people, were less likely to return any faces ($r = -0.25$ between concreteness
and whether a term returned at least one face; 95\% CI $= -0.25$ to $-0.24$;
\textit{P}~$< 2.2\times10^{-16}$; $n = 39{,}956$; two-tailed Pearson's correlation). Only 2.5\% of
terms in the most abstract fifth of the vocabulary returned no faces, compared with 23.8\% of terms
in the most concrete fifth. The gradient in Fig.~3 therefore combines two patterns: more abstract
concepts are more likely to be depicted with people at all, and, when people are depicted, they are
more likely to be men. Here we isolate the second pattern and show that it is robust.

Supplementary Table~\ref{tab:S-image-agg} summarizes the association in the form in which Fig.~3
presents it, under three specifications: the correlation and slope of the fraction of male faces
on each measure, and the mean fraction of male faces in the most abstract and the most concrete
twentieth of terms, that is, the leftmost and rightmost points of a figure drawn as Fig.~3. The
first column is the association shown in Fig.~3; the second examines the terms whose images
contained at least one face; and the third adjusts the fraction of male faces of those terms for
the number of faces, that is, the number of a term's images in which a face was detected.
Supplementary Table~\ref{tab:S-image} then reports a series of regressions that predict the fraction of
male faces associated with each term from either human judgements of its concreteness (1--5) or its
perceptual strength (0--5). Specification (1) reproduces the bivariate association shown in
Fig.~3. Specification (2) examines only the terms whose images contained at least one face
($n = 35{,}913$ for concreteness; $n = 35{,}677$ for perceptual strength). Among these terms, a
one-point increase in concreteness was associated with a decrease of 1.3 percentage points in the
share of male faces ($r = -0.07$; 95\% CI $= -0.08$ to $-0.06$), and the share of male faces
declined from 66.9\% in the most abstract twentieth of terms to 62.1\% in the most concrete
twentieth; for perceptual strength, the corresponding decline was from 65.3\% to 62.3\%
($r = -0.05$; 95\% CI $= -0.06$ to $-0.04$). Men were thus the majority of faces at every level of
concreteness (67.3\% of the 1,211,815 faces detected overall), and the association between
abstraction and male faces, although considerably smaller than in specification (1), remained
highly significant. Specifications (3)--(5) show that this association is robust to controlling for
the number of faces associated with each term (log-transformed), the frequency of the term in
everyday language (log-transformed SUBTLEX-US frequency, as distributed with the concreteness
norms\cite{brysbaert2014}), the mean confidence of
the gender classifications for the term and fixed effects for the term's dominant part of speech.
Specification (6) shows that it holds when examining only terms associated with at least ten faces,
for which the fraction of male faces is measured more reliably. Finally, specification (7) models
the counts of male and female faces for each term directly, using a quasi-binomial regression that
weights each term by its number of faces and allows for overdispersion; coefficients for this model
are in log-odds. Across all specifications, and for both measures, more abstract terms were
associated with a significantly higher fraction of male faces. The last column of Supplementary
Table~\ref{tab:S-image-agg} shows the same for the quantities plotted in Fig.~3: after the fraction
of male faces is adjusted for the number of faces, it still declines from 66.5\% to 63.6\% across
the twentieths of concreteness (partial $r = -0.05$; 95\% CI $= -0.06$ to $-0.04$) and from 65.7\%
to 62.8\% across the twentieths of perceptual strength (partial $r = -0.05$; 95\% CI $= -0.06$ to
$-0.04$).

\begin{table}[!htbp]
  \centering
  \caption{\textbf{The association between conceptual abstraction and male faces in Google Images survives the
  exclusion of search terms without faces and adjustment for the number of faces.} Each column is a specification of
  the association between the fraction of male faces in a search term's images (0--1) and human judgements of its
  concreteness (upper block) or perceptual strength (lower block). $r$ is the two-tailed Pearson's correlation (in the
  adjusted specification, the partial correlation controlling for the number of faces); $\beta$ is the unstandardized
  ordinary least squares slope per scale point of the predictor, with heteroskedasticity-robust (HC1) 95\% confidence
  intervals and \textit{P} values; the twentieth means are the mean fraction of male faces in the lowest and highest of
  20 equal-count bins of the predictor, that is, the leftmost and rightmost points of a figure drawn as Fig.~3. $n$ is
  the number of search terms. The terms with at least one face account for 1{,}211{,}815 faces (1{,}206{,}055 among the terms with perceptual-strength ratings). $^{a}$The 4{,}043 terms whose images contained no detectable face take the value zero, as in Fig.~3. $^{b}$Adjusted for the number of faces, that is, the number of a term's images in which a face was detected: $\beta$ is the
  coefficient on the predictor in a regression that includes $\log(1+\text{number of faces})$, and the twentieth means are of
  the fraction of male faces net of its linear relation to $\log(1+\text{number of faces})$, re-centred on the grand mean.}
  \label{tab:S-image-agg}
  \begingroup\footnotesize\setlength{\tabcolsep}{6pt}
  \begin{tabular}{@{}lccc@{}}
    \toprule
     & \shortstack[c]{All terms\\(as in Fig.~3)$^{a}$} & \shortstack[c]{Terms with\\$\geq$1 face} & \shortstack[c]{Terms with $\geq$1 face,\\adjusted for faces$^{b}$} \\
    \midrule
    \multicolumn{4}{@{}l}{\textit{Human judgements of concreteness (1--5; Fig.~3b)}} \\
    Pearson's $r$ & $-0.225$ & $-0.070$ & $-0.050$ \\
    \quad 95\% CI & [$-0.234$, $-0.215$] & [$-0.081$, $-0.060$] & [$-0.060$, $-0.040$] \\
    OLS slope $\beta$ (per scale point) & $-0.0577$ & $-0.0133$ & $-0.0095$ \\
    \quad 95\% CI (HC1) & [$-0.0603$, $-0.0550$] & [$-0.0154$, $-0.0111$] & [$-0.0115$, $-0.0075$] \\
    \quad \textit{P} & $<2.2\times10^{-16}$ & $<2.2\times10^{-16}$ & $<2.2\times10^{-16}$ \\
    Male fraction, most abstract twentieth & 0.649 & 0.669 & 0.665 \\
    Male fraction, most concrete twentieth & 0.413 & 0.621 & 0.636 \\
    $n$ search terms & 39{,}956 & 35{,}913 & 35{,}913 \\
    \addlinespace[3pt]
    \multicolumn{4}{@{}l}{\textit{Human judgements of perceptual strength (0--5; Fig.~3c)}} \\
    Pearson's $r$ & $-0.084$ & $-0.049$ & $-0.051$ \\
    \quad 95\% CI & [$-0.094$, $-0.074$] & [$-0.059$, $-0.038$] & [$-0.061$, $-0.040$] \\
    OLS slope $\beta$ (per scale point) & $-0.0595$ & $-0.0247$ & $-0.0255$ \\
    \quad 95\% CI (HC1) & [$-0.0668$, $-0.0522$] & [$-0.0304$, $-0.0191$] & [$-0.0311$, $-0.0199$] \\
    \quad \textit{P} & $<2.2\times10^{-16}$ & $<2.2\times10^{-16}$ & $<2.2\times10^{-16}$ \\
    Male fraction, least perceptual twentieth & 0.607 & 0.653 & 0.657 \\
    Male fraction, most perceptual twentieth & 0.524 & 0.623 & 0.628 \\
    $n$ search terms & 39{,}707 & 35{,}677 & 35{,}677 \\
    \bottomrule
  \end{tabular}\endgroup
\end{table}

\begin{table}[!htbp]
  \centering
  \caption{\textbf{The association between conceptual abstraction and male faces in Google Images
  is robust to the treatment of search terms without faces and to statistical controls.} Each row
  is a separate regression predicting the male association of a search term from human judgements
  of its concreteness (1--5; left) or its perceptual strength (Lancaster Sensorimotor Norms; 0--5;
  right); $\beta$ is the unstandardized coefficient for that predictor. Specifications (1)--(6) are
  ordinary least squares regressions of the fraction of male faces (0--1) with
  heteroskedasticity-robust (HC1) standard errors, for which the test statistic is $t$.
  Specification (7) is a quasi-binomial regression of the proportion of male faces weighted by the
  number of faces, for which $\beta$ is in log-odds and the test statistic is $z$. All \textit{P}
  values are two-tailed; $n$ is the number of search terms.}
  \label{tab:S-image}
  \begingroup\scriptsize\setlength{\tabcolsep}{2.5pt}
  \begin{tabular}{@{}p{3.7cm}rrrcrrrrcr@{}}
    \toprule
    & \multicolumn{5}{c}{Concreteness} & \multicolumn{5}{c}{Perceptual strength} \\
    \cmidrule(lr){2-6}\cmidrule(lr){7-11}
    Specification & $\beta$ & s.e. & $t$/$z$ & \textit{P} & $n$ & $\beta$ & s.e. & $t$/$z$ & \textit{P} & $n$ \\
    \midrule
    (1) All terms; terms without faces set to zero (as in Fig.~3); no controls & $-0.0577$ & $0.0014$ & $-42.27$ & $<2.2\times10^{-16}$ & 39{,}956 & $-0.0595$ & $0.0037$ & $-16.05$ & $<2.2\times10^{-16}$ & 39{,}707 \\[2pt]
    (2) Terms with $\geq$1 face; no controls & $-0.0133$ & $0.0011$ & $-12.28$ & $<2.2\times10^{-16}$ & 35{,}913 & $-0.0247$ & $0.0029$ & $-8.59$ & $<2.2\times10^{-16}$ & 35{,}677 \\[2pt]
    (3) As (2), controlling for the number of faces (log) & $-0.0095$ & $0.0010$ & $-9.12$ & $<2.2\times10^{-16}$ & 35{,}913 & $-0.0255$ & $0.0029$ & $-8.91$ & $<2.2\times10^{-16}$ & 35{,}677 \\[2pt]
    (4) As (3), plus word frequency (log) and classifier confidence & $-0.0075$ & $0.0010$ & $-7.18$ & $6.8\times10^{-13}$ & 35{,}913 & $-0.0208$ & $0.0028$ & $-7.32$ & $2.4\times10^{-13}$ & 35{,}677 \\[2pt]
    (5) As (4), plus part-of-speech fixed effects & $-0.0104$ & $0.0012$ & $-8.99$ & $<2.2\times10^{-16}$ & 35{,}913 & $-0.0214$ & $0.0029$ & $-7.46$ & $8.5\times10^{-14}$ & 35{,}677 \\[2pt]
    (6) As (5), terms with $\geq$10 faces only & $-0.0077$ & $0.0010$ & $-7.94$ & $2.0\times10^{-15}$ & 26{,}826 & $-0.0206$ & $0.0023$ & $-8.97$ & $<2.2\times10^{-16}$ & 26{,}681 \\[2pt]
    (7) Quasi-binomial model of male versus female face counts; controls as in (5) & $-0.0261$ & $0.0037$ & $-7.12$ & $1.1\times10^{-12}$ & 35{,}913 & $-0.1054$ & $0.0089$ & $-11.88$ & $<2.2\times10^{-16}$ & 35{,}677 \\[2pt]
    \bottomrule
  \end{tabular}\endgroup
\end{table}

\subsection{Robustness to the number of detected faces}
\label{si:image-face-controls}

The main-text analysis of Google Images measures the gender association of each search term as the fraction of the faces detected in its images that were classified as male. The image set comprises approximately 12 million images retrieved for 39,956 search terms, in which 1,211,815 faces were detected (67.3\% classified as male). Not every search term returns faces: 4,043 terms (10.1\%) returned none, and these terms enter the main-text analysis with a male fraction of zero. Terms without faces are not distributed evenly. Sorting terms into 20 equal-sized bins, the share of terms without a detected face rises from 3.0\% in the most abstract bin to 33.9\% in the most concrete bin, and from 7.1\% to 15.8\% across bins of perceptual strength; each standard deviation of concreteness multiplies the odds that a term returns no face by 2.29 (95\% CI 2.21--2.38), and each standard deviation of perceptual strength by 1.25 (1.21--1.28; Supplementary Table~\ref{tab:S-image-face-controls}, row E1). The association shown in the main text therefore combines two components: more concrete terms are less likely to return images of people at all, and, when people are shown, a smaller fraction of them are male. This section tests whether the second component holds on its own, that is, whether the negative association between concreteness and male representation is robust to how terms without faces are treated and to variation in the number of faces per term.

We estimated five families of models, each separately for human judgments of concreteness (Brysbaert norms; $N = 39{,}956$ terms) and for perceptual strength (Lancaster sensorimotor norms, mean of the six perceptual modalities; $N = 39{,}707$ terms). (A) Retaining all search terms, as in the main text, we added as covariates the number of detected faces, $\log(1 + \text{faces})$, an indicator for whether the term returned any face, and word frequency ($\ln(1 + \text{SUBTLEX count})$). (B) We restricted the sample to terms with at least one detected face, for which the male fraction is defined, and then to terms with at least 5, 10 and 20 faces, for which it is measured with increasing precision. (C) Among terms with at least one face, we used three alternative estimators: ordinary least squares weighting each term by its number of faces; a quasi-binomial logistic regression of the counts of male and female faces, which treats the face rather than the search term as the unit of observation and estimates the dispersion from the data; and the Spearman rank correlation. (D) Retaining all terms, we replaced the outcome with the difference between the male and female fractions of faces, which equals zero for terms without faces and so treats them as neutral rather than as 0\% male. (E) We modelled the probability that a term returns no face. In all least-squares models the outcome and all continuous predictors were $z$-scored within the estimation sample, so coefficients are standardized ($\beta$); standard errors are heteroskedasticity-robust (HC1); all tests are two-sided.

Supplementary Table~\ref{tab:S-image-face-controls} reports all estimates. The association was negative in every specification, for both measures, and every 95\% confidence interval excluded the null value (zero for $\beta$ and $\rho$, one for odds ratios). Its magnitude, however, depended substantially on the treatment of terms without faces.

For concreteness, the main-text association was $\beta = -0.225$ (95\% CI $-0.235$ to $-0.214$). A linear control for the number of faces changed this little ($\beta = -0.204$), because a linear term does not absorb the discontinuity at zero faces; controlling for $\log(1 + \text{faces})$ reduced it to $\beta = -0.074$, an indicator for any face to $\beta = -0.047$, and the indicator together with $\log(1 + \text{faces})$ and word frequency to $\beta = -0.031$ ($-0.039$ to $-0.024$). Restricting the sample to terms with at least one face gave $\beta = -0.070$ ($-0.082$ to $-0.059$), and $\beta = -0.048$ ($-0.059$ to $-0.037$) with controls for $\log(1 + \text{faces})$ and word frequency; estimates remained between $-0.041$ and $-0.056$ when requiring at least 5, 10 or 20 faces. Roughly two-thirds of the main-text association with concreteness is thus attributable to terms that return no faces, and the remainder to a lower male fraction among the faces that are shown: among terms with at least one face, the mean male fraction declined from 0.670 in the most abstract bin to 0.624 in the most concrete bin.

For perceptual strength, the main-text association was $\beta = -0.084$ ($-0.094$ to $-0.074$), and $\beta = -0.049$ ($-0.060$ to $-0.038$) among terms with at least one face ($\beta = -0.048$ with controls). In contrast to concreteness, the association strengthened as the minimum number of faces increased, to $\beta = -0.062$, $-0.077$ and $-0.097$ ($-0.111$ to $-0.082$) for at least 5, 10 and 20 faces.

The estimators that weight terms by their number of faces yielded the smallest estimates for concreteness: $\beta = -0.015$ ($-0.028$ to $-0.002$; $P = 0.020$) in the face-weighted regression, and an odds ratio of 0.987 per standard deviation (0.980 to 0.994; $P = 1.8 \times 10^{-4}$) in the face-level quasi-binomial model, indicating that the association with concreteness is weaker among terms that return many faces. The corresponding estimates for perceptual strength were $\beta = -0.065$ ($-0.077$ to $-0.053$) and an odds ratio of 0.946 (0.940 to 0.953). Rank correlations among terms with faces were $\rho = -0.047$ for concreteness and $\rho = -0.048$ for perceptual strength. Treating terms without faces as neutral rather than as 0\% male gave $\beta = -0.123$ ($-0.133$ to $-0.113$) for concreteness and $\beta = -0.062$ ($-0.072$ to $-0.052$) for perceptual strength.

Taken together, the direction of the association shown in the main text is robust to the number of detected faces, but its magnitude is an upper bound: conditional on a term returning faces, a one-standard-deviation increase in either concreteness or perceptual strength is associated with a decrease of approximately 0.05 standard deviations in the fraction of faces that are male.

\begin{table}[p]
\centering
\caption{\textbf{Robustness of the association between concreteness and male representation in Google Images to the number of detected faces.} Each row is a separate model, estimated once with human judgments of concreteness (Brysbaert norms) and once with perceptual strength (Lancaster norms) as the predictor. The outcome is the fraction of a search term's detected faces classified as male unless stated otherwise. Estimates are standardized coefficients ($\beta$) with heteroskedasticity-robust (HC1) 95\% confidence intervals, except where rows indicate an odds ratio (OR) per standard deviation of the predictor or a Spearman correlation ($\rho$; Fisher-$z$ interval). In family A, terms with no detected faces have a male fraction of zero, as in the main text. $P$ values are two-sided. Faces, number of faces detected across a term's images; word frequency, $\ln(1 + \text{SUBTLEX count})$.}
\label{tab:S-image-face-controls}
\vspace{4pt}
\resizebox{\textwidth}{!}{%
\begin{tabular}{llrrrrrr}
\toprule
 & & \multicolumn{3}{c}{Concreteness (human judgments)} & \multicolumn{3}{c}{Perceptual strength (Lancaster)} \\
\cmidrule(lr){3-5}\cmidrule(lr){6-8}
 & Specification & Estimate [95\% CI] & $P$ & $N$ terms & Estimate [95\% CI] & $P$ & $N$ terms \\
\midrule
\multicolumn{8}{l}{\textit{A. All search terms; terms without faces coded 0 (standardized $\beta$)}} \\
A1 & No controls (main-text figure) & $-0.225$ [$-0.235$, $-0.214$] & $<10^{-300}$ & 39,956 & $-0.084$ [$-0.094$, $-0.074$] & $9.0\times10^{-58}$ & 39,707 \\
A2 & + number of faces & $-0.204$ [$-0.214$, $-0.194$] & $<10^{-300}$ & 39,956 & $-0.090$ [$-0.099$, $-0.080$] & $1.3\times10^{-69}$ & 39,707 \\
A3 & + log(1 + faces) & $-0.074$ [$-0.082$, $-0.065$] & $1.4\times10^{-62}$ & 39,956 & $-0.061$ [$-0.069$, $-0.052$] & $3.2\times10^{-44}$ & 39,707 \\
A4 & + any-face indicator & $-0.047$ [$-0.054$, $-0.039$] & $1.4\times10^{-34}$ & 39,956 & $-0.031$ [$-0.038$, $-0.024$] & $9.8\times10^{-18}$ & 39,707 \\
A5 & + any-face indicator, log(1 + faces) & $-0.033$ [$-0.040$, $-0.026$] & $8.3\times10^{-20}$ & 39,956 & $-0.032$ [$-0.039$, $-0.025$] & $5.8\times10^{-19}$ & 39,707 \\
A6 & + any-face indicator, log(1 + faces), word frequency & $-0.031$ [$-0.039$, $-0.024$] & $2.9\times10^{-17}$ & 39,956 & $-0.030$ [$-0.037$, $-0.023$] & $1.2\times10^{-16}$ & 39,707 \\
\addlinespace
\multicolumn{8}{l}{\textit{B. Terms with detected faces only (standardized $\beta$)}} \\
B1 & $\geq 1$ face; no controls & $-0.070$ [$-0.082$, $-0.059$] & $1.3\times10^{-34}$ & 35,913 & $-0.049$ [$-0.060$, $-0.038$] & $9.2\times10^{-18}$ & 35,677 \\
B2 & $\geq 1$ face; + log(1 + faces) & $-0.050$ [$-0.061$, $-0.040$] & $8.2\times10^{-20}$ & 35,913 & $-0.050$ [$-0.061$, $-0.039$] & $5.5\times10^{-19}$ & 35,677 \\
B3 & $\geq 1$ face; + log(1 + faces), word frequency & $-0.048$ [$-0.059$, $-0.037$] & $5.4\times10^{-17}$ & 35,913 & $-0.048$ [$-0.059$, $-0.036$] & $1.8\times10^{-16}$ & 35,677 \\
B4 & $\geq 5$ faces; + log(1 + faces), word frequency & $-0.056$ [$-0.068$, $-0.044$] & $2.2\times10^{-20}$ & 31,760 & $-0.062$ [$-0.073$, $-0.050$] & $7.5\times10^{-25}$ & 31,561 \\
B5 & $\geq 10$ faces; + log(1 + faces), word frequency & $-0.053$ [$-0.066$, $-0.040$] & $1.1\times10^{-15}$ & 26,826 & $-0.077$ [$-0.090$, $-0.065$] & $4.2\times10^{-33}$ & 26,681 \\
B6 & $\geq 20$ faces; + log(1 + faces), word frequency & $-0.041$ [$-0.056$, $-0.025$] & $4.6\times10^{-7}$ & 19,122 & $-0.097$ [$-0.111$, $-0.082$] & $2.3\times10^{-37}$ & 19,034 \\
\addlinespace
\multicolumn{8}{l}{\textit{C. Alternative estimators; terms with $\geq 1$ face}} \\
C1 & OLS weighted by number of faces ($\beta$) & $-0.015$ [$-0.028$, $-0.002$] & 0.020 & 35,913 & $-0.065$ [$-0.077$, $-0.053$] & $1.9\times10^{-27}$ & 35,677 \\
C2 & Face-level quasi-binomial logit (OR per s.d.) & $0.987$ [$0.980$, $0.994$] & $1.8\times10^{-4}$ & 35,913 & $0.946$ [$0.940$, $0.953$] & $2.1\times10^{-56}$ & 35,677 \\
C3 & Spearman rank correlation ($\rho$) & $-0.047$ [$-0.058$, $-0.037$] & $3.1\times10^{-19}$ & 35,913 & $-0.048$ [$-0.058$, $-0.038$] & $1.4\times10^{-19}$ & 35,677 \\
\addlinespace
\multicolumn{8}{l}{\textit{D. Outcome in which terms without faces are neutral}} \\
D1 & Male minus female fraction, all terms ($\beta$) & $-0.123$ [$-0.133$, $-0.113$] & $4.7\times10^{-126}$ & 39,956 & $-0.062$ [$-0.072$, $-0.052$] & $5.5\times10^{-32}$ & 39,707 \\
\addlinespace
\multicolumn{8}{l}{\textit{E. Probability that a term returns no face}} \\
E1 & No face detected, all terms (logit OR per s.d.) & $2.293$ [$2.214$, $2.376$] & $<10^{-300}$ & 39,956 & $1.246$ [$1.209$, $1.284$] & $1.2\times10^{-46}$ & 39,707 \\
\bottomrule
\end{tabular}}
\end{table}

\section{Robustness of the unique-object results to the number of images per term}
\label{si:many-robustness}

In Fig.~4 of the main text, the visual diversity of a search term is the number of unique objects
detected across its Google Images. Up to 300 images were retrieved per term, but the number
available varied: among the 36,124 terms analysed, the median was 265 images (mean $= 256$;
s.d.\ $= 39$; range $= 3$ to $300$), and 1,583 terms (4.4\%) had fewer than 200. Because each
additional image can only add to the set of distinct objects, the number of unique objects rises
with the number of images ($r = 0.47$ with the number of images and $0.46$ with its logarithm), so
an association between a predictor and visual diversity could in principle arise if the predictor
were correlated with the number of images retrieved. In practice the number of images was only
weakly related to the three predictors ($r = 0.05$ with concreteness, $0.05$ with the fraction of
male faces and $-0.02$ with the text-based gender association), but the terms with the most images
(280 or more; 14\% of terms) were disproportionately concrete (mean concreteness $= 3.59$, compared
with $2.86$ for the remaining terms) and yet had fewer unique objects (755 compared with 783), so
the bivariate association in Fig.~4a slightly understates the association at a fixed number of
images. For this reason, Fig.~4 of the main text controls for the number of images, and this
section reports how each of its three panels depends on that choice.

For each panel, we estimated an ordinary least squares regression of the number of unique objects
on the predictor together with a control for the number of images; the main text and Fig.~4 use
the log-transformed number of images (specification (2) of Supplementary Table~\ref{tab:S-many}).
The partial correlation reported is the Pearson correlation between the residuals of the outcome
and of the predictor after each has been regressed on the control, with a 95\% confidence interval
from the Fisher $z$ transformation; the partial Spearman's $\rho$ is the rank correlation of the
same residuals; and the slope is the regression coefficient for the predictor, in objects per unit
of the predictor, with its conventional standard error and $t$ statistic. In Fig.~4, each term's
number of unique objects is displayed adjusted to the mean number of images, that is, its observed
count minus the product of the control's coefficient and the term's deviation from the mean log
number of images, and the fitted line is the regression's prediction at the mean number of images.
Specification (1) is the bivariate association without the control. Specifications (3)--(5) vary the functional form of the control: the
untransformed number of images, a quadratic in the number of images and fixed effects for deciles
of the number of images, which make no assumption about functional form. Specification (6) adds
the log-transformed total number of object tokens detected for the term; this is a stricter
control, because the number of objects detected per image is itself part of what the diversity
measure captures. Specifications (7) and (8) restrict the analysis to the 34,541 terms with at
least 200 images (34,149 for panel (b)), without and with the log control.

Supplementary Table~\ref{tab:S-many} shows that all three associations are robust. The association
between concreteness and visual diversity strengthened under every control, from $r = -0.38$
without the control to between $-0.39$ (decile fixed effects) and $-0.48$ (log images and object
tokens), with the slope changing from 83.5 to between 55.2 and 90.0 additional unique objects per
point of concreteness. The association with the fraction of male faces was essentially unchanged by
the number of images ($r = 0.11$ without the control; $0.10$ under specifications (2)--(5)) and
remained positive and highly significant under the stricter object-token control ($r = 0.06$).
The association with the text-based gender association rose slightly under the image controls
($r = 0.07$ without the control; $0.09$ under specifications (2)--(4) and $0.07$ with decile fixed
effects) and was larger still under the object-token control ($r = 0.15$). Restricting the analysis
to terms with at least 200 images gave the same picture. Across all 24 estimates, every association
retained its sign, and every \textit{P} value was below $2.2\times10^{-16}$.

\begin{table}[!htbp]
  \centering
  \caption{\textbf{The associations between visual diversity and concreteness, image gender and
  text gender are robust to the number of images per term.} Each row is a separate ordinary least
  squares regression predicting the number of unique objects detected across a search term's Google
  Images from one predictor: (a) human judgements of the term's concreteness, (b) the fraction of
  male faces among the term's images (terms with at least one face) or (c) the term's text-based
  gender association (mean $z$ across the 12 language models). $r$ is the partial Pearson's
  correlation between the predictor and the number of unique objects after removing the controls
  listed (the bivariate correlation in specifications (1) and (7)), with a 95\% confidence
  interval; $\rho$ is the corresponding Spearman's rank correlation; slope is the unstandardized
  coefficient for the predictor, in objects per unit of the predictor (one point of concreteness,
  the full range of the fraction of male faces, or one unit of the mean $z$-score), with its
  standard error and $t$ statistic. All \textit{P} values are two-tailed; $n$ is the number of
  search terms. Specification (2) is the one reported in Fig.~4 of the main text.}
  \label{tab:S-many}
  \begingroup\scriptsize\setlength{\tabcolsep}{3pt}
  \begin{tabular}{@{}p{5.3cm}rlrrrrlr@{}}
    \toprule
    Specification & $r$ & 95\% CI & $\rho$ & slope & s.e. & $t$ & \textit{P} & $n$ \\
    \midrule
    \multicolumn{8}{@{}l}{\textit{(a) Concreteness (1--5)}} \\[1pt]
    (1) No controls & $-0.381$ & $-0.390$ to $-0.372$ & $-0.376$ & $-83.5$ & $1.1$ & $-78.3$ & $<2.2\times10^{-16}$ & 36{,}124 \\
    (2) Controlling for the number of images (log); as in Fig.~4 & $-0.440$ & $-0.448$ to $-0.432$ & $-0.412$ & $-85.8$ & $0.9$ & $-93.1$ & $<2.2\times10^{-16}$ & 36{,}124 \\
    (3) Controlling for the number of images (linear) & $-0.457$ & $-0.465$ to $-0.449$ & $-0.433$ & $-88.6$ & $0.9$ & $-97.6$ & $<2.2\times10^{-16}$ & 36{,}124 \\
    (4) Controlling for the number of images (linear and quadratic) & $-0.452$ & $-0.460$ to $-0.444$ & $-0.425$ & $-88.1$ & $0.9$ & $-96.3$ & $<2.2\times10^{-16}$ & 36{,}124 \\
    (5) Fixed effects for deciles of the number of images & $-0.391$ & $-0.400$ to $-0.382$ & $-0.364$ & $-79.3$ & $1.0$ & $-80.8$ & $<2.2\times10^{-16}$ & 36{,}124 \\
    (6) As (2), plus the number of object tokens (log) & $-0.479$ & $-0.487$ to $-0.471$ & $-0.473$ & $-55.2$ & $0.5$ & $-103.8$ & $<2.2\times10^{-16}$ & 36{,}124 \\
    (7) As (1), terms with $\geq$200 images only & $-0.426$ & $-0.434$ to $-0.417$ & $-0.396$ & $-84.7$ & $1.0$ & $-87.4$ & $<2.2\times10^{-16}$ & 34{,}541 \\
    (8) As (2), terms with $\geq$200 images only & $-0.456$ & $-0.464$ to $-0.447$ & $-0.428$ & $-90.0$ & $0.9$ & $-95.1$ & $<2.2\times10^{-16}$ & 34{,}541 \\
    \addlinespace[3pt]
    \multicolumn{8}{@{}l}{\textit{(b) Fraction of male faces (0--1; terms with $\geq$1 face)}} \\[1pt]
    (1) No controls & $+0.113$ & $+0.103$ to $+0.123$ & $+0.080$ & $+131.8$ & $6.1$ & $+21.5$ & $<2.2\times10^{-16}$ & 35{,}603 \\
    (2) Controlling for the number of images (log); as in Fig.~4 & $+0.102$ & $+0.092$ to $+0.113$ & $+0.075$ & $+107.4$ & $5.5$ & $+19.4$ & $<2.2\times10^{-16}$ & 35{,}603 \\
    (3) Controlling for the number of images (linear) & $+0.099$ & $+0.089$ to $+0.110$ & $+0.072$ & $+103.7$ & $5.5$ & $+18.8$ & $<2.2\times10^{-16}$ & 35{,}603 \\
    (4) Controlling for the number of images (linear and quadratic) & $+0.101$ & $+0.091$ to $+0.111$ & $+0.074$ & $+105.1$ & $5.5$ & $+19.1$ & $<2.2\times10^{-16}$ & 35{,}603 \\
    (5) Fixed effects for deciles of the number of images & $+0.095$ & $+0.085$ to $+0.106$ & $+0.070$ & $+99.5$ & $5.5$ & $+18.1$ & $<2.2\times10^{-16}$ & 35{,}603 \\
    (6) As (2), plus the number of object tokens (log) & $+0.057$ & $+0.047$ to $+0.067$ & $+0.031$ & $+34.5$ & $3.2$ & $+10.8$ & $<2.2\times10^{-16}$ & 35{,}603 \\
    (7) As (1), terms with $\geq$200 images only & $+0.108$ & $+0.097$ to $+0.118$ & $+0.078$ & $+117.8$ & $5.9$ & $+20.0$ & $<2.2\times10^{-16}$ & 34{,}149 \\
    (8) As (2), terms with $\geq$200 images only & $+0.101$ & $+0.091$ to $+0.112$ & $+0.074$ & $+109.4$ & $5.8$ & $+18.8$ & $<2.2\times10^{-16}$ & 34{,}149 \\
    \addlinespace[3pt]
    \multicolumn{8}{@{}l}{\textit{(c) Text-based gender association (mean $z$ across 12 models)}} \\[1pt]
    (1) No controls & $+0.074$ & $+0.063$ to $+0.084$ & $+0.099$ & $+30.1$ & $2.1$ & $+14.0$ & $<2.2\times10^{-16}$ & 36{,}124 \\
    (2) Controlling for the number of images (log); as in Fig.~4 & $+0.089$ & $+0.079$ to $+0.099$ & $+0.105$ & $+32.4$ & $1.9$ & $+16.9$ & $<2.2\times10^{-16}$ & 36{,}124 \\
    (3) Controlling for the number of images (linear) & $+0.091$ & $+0.081$ to $+0.102$ & $+0.107$ & $+33.1$ & $1.9$ & $+17.4$ & $<2.2\times10^{-16}$ & 36{,}124 \\
    (4) Controlling for the number of images (linear and quadratic) & $+0.090$ & $+0.080$ to $+0.101$ & $+0.106$ & $+32.6$ & $1.9$ & $+17.2$ & $<2.2\times10^{-16}$ & 36{,}124 \\
    (5) Fixed effects for deciles of the number of images & $+0.068$ & $+0.058$ to $+0.078$ & $+0.075$ & $+24.8$ & $1.9$ & $+13.0$ & $<2.2\times10^{-16}$ & 36{,}124 \\
    (6) As (2), plus the number of object tokens (log) & $+0.149$ & $+0.139$ to $+0.159$ & $+0.154$ & $+31.4$ & $1.1$ & $+28.7$ & $<2.2\times10^{-16}$ & 36{,}124 \\
    (7) As (1), terms with $\geq$200 images only & $+0.088$ & $+0.077$ to $+0.098$ & $+0.106$ & $+32.4$ & $2.0$ & $+16.4$ & $<2.2\times10^{-16}$ & 34{,}541 \\
    (8) As (2), terms with $\geq$200 images only & $+0.092$ & $+0.081$ to $+0.102$ & $+0.106$ & $+33.3$ & $2.0$ & $+17.1$ & $<2.2\times10^{-16}$ & 34{,}541 \\
    \bottomrule
  \end{tabular}\endgroup
\end{table}

%% file: stats_si_text.tex
\newcommand{\TabTextCovWords}{%
\begingroup\scriptsize\setlength{\tabcolsep}{3pt}%
\begin{tabular}{@{}p{6.4cm}cccccr@{}}
\toprule
Specification & Mean $\beta$ & 95\% CI & Range & Negative & Sig. & $n$ per model \\
\midrule
\multicolumn{7}{l}{\textit{A. Lexical covariates}} \\
No controls (as in Fig.~1a; Supplementary Table~\ref{tab:S-fig1-human}) & $-0.085$ & [$-0.131$, $-0.039$] & $-0.209$ to $0.037$ & 11/12 & 11 & 31,609--39,953 \\
+ word frequency & $-0.091$ & [$-0.143$, $-0.039$] & $-0.219$ to $0.035$ & 10/12 & 9 & 31,609--39,953 \\
+ length (characters and syllables) & $-0.117$ & [$-0.169$, $-0.064$] & $-0.238$ to $0.030$ & 10/12 & 10 & 31,609--39,953 \\
+ polysemy (number of WordNet senses) & $-0.086$ & [$-0.132$, $-0.039$] & $-0.209$ to $0.042$ & 11/12 & 9 & 28,958--34,026 \\
\quad no controls, same words & $-0.083$ & [$-0.127$, $-0.038$] & $-0.204$ to $0.043$ & 11/12 & 10 & 28,958--34,026 \\
+ age of acquisition\cite{kuperman2012} & $-0.072$ & [$-0.127$, $-0.017$] & $-0.194$ to $0.064$ & 8/12 & 8 & 23,022--23,860 \\
\quad no controls, same words & $-0.065$ & [$-0.110$, $-0.020$] & $-0.188$ to $0.038$ & 10/12 & 10 & 23,022--23,860 \\
+ valence, arousal and dominance\cite{warriner2013} & $-0.052$ & [$-0.094$, $-0.010$] & $-0.129$ to $0.050$ & 9/12 & 8 & 13,312--13,385 \\
\quad no controls, same words & $-0.063$ & [$-0.108$, $-0.017$] & $-0.157$ to $0.053$ & 9/12 & 9 & 13,312--13,385 \\
+ part of speech (fixed effects) & $-0.068$ & [$-0.115$, $-0.021$] & $-0.207$ to $0.024$ & 9/12 & 8 & 30,375--35,760 \\
+ frequency, length, polysemy and part of speech & $-0.101$ & [$-0.149$, $-0.053$] & $-0.229$ to $0.005$ & 11/12 & 10 & 28,958--34,026 \\
+ all of the above (frequency, length, polysemy, age of acquisition, valence, arousal, dominance, part of speech) & $-0.099$ & [$-0.143$, $-0.054$] & $-0.220$ to $0.004$ & 11/12 & 10 & 13,069--13,082 \\
\quad no controls, same words & $-0.063$ & [$-0.109$, $-0.017$] & $-0.158$ to $0.053$ & 9/12 & 9 & 13,069--13,082 \\
\addlinespace
\multicolumn{7}{l}{\textit{B. Reliably rated words only}} \\
Known by $\geq$95\% of raters & $-0.089$ & [$-0.133$, $-0.044$] & $-0.209$ to $0.039$ & 11/12 & 11 & 24,261--29,066 \\
Rater s.d.\ of concreteness $\leq$ median & $-0.115$ & [$-0.174$, $-0.055$] & $-0.263$ to $0.056$ & 11/12 & 11 & 15,471--20,185 \\
Both restrictions & $-0.114$ & [$-0.170$, $-0.058$] & $-0.253$ to $0.062$ & 11/12 & 10 & 12,298--15,145 \\
\addlinespace
\multicolumn{7}{l}{\textit{C. Within quintiles of word frequency}} \\
Quintile 1 (least frequent) & $-0.127$ & [$-0.182$, $-0.073$] & $-0.261$ to $0.043$ & 11/12 & 10 & 2,558--7,991 \\
Quintile 2 & $-0.096$ & [$-0.160$, $-0.032$] & $-0.271$ to $0.061$ & 10/12 & 9 & 5,216--7,990 \\
Quintile 3 & $-0.071$ & [$-0.129$, $-0.013$] & $-0.236$ to $0.048$ & 9/12 & 7 & 7,874--7,991 \\
Quintile 4 & $-0.065$ & [$-0.122$, $-0.007$] & $-0.183$ to $0.057$ & 8/12 & 8 & 7,972--7,990 \\
Quintile 5 (most frequent) & $-0.089$ & [$-0.141$, $-0.037$] & $-0.179$ to $0.044$ & 9/12 & 8 & 7,972--7,991 \\
\bottomrule
\end{tabular}\endgroup}

\newcommand{\TabTextCovPhrases}{%
\begingroup\scriptsize\setlength{\tabcolsep}{3pt}%
\begin{tabular}{@{}p{6.4cm}cccccr@{}}
\toprule
Specification & Mean $\beta$ & 95\% CI & Range & Negative & Sig. & $n$ per model \\
\midrule
No controls (as in Fig.~1a; Supplementary Table~\ref{tab:S-fig1-human}) & $-0.252$ & [$-0.295$, $-0.210$] & $-0.324$ to $-0.108$ & 12/12 & 12 & 51,266--62,889 \\
+ word frequency (mean of constituent words) & $-0.146$ & [$-0.185$, $-0.108$] & $-0.223$ to $-0.026$ & 12/12 & 12 & 50,839--61,308 \\
+ length (characters and number of words) & $-0.181$ & [$-0.223$, $-0.139$] & $-0.251$ to $-0.022$ & 12/12 & 12 & 51,266--62,889 \\
+ polysemy (WordNet senses of the phrase; indicator for a WordNet entry) & $-0.234$ & [$-0.276$, $-0.192$] & $-0.306$ to $-0.080$ & 12/12 & 12 & 51,266--62,889 \\
+ age of acquisition (mean of constituent words)\cite{kuperman2012} & $-0.231$ & [$-0.271$, $-0.190$] & $-0.301$ to $-0.093$ & 12/12 & 12 & 50,713--61,157 \\
\quad no controls, same phrases & $-0.255$ & [$-0.299$, $-0.212$] & $-0.335$ to $-0.109$ & 12/12 & 12 & 50,713--61,157 \\
+ valence, arousal and dominance (mean of constituent words)\cite{warriner2013} & $-0.249$ & [$-0.292$, $-0.206$] & $-0.331$ to $-0.106$ & 12/12 & 12 & 48,291--57,903 \\
\quad no controls, same phrases & $-0.247$ & [$-0.290$, $-0.205$] & $-0.329$ to $-0.102$ & 12/12 & 12 & 48,291--57,903 \\
+ part of speech (fixed effects) & $-0.196$ & [$-0.235$, $-0.157$] & $-0.255$ to $-0.050$ & 12/12 & 12 & 50,953--61,903 \\
+ all of the above & $-0.106$ & [$-0.150$, $-0.062$] & $-0.185$ to $0.041$ & 11/12 & 10 & 48,073--57,504 \\
\quad no controls, same phrases & $-0.247$ & [$-0.290$, $-0.204$] & $-0.330$ to $-0.101$ & 12/12 & 12 & 48,073--57,504 \\
\bottomrule
\end{tabular}\endgroup}

\newcommand{\TabTextCovPerModel}{%
\begingroup\scriptsize\setlength{\tabcolsep}{2.5pt}%
\begin{tabular}{@{}lcccccccccc@{}}
\toprule
 & \multicolumn{5}{c}{Words} & \multicolumn{5}{c}{Phrases} \\
\cmidrule(lr){2-6}\cmidrule(lr){7-11}
Model & None & + Freq. & + Length & + VAD & + All & None & + Freq. & + Length & + POS & + All \\
\midrule
FastText & $-0.070$ & $-0.089$ & $-0.140$ & $-0.033$ & $-0.136$ & $-0.317$ & $-0.153$ & $-0.223$ & $-0.190$ & $-0.107$ \\
word2vec & $-0.115$ & $-0.158$ & $-0.180$ & $-0.129$ & $-0.220$ & $-0.322$ & $-0.178$ & $-0.251$ & $-0.245$ & $-0.157$ \\
Wikipedia GloVe & $-0.076$ & $-0.137$ & $-0.149$ & $-0.102$ & $-0.156$ & $-0.324$ & $-0.150$ & $-0.238$ & $-0.252$ & $-0.106$ \\
ConceptNet & $0.037$ & $0.035$ & $0.030$ & $0.050$ & $-0.032$ & $-0.189$ & $-0.026$ & $-0.101$ & $-0.106$ & $-0.011$$^{\dagger}$ \\
BERT & $-0.113$ & $-0.033$ & $-0.060$ & $0.025$ & $-0.037$ & $-0.267$ & $-0.120$ & $-0.195$ & $-0.198$ & $-0.067$ \\
distilBERT & $-0.042$ & $-0.104$ & $-0.159$ & $-0.067$ & $-0.101$ & $-0.278$ & $-0.184$ & $-0.206$ & $-0.226$ & $-0.149$ \\
RoBERTa & $-0.026$ & $0.019$ & $0.021$ & $-0.020$$^{\dagger}$ & $0.004$$^{\dagger}$ & $-0.108$ & $-0.042$ & $-0.022$ & $-0.050$ & $0.041$ \\
MPNet & $-0.057$ & $-0.064$ & $-0.143$ & $-0.049$ & $-0.140$ & $-0.235$ & $-0.187$ & $-0.186$ & $-0.199$ & $-0.185$ \\
GPT4 & $-0.150$ & $-0.144$ & $-0.161$ & $-0.095$ & $-0.084$ & $-0.180$ & $-0.150$ & $-0.134$ & $-0.173$ & $-0.122$ \\
Qwen3 & $-0.209$ & $-0.219$ & $-0.238$ & $-0.124$ & $-0.173$ & $-0.232$ & $-0.223$ & $-0.217$ & $-0.220$ & $-0.185$ \\
Nomic v2 MoE & $-0.182$ & $-0.185$ & $-0.170$ & $-0.126$ & $-0.098$ & $-0.267$ & $-0.211$ & $-0.222$ & $-0.255$ & $-0.156$ \\
GPT2 & $-0.017$ & $-0.014$$^{\dagger}$ & $-0.051$ & $0.047$ & $-0.011$$^{\dagger}$ & $-0.307$ & $-0.130$ & $-0.178$ & $-0.237$ & $-0.071$ \\
\midrule
\textbf{Mean across models} & $-0.085$ & $-0.091$ & $-0.117$ & $-0.052$ & $-0.099$ & $-0.252$ & $-0.146$ & $-0.181$ & $-0.196$ & $-0.106$ \\
\bottomrule
\end{tabular}\endgroup}

\newcommand{\TabTextComposition}{%
\begingroup\scriptsize\setlength{\tabcolsep}{3pt}%
\begin{tabular}{@{}p{5.6cm}cccccc@{}}
\toprule
 & \multicolumn{3}{c}{Words (12 models)} & \multicolumn{3}{c}{Phrases (12 models)} \\
\cmidrule(lr){2-4}\cmidrule(lr){5-7}
Specification & Mean $\beta$ & 95\% CI & Neg. (sig.) & Mean $\beta$ & 95\% CI & Neg. (sig.) \\
\midrule
All terms (as in Fig.~1a) & $-0.085$ & [$-0.131$, $-0.039$] & 11/12 (11) & $-0.252$ & [$-0.295$, $-0.210$] & 12/12 (12) \\
Excluding person nouns, lexically gendered terms, function words and names & $-0.092$ & [$-0.147$, $-0.037$] & 11/12 (10) & $-0.269$ & [$-0.317$, $-0.222$] & 12/12 (12) \\
As above, controlling for frequency, length and part of speech & $-0.103$ & [$-0.160$, $-0.046$] & 10/12 (8) & $-0.126$ & [$-0.184$, $-0.068$] & 10/12 (10) \\
Excluding the 5\% most strongly gendered terms in each model & $-0.072$ & [$-0.122$, $-0.022$] & 10/12 (10) & $-0.258$ & [$-0.302$, $-0.215$] & 12/12 (12) \\
Excluding the 10\% most strongly gendered terms in each model & $-0.068$ & [$-0.116$, $-0.019$] & 10/12 (10) & $-0.246$ & [$-0.288$, $-0.205$] & 12/12 (12) \\
Excluding person, gendered and function words and the 5\% most gendered & $-0.079$ & [$-0.133$, $-0.026$] & 10/12 (9) & $-0.262$ & [$-0.309$, $-0.216$] & 12/12 (12) \\
Spearman rank correlation, all terms & $-0.081$ & [$-0.131$, $-0.030$] & 10/12 (10) & $-0.273$ & [$-0.318$, $-0.229$] & 12/12 (12) \\
Leave-one-domain-out: least negative estimate for each model & $-0.067$ & [$-0.112$, $-0.021$] & 11/12 (9) & $-0.243$ & [$-0.285$, $-0.201$] & 12/12 (12) \\
\bottomrule
\end{tabular}\endgroup}

%% file: references.bib
@article{rosch1976basic,
  author  = {Rosch, Eleanor and Mervis, Carolyn B. and Gray, Wayne D. and Johnson, David M. and Boyes-Braem, Penny},
  title   = {Basic Objects in Natural Categories},
  journal = {Cogn. Psychol.},
  year    = {1976},
  volume  = {8},
  number  = {3},
  pages   = {382--439},
  doi     = {10.1016/0010-0285(76)90013-X}
}

@incollection{rosch1978principles,
  author    = {Rosch, Eleanor},
  title     = {Principles of Categorization},
  booktitle = {Cognition and Categorization},
  editor    = {Rosch, Eleanor and Lloyd, Barbara B.},
  publisher = {Lawrence Erlbaum Associates},
  address   = {Hillsdale, NJ},
  year      = {1978},
  pages     = {27--48}
}

@article{murphy1982basic,
  author  = {Murphy, Gregory L. and Smith, Edward E.},
  title   = {Basic-Level Superiority in Picture Categorization},
  journal = {J. Verbal Learn. Verbal Behav.},
  year    = {1982},
  volume  = {21},
  number  = {1},
  pages   = {1--20}
}

@article{tenenbaum2011,
  author  = {Tenenbaum, Joshua B. and Kemp, Charles and Griffiths, Thomas L. and Goodman, Noah D.},
  title   = {How to Grow a Mind: Statistics, Structure, and Abstraction},
  journal = {Science},
  year    = {2011},
  volume  = {331},
  number  = {6022},
  pages   = {1279--1285},
  doi     = {10.1126/science.1192788}
}

@article{lawson2022hiring,
  title   = {Hiring women into senior leadership positions is associated with a reduction in gender stereotypes in organizational language},
  author  = {Lawson, M. Asher and Martin, Ashley E. and Huda, Imrul and Matz, Sandra C.},
  journal = {Proc. Natl Acad. Sci. USA},
  volume  = {119},
  number  = {9},
  pages   = {e2026443119},
  year    = {2022},
  doi     = {10.1073/pnas.2026443119}
}

@book{griffiths2026,
  author    = {Griffiths, Thomas L.},
  title     = {The Laws of Thought: The Quest for a Mathematical Theory of the Mind},
  publisher = {Henry Holt and Company},
  address   = {New York},
  year      = {2026},
  isbn      = {9781250358356}
}

@article{corter1992explaining,
  author  = {Corter, James E. and Gluck, Mark A.},
  title   = {Explaining Basic Categories: Feature Predictability and Information},
  journal = {Psychol. Bull.},
  year    = {1992},
  volume  = {111},
  number  = {2},
  pages   = {291--303}
}

@article{iordan2015basic,
  author  = {Iordan, Marius Catalin and Greene, Michelle R. and Beck, Diane M. and Fei-Fei, Li},
  title   = {Basic Level Category Structure Emerges Gradually across Human Ventral Visual Cortex},
  journal = {J. Cogn. Neurosci.},
  year    = {2015},
  volume  = {27},
  number  = {7},
  pages   = {1427--1446}
}

@book{murphy2002,
  author    = {Murphy, G. L.},
  title     = {The Big Book of Concepts},
  publisher = {MIT Press},
  year      = {2002}
}

@article{borghi2025,
  author  = {Borghi, A. M. and Mazzuca, C. and Tummolini, L.},
  title   = {The role of social interaction in the formation and use of abstract concepts},
  journal = {Nat. Rev. Psychol.},
  volume  = {4},
  pages   = {470--483},
  year    = {2025}
}

@article{wood2002,
  author  = {Wood, W. and Eagly, A. H.},
  title   = {A cross-cultural analysis of the behavior of women and men: implications for the origins of sex differences},
  journal = {Psychol. Bull.},
  volume  = {128},
  pages   = {699--727},
  year    = {2002}
}

@book{ridgeway2011,
  author    = {Ridgeway, C. L.},
  title     = {Framed by Gender: How Gender Inequality Persists in the Modern World},
  publisher = {Oxford Univ. Press},
  year      = {2011}
}

@article{martin2021,
  author  = {Martin, A. E. and Slepian, M. L.},
  title   = {The primacy of gender: gendered cognition underlies the {Big Two} dimensions of social cognition},
  journal = {Perspect. Psychol. Sci.},
  volume  = {16},
  pages   = {1143--1158},
  year    = {2021}
}

@article{bian2017,
  author  = {Bian, L. and Leslie, S.-J. and Cimpian, A.},
  title   = {Gender stereotypes about intellectual ability emerge early and influence children's interests},
  journal = {Science},
  volume  = {355},
  pages   = {389--391},
  year    = {2017}
}

@article{plant2000,
  author  = {Plant, E. A. and Hyde, J. S. and Keltner, D. and Devine, P. G.},
  title   = {The gender stereotyping of emotions},
  journal = {Psychol. Women Q.},
  volume  = {24},
  pages   = {81--92},
  year    = {2000}
}

@article{wardle2022,
  author  = {Wardle, S. G. and Paranjape, S. and Taubert, J. and Baker, C. I.},
  title   = {Illusory faces are more likely to be perceived as male than female},
  journal = {Proc. Natl Acad. Sci. USA},
  volume  = {119},
  pages   = {e2117413119},
  year    = {2022}
}

@article{collins1969,
  author  = {Collins, A. M. and Quillian, M. R.},
  title   = {Retrieval time from semantic memory},
  journal = {J. Verbal Learn. Verbal Behav.},
  volume  = {8},
  pages   = {240--247},
  year    = {1969}
}

@article{rosch1976,
  author  = {Rosch, E. and Mervis, C. B. and Gray, W. D. and Johnson, D. M. and Boyes-Braem, P.},
  title   = {Basic objects in natural categories},
  journal = {Cogn. Psychol.},
  volume  = {8},
  pages   = {382--439},
  year    = {1976}
}

@article{bolognesi2020,
  author  = {Bolognesi, M. and Burgers, C. and Caselli, T.},
  title   = {On abstraction: decoupling conceptual concreteness and categorical specificity},
  journal = {Cogn. Process.},
  volume  = {21},
  pages   = {365--381},
  year    = {2020}
}

@article{muraki2026,
  author  = {Muraki, E. J. and Pexman, P. M.},
  title   = {Distinguishing abstraction from abstractness: specificity norms for 8,500 {English} words},
  journal = {Behav. Res. Methods},
  volume  = {58},
  pages   = {60},
  year    = {2026}
}

@article{iliev2017,
  author  = {Iliev, R. and Axelrod, R.},
  title   = {The paradox of abstraction: precision versus concreteness},
  journal = {J. Psycholinguist. Res.},
  volume  = {46},
  pages   = {715--729},
  year    = {2017}
}

@article{brysbaert2014,
  author  = {Brysbaert, M. and Warriner, A. B. and Kuperman, V.},
  title   = {Concreteness ratings for 40 thousand generally known {English} word lemmas},
  journal = {Behav. Res. Methods},
  volume  = {46},
  pages   = {904--911},
  year    = {2014}
}

@book{paivio1986,
  author    = {Paivio, A.},
  title     = {Mental Representations: A Dual Coding Approach},
  publisher = {Oxford Univ. Press},
  year      = {1986}
}

@incollection{barsalou2005,
  author    = {Barsalou, L. W. and Wiemer-Hastings, K.},
  title     = {Situating abstract concepts},
  booktitle = {Grounding Cognition: The Role of Perception and Action in Memory, Language, and Thought},
  editor    = {Pecher, D. and Zwaan, R. A.},
  pages     = {129--163},
  publisher = {Cambridge Univ. Press},
  year      = {2005}
}

@article{borghi2017,
  author  = {Borghi, A. M. and Binkofski, F. and Castelfranchi, C. and Cimatti, F. and Scorolli, C. and Tummolini, L.},
  title   = {The challenge of abstract concepts},
  journal = {Psychol. Bull.},
  volume  = {143},
  pages   = {263--292},
  year    = {2017}
}

@article{pollock2018,
  author  = {Pollock, L.},
  title   = {Statistical and methodological problems with concreteness and other semantic variables: a list memory experiment case study},
  journal = {Behav. Res. Methods},
  volume  = {50},
  pages   = {1198--1216},
  year    = {2018}
}

@article{brysbaert2014dutch,
  author  = {Brysbaert, M. and Stevens, M. and De Deyne, S. and Voorspoels, W. and Storms, G.},
  title   = {Norms of age of acquisition and concreteness for 30,000 {Dutch} words},
  journal = {Acta Psychol.},
  volume  = {150},
  pages   = {80--84},
  year    = {2014}
}

@inproceedings{thompson2018,
  author    = {Thompson, B. and Lupyan, G.},
  title     = {Automatic estimation of lexical concreteness in 77 languages},
  booktitle = {Proc. 40th Annual Conference of the Cognitive Science Society},
  pages     = {1122--1127},
  publisher = {Cognitive Science Society},
  year      = {2018}
}

@article{martinez2025,
  author  = {Mart{\'i}nez, G. and Molero, J. D. and Gonz{\'a}lez, S. and Conde, J. and Brysbaert, M. and Reviriego, P.},
  title   = {Using large language models to estimate features of multi-word expressions: concreteness, valence, arousal},
  journal = {Behav. Res. Methods},
  volume  = {57},
  pages   = {5},
  year    = {2025}
}

@article{gleitman2005,
  author  = {Gleitman, L. R. and Cassidy, K. and Nappa, R. and Papafragou, A. and Trueswell, J. C.},
  title   = {Hard words},
  journal = {Lang. Learn. Dev.},
  volume  = {1},
  pages   = {23--64},
  year    = {2005}
}

@article{schwanenflugel1983,
  author  = {Schwanenflugel, P. J. and Shoben, E. J.},
  title   = {Differential context effects in the comprehension of abstract and concrete verbal materials},
  journal = {J. Exp. Psychol. Learn. Mem. Cogn.},
  volume  = {9},
  pages   = {82--102},
  year    = {1983}
}

@article{kousta2011,
  author  = {Kousta, S.-T. and Vigliocco, G. and Vinson, D. P. and Andrews, M. and Del Campo, E.},
  title   = {The representation of abstract words: why emotion matters},
  journal = {J. Exp. Psychol. Gen.},
  volume  = {140},
  pages   = {14--34},
  year    = {2011}
}

@book{lakoff1980,
  author    = {Lakoff, G. and Johnson, M.},
  title     = {Metaphors We Live By},
  publisher = {Univ. Chicago Press},
  year      = {1980}
}

@article{jamrozik2016,
  author  = {Jamrozik, A. and McQuire, M. and Cardillo, E. R. and Chatterjee, A.},
  title   = {Metaphor: bridging embodiment to abstraction},
  journal = {Psychon. Bull. Rev.},
  volume  = {23},
  pages   = {1080--1089},
  year    = {2016}
}

@article{hansen2010,
  author  = {Hansen, J. and W{\"a}nke, M.},
  title   = {Truth from language and truth from fit: the impact of linguistic concreteness and level of construal on subjective truth},
  journal = {Pers. Soc. Psychol. Bull.},
  volume  = {36},
  pages   = {1576--1588},
  year    = {2010}
}

@article{packard2021,
  author  = {Packard, G. and Berger, J.},
  title   = {How concrete language shapes customer satisfaction},
  journal = {J. Consum. Res.},
  volume  = {47},
  pages   = {787--806},
  year    = {2021}
}

@article{eagly1984,
  author  = {Eagly, A. H. and Steffen, V. J.},
  title   = {Gender stereotypes stem from the distribution of women and men into social roles},
  journal = {J. Pers. Soc. Psychol.},
  volume  = {46},
  pages   = {735--754},
  year    = {1984}
}

@article{fiske2002,
  author  = {Fiske, S. T. and Cuddy, A. J. C. and Glick, P. and Xu, J.},
  title   = {A model of (often mixed) stereotype content: competence and warmth respectively follow from perceived status and competition},
  journal = {J. Pers. Soc. Psychol.},
  volume  = {82},
  pages   = {878--902},
  year    = {2002}
}

@article{ellemers2018,
  author  = {Ellemers, N.},
  title   = {Gender stereotypes},
  journal = {Annu. Rev. Psychol.},
  volume  = {69},
  pages   = {275--298},
  year    = {2018}
}

@article{leshin2025,
  author  = {Leshin, R. A. and Rhodes, M.},
  title   = {Girls as objects, boys as humans: young children tend to be objectified along gender lines},
  journal = {J. Pers. Soc. Psychol.},
  volume  = {129},
  pages   = {266--285},
  year    = {2025}
}

@article{leslie2015,
  author  = {Leslie, S.-J. and Cimpian, A. and Meyer, M. and Freeland, E.},
  title   = {Expectations of brilliance underlie gender distributions across academic disciplines},
  journal = {Science},
  volume  = {347},
  pages   = {262--265},
  year    = {2015}
}

@article{nosek2002,
  author  = {Nosek, B. A. and Banaji, M. R. and Greenwald, A. G.},
  title   = {Math = male, me = female, therefore math $\neq$ me},
  journal = {J. Pers. Soc. Psychol.},
  volume  = {83},
  pages   = {44--59},
  year    = {2002}
}

@article{hyde2008,
  author  = {Hyde, J. S. and Lindberg, S. M. and Linn, M. C. and Ellis, A. B. and Williams, C. C.},
  title   = {Gender similarities characterize math performance},
  journal = {Science},
  volume  = {321},
  pages   = {494--495},
  year    = {2008}
}

@article{mossracusin2012,
  author  = {Moss-Racusin, C. A. and Dovidio, J. F. and Brescoll, V. L. and Graham, M. J. and Handelsman, J.},
  title   = {Science faculty's subtle gender biases favor male students},
  journal = {Proc. Natl Acad. Sci. USA},
  volume  = {109},
  pages   = {16474--16479},
  year    = {2012}
}

@article{trope2010,
  author  = {Trope, Y. and Liberman, N.},
  title   = {Construal-level theory of psychological distance},
  journal = {Psychol. Rev.},
  volume  = {117},
  pages   = {440--463},
  year    = {2010}
}

@article{semin1988,
  author  = {Semin, G. R. and Fiedler, K.},
  title   = {The cognitive functions of linguistic categories in describing persons: social cognition and language},
  journal = {J. Pers. Soc. Psychol.},
  volume  = {54},
  pages   = {558--568},
  year    = {1988}
}

@article{joshi2020,
  author  = {Joshi, P. D. and Wakslak, C. J. and Appel, G. and Huang, L.},
  title   = {Gender differences in communicative abstraction},
  journal = {J. Pers. Soc. Psychol.},
  volume  = {118},
  pages   = {417--435},
  year    = {2020}
}

@article{bailey2026,
  author  = {Bailey, A. H. and Leshin, R. A.},
  title   = {People think of women as one thing, men as many},
  journal = {Trends Cogn. Sci.},
  year    = {2026},
  doi     = {10.1016/j.tics.2026.02.007},
  url     = {https://doi.org/10.1016/j.tics.2026.02.007}
}

@article{bailey2022,
  author  = {Bailey, A. H. and Williams, A. and Cimpian, A.},
  title   = {Based on billions of words on the internet, people = men},
  journal = {Sci. Adv.},
  volume  = {8},
  pages   = {eabm2463},
  year    = {2022}
}

@article{dodson2025,
  author  = {Dodson, S. J. and Goodwin, R. D. and Wakslak, C. J. and Diekmann, K. A. and Graham, J.},
  title   = {She sees the trees, he sees the forest: descriptive gender stereotypes of concreteness and abstractness},
  journal = {J. Pers. Soc. Psychol.},
  volume  = {129},
  pages   = {1054--1082},
  year    = {2025}
}

@article{barsalou1999,
  author  = {Barsalou, L. W.},
  title   = {Perceptual symbol systems},
  journal = {Behav. Brain Sci.},
  volume  = {22},
  pages   = {577--660},
  year    = {1999}
}

@article{barsalou2008,
  author  = {Barsalou, L. W.},
  title   = {Grounded cognition},
  journal = {Annu. Rev. Psychol.},
  volume  = {59},
  pages   = {617--645},
  year    = {2008}
}

@article{guilbeault2024,
  author  = {Guilbeault, D. and Delecourt, S. and Hull, T. and Desikan, B. S. and Chu, M. and Nadler, E.},
  title   = {Online images amplify gender bias},
  journal = {Nature},
  volume  = {626},
  pages   = {1049--1055},
  year    = {2024}
}

@article{vlasceanu2022,
  author  = {Vlasceanu, M. and Amodio, D. M.},
  title   = {Propagation of societal gender inequality by internet search algorithms},
  journal = {Proc. Natl Acad. Sci. USA},
  volume  = {119},
  pages   = {e2204529119},
  year    = {2022}
}

@article{guilbeault2025,
  author  = {Guilbeault, D. and Delecourt, S. and {Srinivasa Desikan}, B.},
  title   = {Age and gender distortion in online media and large language models},
  journal = {Nature},
  volume  = {646},
  pages   = {1129--1137},
  year    = {2025}
}

@article{caliskan2017,
  author  = {Caliskan, A. and Bryson, J. J. and Narayanan, A.},
  title   = {Semantics derived automatically from language corpora contain human-like biases},
  journal = {Science},
  volume  = {356},
  pages   = {183--186},
  year    = {2017}
}

@article{garg2018,
  author  = {Garg, N. and Schiebinger, L. and Jurafsky, D. and Zou, J.},
  title   = {Word embeddings quantify 100 years of gender and ethnic stereotypes},
  journal = {Proc. Natl Acad. Sci. USA},
  volume  = {115},
  pages   = {E3635--E3644},
  year    = {2018}
}

@article{lewis2020,
  author  = {Lewis, M. and Lupyan, G.},
  title   = {Gender stereotypes are reflected in the distributional structure of 25 languages},
  journal = {Nat. Hum. Behav.},
  volume  = {4},
  pages   = {1021--1028},
  year    = {2020}
}

@article{charlesworth2021,
  author  = {Charlesworth, T. E. S. and Yang, V. and Mann, T. C. and Kurdi, B. and Banaji, M. R.},
  title   = {Gender stereotypes in natural language: word embeddings show robust consistency across child and adult language corpora of more than 65 million words},
  journal = {Psychol. Sci.},
  volume  = {32},
  pages   = {218--240},
  year    = {2021}
}

@article{dedeyne2019,
  author  = {De Deyne, S. and Navarro, D. J. and Perfors, A. and Brysbaert, M. and Storms, G.},
  title   = {The ``{Small World of Words}'' {English} word association norms for over 12,000 cue words},
  journal = {Behav. Res. Methods},
  volume  = {51},
  pages   = {987--1006},
  year    = {2019}
}

@article{abramski2025,
  author  = {Abramski, K. and Improta, R. and Rossetti, G. and Stella, M.},
  title   = {The ``{LLM World of Words}'' {English} free association norms generated by large language models},
  journal = {Sci. Data},
  volume  = {12},
  pages   = {803},
  year    = {2025}
}

@article{kozlowski2019,
  author  = {Kozlowski, A. C. and Taddy, M. and Evans, J. A.},
  title   = {The geometry of culture: analyzing the meanings of class through word embeddings},
  journal = {Am. Sociol. Rev.},
  volume  = {84},
  pages   = {905--949},
  year    = {2019}
}

@article{muraki2023,
  author  = {Muraki, E. J. and Abdalla, S. and Brysbaert, M. and Pexman, P. M.},
  title   = {Concreteness ratings for 62,000 {English} multiword expressions},
  journal = {Behav. Res. Methods},
  volume  = {55},
  pages   = {2522--2531},
  year    = {2023}
}

@article{lynott2020,
  author  = {Lynott, D. and Connell, L. and Brysbaert, M. and Brand, J. and Carney, J.},
  title   = {The {Lancaster Sensorimotor Norms}: multidimensional measures of perceptual and action strength for 40,000 {English} words},
  journal = {Behav. Res. Methods},
  volume  = {52},
  pages   = {1271--1291},
  year    = {2020}
}

@article{miller1995,
  author  = {Miller, G. A.},
  title   = {{WordNet}: a lexical database for {English}},
  journal = {Commun. ACM},
  volume  = {38},
  pages   = {39--41},
  year    = {1995}
}

@book{lakoff1989more,
  author    = {Lakoff, George and Turner, Mark},
  title     = {More than Cool Reason: A Field Guide to Poetic Metaphor},
  publisher = {University of Chicago Press},
  address   = {Chicago},
  year      = {1989},
  doi       = {10.7208/chicago/9780226470986.001.0001}
}

@article{merton1948selffulfilling,
  author  = {Merton, Robert K.},
  title   = {The Self-Fulfilling Prophecy},
  journal = {Antioch Rev.},
  year    = {1948},
  volume  = {8},
  number  = {2},
  pages   = {193--210},
  doi     = {10.2307/4609267}
}

@misc{guilbeault2026review,
  author = {Guilbeault, D.},
  title  = {TITLE OF SOCARXIV REVIEW PREPRINT},
  url    = {https://osf.io/preprints/socarxiv/ujnkx_v1},
  year   = {2026}
}

@article{chen2024,
  author  = {Chen, Y. and Zhai, Y. and Sun, S.},
  title   = {The gendered lens of {AI}: examining news imagery across digital spaces},
  journal = {J. Comput.-Mediat. Commun.},
  volume  = {29},
  pages   = {zmad047},
  year    = {2024}
}

@article{han2025,
  author  = {Han, H. and Lee, Y. and Zhang, C. and Lu, J. and Wang, L.},
  title   = {He or she? A male-default bias in chatbot gender attribution across explicit and implicit measures},
  journal = {Comput. Hum. Behav. Rep.},
  volume  = {20},
  pages   = {100860},
  year    = {2025}
}

@misc{wongkim2023,
  author = {Wong, J. and Kim, J.},
  title  = {{ChatGPT} is more likely to be perceived as male than female},
  url    = {https://arxiv.org/abs/2305.12564},
  year   = {2023}
}

@book{adam1998,
  author    = {Adam, A.},
  title     = {Artificial Knowing: Gender and the Thinking Machine},
  publisher = {Routledge},
  year      = {1998}
}

@article{cavedihal2020,
  author  = {Cave, S. and Dihal, K.},
  title   = {The whiteness of {AI}},
  journal = {Philos. Technol.},
  volume  = {33},
  pages   = {685--703},
  year    = {2020}
}

@article{craiut2022,
  author  = {Craiut, M.-V. and Iancu, I. R.},
  title   = {Is technology gender neutral? A systematic literature review on gender stereotypes attached to artificial intelligence},
  journal = {Hum. Technol.},
  volume  = {18},
  pages   = {297--315},
  year    = {2022}
}

@book{bakan1966,
  author    = {Bakan, D.},
  title     = {The Duality of Human Existence: An Essay on Psychology and Religion},
  publisher = {Rand McNally},
  year      = {1966}
}

@article{abele2014,
  author  = {Abele, A. E. and Wojciszke, B.},
  title   = {Communal and agentic content in social cognition: a dual perspective model},
  journal = {Adv. Exp. Soc. Psychol.},
  volume  = {50},
  pages   = {195--255},
  year    = {2014}
}

@book{lloyd1984,
  author    = {Lloyd, G.},
  title     = {The Man of Reason: `Male' and `Female' in {Western} Philosophy},
  publisher = {Methuen},
  year      = {1984}
}

@book{schiebinger1989,
  author    = {Schiebinger, L.},
  title     = {The Mind Has No Sex? Women in the Origins of Modern Science},
  publisher = {Harvard Univ. Press},
  year      = {1989}
}

@incollection{ortner1974,
  author    = {Ortner, S. B.},
  title     = {Is female to male as nature is to culture?},
  booktitle = {Woman, Culture, and Society},
  editor    = {Rosaldo, M. Z. and Lamphere, L.},
  pages     = {68--87},
  publisher = {Stanford Univ. Press},
  year      = {1974}
}

@book{bourdieu2001,
  author    = {Bourdieu, P.},
  title     = {Masculine Domination},
  publisher = {Stanford Univ. Press},
  year      = {2001}
}

@book{varela1991,
  author    = {Varela, F. J. and Thompson, E. and Rosch, E.},
  title     = {The Embodied Mind: Cognitive Science and Human Experience},
  publisher = {MIT Press},
  year      = {1991}
}

@book{thelensmith1994,
  author    = {Thelen, E. and Smith, L. B.},
  title     = {A Dynamic Systems Approach to the Development of Cognition and Action},
  publisher = {MIT Press},
  year      = {1994}
}

@article{nadler2025,
  author  = {Nadler, E. O. and Guilbeault, D. and Ringold, S. M. and Williamson, T. R. and Bellemare-Pepin, A. and Com{\c{s}}a, I. M. and Jerbi, K. and Narayanan, S. and Aziz-Zadeh, L.},
  title   = {Statistical or embodied? {C}omparing colorseeing, colorblind, painters, and large language models in their processing of color metaphors},
  journal = {Cogn. Sci.},
  volume  = {49},
  pages   = {e70083},
  year    = {2025}
}

@article{guilbeault2020,
  author  = {Guilbeault, D. and Nadler, E. O. and Chu, M. and {Lo Sardo}, D. R. and Kar, A. A. and Desikan, B. S.},
  title   = {Color associations in abstract semantic space},
  journal = {Cognition},
  volume  = {201},
  pages   = {104306},
  year    = {2020}
}

@article{cheryan2017,
  author  = {Cheryan, S. and Ziegler, S. A. and Montoya, A. K. and Jiang, L.},
  title   = {Why are some {STEM} fields more gender balanced than others?},
  journal = {Psychol. Bull.},
  volume  = {143},
  pages   = {1--35},
  year    = {2017}
}

@book{lerner1986,
  author    = {Lerner, G.},
  title     = {The Creation of Patriarchy},
  publisher = {Oxford Univ. Press},
  year      = {1986}
}

@incollection{boroditsky2003,
  author    = {Boroditsky, L. and Schmidt, L. A. and Phillips, W.},
  title     = {Sex, syntax, and semantics},
  booktitle = {Language in Mind: Advances in the Study of Language and Thought},
  editor    = {Gentner, D. and Goldin-Meadow, S.},
  pages     = {61--79},
  publisher = {MIT Press},
  year      = {2003}
}

@article{stuhler2024gender,
  author  = {Stuhler, Oscar},
  title   = {The gender agency gap in fiction writing (1850 to 2010)},
  journal = {Proc. Natl Acad. Sci. USA},
  year    = {2024},
  volume  = {121},
  number  = {29},
  pages   = {e2319514121},
  doi     = {10.1073/pnas.2319514121}
}

@article{eagly1984gender,
  author  = {Eagly, Alice H. and Steffen, Valerie J.},
  title   = {Gender stereotypes stem from the distribution of women and men into social roles},
  journal = {J. Pers. Soc. Psychol.},
  year    = {1984},
  volume  = {46},
  number  = {4},
  pages   = {735--754},
  doi     = {10.1037/0022-3514.46.4.735}
}

@article{diekman2000stereotypes,
  author  = {Diekman, Amanda B. and Eagly, Alice H.},
  title   = {Stereotypes as dynamic constructs: Women and men of the past, present, and future},
  journal = {Pers. Soc. Psychol. Bull.},
  year    = {2000},
  volume  = {26},
  number  = {10},
  pages   = {1171--1188},
  doi     = {10.1177/0146167200262001}
}

@article{rudman2001prescriptive,
  author  = {Rudman, Laurie A. and Glick, Peter},
  title   = {Prescriptive gender stereotypes and backlash toward agentic women},
  journal = {J. Soc. Issues},
  year    = {2001},
  volume  = {57},
  number  = {4},
  pages   = {743--762},
  doi     = {10.1111/0022-4537.00239}
}

@article{abele2007agency,
  author  = {Abele, Andrea E. and Wojciszke, Bogdan},
  title   = {Agency and communion from the perspective of self versus others},
  journal = {J. Pers. Soc. Psychol.},
  year    = {2007},
  volume  = {93},
  number  = {5},
  pages   = {751--763},
  doi     = {10.1037/0022-3514.93.5.751}
}

@article{fedorenko2024language,
  author  = {Fedorenko, Evelina and Piantadosi, Steven T. and Gibson, Edward A. F.},
  title   = {Language is primarily a tool for communication rather than thought},
  journal = {Nature},
  year    = {2024},
  volume  = {630},
  number  = {8017},
  pages   = {575--586},
  doi     = {10.1038/s41586-024-07522-w}
}

@misc{meta2024llama32,
  author       = {{Meta AI}},
  title        = {Llama 3.2: Revolutionizing edge {AI} and vision with open, customizable models},
  year         = {2024},
  month        = sep,
  howpublished = {Meta AI Blog},
  url          = {https://ai.meta.com/blog/llama-3-2-connect-2024-vision-edge-mobile-devices/},
  note         = {Model: \texttt{meta-llama/Llama-3.2-11B-Vision-Instruct}. Model card: \url{https://github.com/meta-llama/llama-models/blob/main/models/llama3_2/MODEL_CARD_VISION.md}}
}

@article{grattafiori2024llama3,
  author  = {Grattafiori, Aaron and Dubey, Abhimanyu and Jauhri, Abhinav and Pandey, Abhinav and Kadian, Abhishek and Al-Dahle, Ahmad and Letman, Aiesha and Mathur, Akhil and Schelten, Alan and Vaughan, Alex and others},
  title   = {The {Llama} 3 Herd of Models},
  journal = {arXiv preprint arXiv:2407.21783},
  year    = {2024},
  doi     = {10.48550/arXiv.2407.21783}
}

@article{koenig2011leader,
  author  = {Koenig, Anne M. and Eagly, Alice H. and Mitchell, Abigail A. and Ristikari, Tiina},
  title   = {Are leader stereotypes masculine? {A} meta-analysis of three research paradigms},
  journal = {Psychol. Bull.},
  year    = {2011},
  volume  = {137},
  number  = {4},
  pages   = {616--642},
  doi     = {10.1037/a0023557}
}

@article{hentschel2019multiple,
  author  = {Hentschel, Tanja and Heilman, Madeline E. and Peus, Claudia V.},
  title   = {The multiple dimensions of gender stereotypes: A current look at men's and women's characterizations of others and themselves},
  journal = {Front. Psychol.},
  year    = {2019},
  volume  = {10},
  pages   = {11},
  doi     = {10.3389/fpsyg.2019.00011}
}

@article{prewitt2012,
  author  = {Prewitt-Freilino, J. L. and Caswell, T. A. and Laakso, E. K.},
  title   = {The gendering of language: a comparison of gender equality in countries with gendered, natural gender, and genderless languages},
  journal = {Sex Roles},
  volume  = {66},
  pages   = {268--281},
  year    = {2012}
}

@article{cave2023,
  author  = {Cave, Stephen and Dihal, Kanta and Drage, Eleanor and McInerney, Kerry},
  title   = {Who makes {AI}? Gender and portrayals of {AI} scientists in popular film, 1920--2020},
  journal = {Public Underst. Sci.},
  volume  = {32},
  number  = {6},
  pages   = {745--760},
  year    = {2023},
  doi     = {10.1177/09636625231153985}
}

@techreport{unesco2019,
  author      = {West, Mark and Kraut, Rebecca and Chew, Han Ei},
  title       = {I'd Blush If {I} Could: Closing Gender Divides in Digital Skills through Education},
  institution = {UNESCO and EQUALS Skills Coalition},
  address     = {Paris},
  year        = {2019},
  url         = {https://unesdoc.unesco.org/ark:/48223/pf0000367416}
}

@inproceedings{feine2020,
  author    = {Feine, Jasper and Gnewuch, Ulrich and Morana, Stefan and Maedche, Alexander},
  title     = {Gender Bias in Chatbot Design},
  booktitle = {Chatbot Research and Design: Third International Workshop, CONVERSATIONS 2019},
  series    = {Lecture Notes in Computer Science},
  pages     = {79--93},
  publisher = {Springer},
  address   = {Cham},
  year      = {2020},
  doi       = {10.1007/978-3-030-39540-7_6}
}

@article{woods2018,
  author  = {Woods, Heather Suzanne},
  title   = {Asking more of {Siri} and {Alexa}: feminine persona in service of surveillance capitalism},
  journal = {Crit. Stud. Media Commun.},
  volume  = {35},
  number  = {4},
  pages   = {334--349},
  year    = {2018},
  doi     = {10.1080/15295036.2018.1488082}
}

@book{strengers2020,
  author    = {Strengers, Yolande and Kennedy, Jenny},
  title     = {The Smart Wife: Why {Siri}, {Alexa}, and Other Smart Home Devices Need a Feminist Reboot},
  publisher = {MIT Press},
  address   = {Cambridge, MA},
  year      = {2020},
  doi       = {10.7551/mitpress/12482.001.0001}
}

@article{borau2021,
  author  = {Borau, Sylvie and Otterbring, Tobias and Laporte, Sandra and Fosso Wamba, Samuel},
  title   = {The most human bot: Female gendering increases humanness perceptions of bots and acceptance of {AI}},
  journal = {Psychol. Mark.},
  volume  = {38},
  number  = {7},
  pages   = {1052--1068},
  year    = {2021},
  doi     = {10.1002/mar.21480}
}

@article{bazazi2025,
  author  = {Bazazi, Sepideh and Karpus, Jurgis and Yasseri, Taha},
  title   = {{AI}'s assigned gender affects human-{AI} cooperation},
  journal = {iScience},
  volume  = {28},
  number  = {12},
  pages   = {113905},
  year    = {2025},
  doi     = {10.1016/j.isci.2025.113905}
}

@article{kuperman2012,
  author  = {Kuperman, V. and Stadthagen-Gonzalez, H. and Brysbaert, M.},
  title   = {Age-of-acquisition ratings for 30,000 {English} words},
  journal = {Behav. Res. Methods},
  volume  = {44},
  pages   = {978--990},
  year    = {2012},
  doi     = {10.3758/s13428-012-0210-4}
}

@article{warriner2013,
  author  = {Warriner, A. B. and Kuperman, V. and Brysbaert, M.},
  title   = {Norms of valence, arousal, and dominance for 13,915 {English} lemmas},
  journal = {Behav. Res. Methods},
  volume  = {45},
  pages   = {1191--1207},
  year    = {2013},
  doi     = {10.3758/s13428-012-0314-x}
}

@article{proos2023,
  author  = {Proos, M. and Aigro, M.},
  title   = {Concreteness ratings for 36,000 {Estonian} words},
  journal = {Behav. Res. Methods},
  volume  = {56},
  pages   = {5178--5189},
  year    = {2024},
  doi     = {10.3758/s13428-023-02257-4}
}

@article{xu2020,
  author  = {Xu, X. and Li, J.},
  title   = {Concreteness/abstractness ratings for two-character {Chinese} words in {MELD-SCH}},
  journal = {PLoS ONE},
  volume  = {15},
  pages   = {e0232133},
  year    = {2020},
  doi     = {10.1371/journal.pone.0232133}
}

@inproceedings{bolukbasi2016,
  author    = {Bolukbasi, T. and Chang, K.-W. and Zou, J. Y. and Saligrama, V. and Kalai, A. T.},
  title     = {Man is to computer programmer as woman is to homemaker? {Debiasing} word embeddings},
  booktitle = {Advances in Neural Information Processing Systems 29},
  pages     = {4349--4357},
  publisher = {Curran Associates},
  year      = {2016}
}

@misc{honnibal2020spacy,
  author = {Honnibal, M. and Montani, I. and Van Landeghem, S. and Boyd, A.},
  title  = {{spaCy}: industrial-strength natural language processing in {Python}},
  url    = {https://doi.org/10.5281/zenodo.1212303},
  year   = {2020}
}

@misc{speer2022wordfreq,
  author = {Speer, R.},
  title  = {rspeer/wordfreq: v3.0},
  url    = {https://doi.org/10.5281/zenodo.7199437},
  year   = {2022}
}

@inproceedings{grave2018,
  author    = {Grave, E. and Bojanowski, P. and Gupta, P. and Joulin, A. and Mikolov, T.},
  title     = {Learning word vectors for 157 languages},
  booktitle = {Proceedings of the Eleventh International Conference on Language Resources and Evaluation (LREC 2018)},
  pages     = {3483--3487},
  publisher = {European Language Resources Association},
  year      = {2018}
}

@article{chan2024,
  author  = {Chan, Y.-L. and Tse, C.-S.},
  title   = {Decoding the essence of two-character {Chinese} words: unveiling valence, arousal, concreteness, familiarity, and imageability through word norming},
  journal = {Behav. Res. Methods},
  volume  = {56},
  pages   = {7574--7601},
  year    = {2024},
  doi     = {10.3758/s13428-024-02437-w}
}

@article{song2021,
  author  = {Song, D. and Li, D.},
  title   = {Psycholinguistic norms for 3,783 two-character words in simplified {Chinese}},
  journal = {SAGE Open},
  volume  = {11},
  pages   = {21582440211054495},
  year    = {2021},
  doi     = {10.1177/21582440211054495}
}

@article{nemati2026,
  author  = {Nemati, F. and Westbury, C. and Rostami, H. and Alavi, F.},
  title   = {Extrapolated {Persian} Lexical Affect Norms ({E-PLAN}) from best--worst judgments of valence, arousal, dominance, and concreteness},
  journal = {Behav. Res. Methods},
  volume  = {58},
  pages   = {111},
  year    = {2026},
  doi     = {10.3758/s13428-026-02963-9}
}

@article{montefinese2014,
  author  = {Montefinese, M. and Ambrosini, E. and Fairfield, B. and Mammarella, N.},
  title   = {The adaptation of the {Affective} {Norms} for {English} {Words} ({ANEW}) for {Italian}},
  journal = {Behav. Res. Methods},
  volume  = {46},
  pages   = {887--903},
  year    = {2014},
  doi     = {10.3758/s13428-013-0405-3}
}

@article{bonin2018,
  author  = {Bonin, P. and M{\'e}ot, A. and Bugaiska, A.},
  title   = {Concreteness norms for 1,659 {French} words: relationships with other psycholinguistic variables and word recognition times},
  journal = {Behav. Res. Methods},
  volume  = {50},
  pages   = {2366--2387},
  year    = {2018},
  doi     = {10.3758/s13428-018-1014-y}
}

@article{goz2017,
  author  = {G{\"o}z, {\.I}. and Tekcan, A. {\.I}. and Erciyes, A. A.},
  title   = {Subjective age-of-acquisition norms for 600 {Turkish} words from four age groups},
  journal = {Behav. Res. Methods},
  volume  = {49},
  pages   = {1736--1746},
  year    = {2017},
  doi     = {10.3758/s13428-016-0817-y}
}

@article{soares2017,
  author  = {Soares, A. P. and Costa, A. S. and Machado, J. and Comesa{\~n}a, M. and Oliveira, H. M.},
  title   = {The {Minho} {Word} {Pool}: norms for imageability, concreteness, and subjective frequency for 3,800 {Portuguese} words},
  journal = {Behav. Res. Methods},
  volume  = {49},
  pages   = {1065--1081},
  year    = {2017},
  doi     = {10.3758/s13428-016-0767-4}
}

@article{sianipar2016,
  author  = {Sianipar, A. and van Groenestijn, P. and Dijkstra, T.},
  title   = {Affective meaning, concreteness, and subjective frequency norms for {Indonesian} words},
  journal = {Front. Psychol.},
  volume  = {7},
  pages   = {1907},
  year    = {2016},
  doi     = {10.3389/fpsyg.2016.01907}
}
